\documentclass[aps,nobibnotes,twocolumn,superscriptaddress,showkeys,thelongbibliography]{revtex4-2}
\usepackage{graphicx}

\usepackage{amssymb}
\usepackage{amsmath}
\usepackage{multirow}
\usepackage{placeins}
\usepackage{xcolor}

\usepackage{etoolbox}

\begin{document}

\title{Robust formation of metachronal waves 
in directional chains of phase oscillators}

\author{A. C. Quillen}
\email{alice.quillen@rochester.edu}
\affiliation{Department of Physics and Astronomy, University of Rochester, Rochester, NY 14627, USA}
%

\begin{abstract}
Biological systems can rely on collective formation of a metachronal wave in an ensemble of oscillators for locomotion and for fluid transport.   We consider one-dimensional chains of phase oscillators with nearest neighbor interactions, connected in a loop and with rotational symmetry, so each oscillator resembles every other oscillator in the chain.  Numerical integrations of the discrete phase oscillator systems and a continuum approximation  show that directional models (those that do not obey reversal symmetry), can exhibit instability to short wavelength perturbations but only in regions where the slope in phase has a particular sign. This causes short wavelength perturbations to develop that can vary the winding number that describes the sum of phase differences across the loop and the resulting metachronal wave speed.  Numerical integrations of stochastic directional phase oscillator models show that even a weak level of noise can seed instabilities that resolve into metachronal wave states.

\end{abstract}


\maketitle

\section{Introduction}

Models of interacting phase oscillators, such as the Kuramoto model, have been used to study the dynamics of synchronization in a wide variety of physical and biological systems \citep{Wiener_1958,Kuramoto_1975, Kuramoto_1987,Pikovsky_2003,Strogatz_2012}. 
The head or tail of an individual flagellum, cilium or nematode moves back and forth with respect to a mean position. This periodic motion can be described with a phase of oscillation, with the collective behavior of the system governed by interactions between neighboring individual bodies.    When the interactions are strong, all oscillators
can lock in phase and beat together in a globally synchronized pattern. 
A metachronal rhythm or metachronal wave refers to a collective state where individuals 
are undergoing periodic motions but synchronization is only local. The motions
of each individual is the same as that of their neighbors but there is a delay between
these motions, giving the appearance of a traveling wave. 
 
Perhaps the most common example of emergent traveling waves are in ciliary carpets. Hydrodynamic interactions between actively beating cilia, spontaneously result in the formation of large-scale metachronal waves \citep{chakrabarti2022multiscale}. Such organized waves are are critical for the motility of ciliated protists (such as the  \textit{Paramecium} \citep{Tamm_1972}), mucus clearance in mammalian airways \citep{Sleigh_1988,Afzelius_2004}, and for fluid transport in the brain \citep{faubel2016cilia}.   Metachronal waves can also form in concentrations  
of swimming nematodes \citep{Peshkov_2022} where they can be mediated by
steric interactions  \citep{Quillen_2021}.

What fraction of possible initial conditions would converge onto a wave-like solution? 
The set of initial conditions that converge onto a particular
solution are called its {\it basin of attraction}.
In many models of interacting phase oscillators, 
the basins of attraction for traveling wave solutions 
are  smaller than that of the synchronous
state  \citep{Wiley_2006,Tilles_2011,Denes_2019}.  
In other words, 
using an ensemble of random generated initial phases 
 for each phase oscillator, a system would be more likely to enter a synchronous rather  than a traveling wave state.  

Because many well studied models 
are more likely to enter a synchronous than a traveling wave state,
or produce waves traveling in either direction, 
they do not capture the behavior illustrated by vinegar eels \citep{Quillen_2021,Peshkov_2022}, or other systems
that exhibit metachronal waves, such as chains of cilia \citep{Niedermayer_2008},
cilia carpets \citep{Solovev_2022} or 
flagella on the surface of {\it Volvox carteri} alga colonies  \citep{Brumley_2012}.
Relevant models for these types of biological systems should exhibit a larger basin of attraction for traveling wave states than for the synchronous state.
Recently \citet{chakrabarti2022multiscale} showed that that in the continuum limit, 
interactions between cilia in a one dimensional loop lead to 
conservation of a type of topological charge or a winding number.  
The conserved quantity 
implies that initial conditions could set the wave speed of attracting solutions. 
To mitigate the role of the constraint imposed by the conserved quantity,  \citet{chakrabarti2022multiscale} proposed that irregularities or gaps in the spacing 
between cilia 
could help account for systems of cila that robustly exhibit metachronal waves. 

A  model with asymptotic behavior dependent upon initial conditions is inconvenient when
trying to model biological systems. However,  
fluctuations are likely to be present in ciliated systems (e.g., \cite{Ma_2014}).
The presence of noise could affect or even determine the statistics
of long-lived states, obviating the need to  
understand the sensitivity to initial conditions.
When coupled to a phase oscillator model for ciliated carpets, 
white noise can cause stochastic transitions between synchronized states
and disordered states  \citep{Solovev_2022}.
 
The focus of this manuscript is to explore properties of interacting phase oscillator systems that allow them to robustly enter wave-like states. 
Building upon the work by \citet{chakrabarti2022multiscale}, we investigate if 
and how model systems can exhibit changes in the winding number. 
In section \ref{sec:states} we describe states for systems
of  interacting phase oscillators. In section \ref{sec:loc} and \ref{sec:rot} we introduce
chains of interacting oscillators and describe what we mean by a directional
model.  In section \ref{sec:winding} we define how we calculate the
phase shift between neighboring oscillators (following \cite{Denes_2019}) 
and the winding number.
In section \ref{sec:cont} we find a partial differential equation
that represents the continuum limit for a loop of oscillators with nearest
neighbor interactions. 
The properties of the associated continuum equations are relevant for
 interpretation of our numerical integrations. 
In section \ref{sec:num} we numerically explore bidirectional, unidirectional
and adjustable directional models to better understand how these 
models exhibit changes in winding number.  In this section we use initial conditions
that are either sinusoidal or drawn from a uniform distribution. 
In section \ref{sec:noise} and following \citet{Solovev_2022}
who found that noise could affect the coherence of wave-like states in ciliary carpets, 
we explore numerically adjustable directional models that are
 perturbed by white noise.   We numerically explore how initial winding number
 and the number oscillators affect the integrated mean phase shift between
 neighboring oscillators and the standard deviation of the phase shifts. 
A summary and discussion follows in section \ref{sec:sum}.

\subsection{Types of states for ensembles of phase oscillators}
\label{sec:states}

We denote each phase oscillator with a non-negative integer $i$. 
The $i$-th oscillator can be described with a phase $\theta_i \in [0, 2 \pi)$ 
that is a function of time  $t$ 
and a frequency of oscillation or a phase velocity 
$\frac{d\theta_i}{dt} =\dot \theta_i = \omega_i $.

Collective phenomena  of an ensemble of interacting phase oscillators
has been described with different nomenclature. 
Following \cite{Acebron_2005,Niedermayer_2008}, 
 a {\it synchronized } state of an ensemble of $N$ oscillators is one where all oscillators 
have identical phases. 
\begin{flalign}
\text{Synchronized:} & \nonumber \\
\qquad \theta_i(t) &= \theta_j(t) \ \  {\rm for\ all} \ \ i,j \in (0, 1, ... N-1) . \end{flalign}
A {\it phase-locked } or {\it frequency synchronized} state \citep{Ermentrout_1986,Ermentrout_1990,Ren_2000} 
is one where all oscillators have  identical phase velocities 
\begin{align}
\text{Phase-locked} & \nonumber \\
  \dot \theta_i(t) &= \dot \theta_j(t) \ \ {\rm  for\ all} \ \  i,j  \in (0, 1, ... N-1). \end{align}
Pairs of oscillators differ by a constant phase difference. 

In a periodic {\it entrained} state, if the oscillators have identical mean or average phase velocities we call the state {\it entrained}; 
\begin{align}
\text{Entrained:} & \nonumber \\
 \tilde \omega_i &=\tilde \omega_j \ \ {\rm  for\ all} \ \  i,j \in (0, 1, ... N-1) . \ \ 
 \end{align}
For a periodic state with period $T$,  the phases satisfy
$\theta_i(t + T) = \theta_i$ for all $i$. 
 The average phase velocity $\tilde \omega_i$ can be computed with an integral
 over the period $T$, \ 
 $\tilde \omega_i = \frac{1}{T} \int_0^T \dot \theta(t) dt$.

For a chain of oscillators, the index $i$ specifies the order in the chain.
One type of traveling wave is a {\it non-synchronous and phase-locked state} characterized by
a constant phase delay or offset between consecutive oscillators in a chain or loop of oscillators.
In other words 
\begin{align}
\text{Contant phase delay:} & \nonumber \\
\theta_{i+1} &= \theta_i + \phi  \qquad \qquad \qquad 
\end{align}
for consecutive oscillators,  where $\phi$ is called the phase delay, phase shift  or
phase difference 
and $\dot \theta_i \ne 0$ for all $i$. 
If individual oscillators undergo similar periodic motions, then another type of traveling wave is a {\it non-synchronous
and entrained state} characterized by a time delay between the motions of consecutive oscillators.
In other words 
\begin{align}
\text{Constant time delay:} & \nonumber \\
\theta_i(t + \tau) &= \theta_{i+1}(t)\qquad \qquad \qquad 
\end{align}
with time delay $\tau$.
In this case the phase velocities need not be constant.
Both types of traveling waves involve periodic oscillator motions and are known in the  literature as metachronal waves (e.g., \cite{Brumley_2012,Elgeti_2013,Quillen_2021,chakrabarti2022multiscale}).

\subsection{Local Kuramoto models}
\label{sec:loc}

The Kuramoto model \citep{Kuramoto_1975,Kuramoto_1987,Acebron_2005}
consists of $N$ phase oscillators, that mutually interact via a sinusoidal interaction term 
\begin{equation}
\frac{d \theta_i}{dt} = \omega_{i,0} +  \sum_{j=1}^N K_{ij} \sin (\theta_j - \theta_i) 
\end{equation}
where $K_{ij}$ are non-negative coefficients giving the strength of the interaction
between a pair of oscillators.  
Here $i \in 0,1, 2, ...., N-1$ and  each angle $\theta_i \in [0, 2\pi]$.  
 In the absence of interaction, the $i$-th oscillator would have a 
constant phase velocity $\omega_{i,0}$ which is called its intrinsic frequency.
The intrinsic frequencies for each oscillator need not be identical. 

With only nearest neighbor interactions, a well studied model,  
sometimes called a local Kuramoto model, is described by
\begin{equation}
\frac{d \theta_i}{dt} = \omega_{i,0} + K \left[ \sin(\theta_{i+1} - \theta_i) + \sin (\theta_{i-1} - \theta_i) \right]
\label{eqn:local_model}
\end{equation}
\citep{Ermentrout_1986,Ermentrout_1990,Ren_2000,Muruganandam_2008,Tilles_2011,Denes_2019}.
Each oscillator only interacts with its nearest neighbors. 
At low values of positive interaction parameter $K$, the oscillators are not affected by their neighbors.
At higher $K$, the oscillators cluster in phase velocity, and the number of clusters 
decreases until they fuse into a single cluster that spans the system.  
At and above a critical value of $K=K_s$ the entire system must enter a global phase-locked state \citep{Aeyels_2004}.
Above the critical value $K>K_s$, there can be multiple stable phase-locked attractors, each with its
own value of global rotation rate $\Omega = \frac{1}{N} \sum_i \omega_i$  \citep{Zheng_1998,Tilles_2011}.


Instead of considering chains of oscillators that have different intrinsic frequencies,
($\omega_{i,0} \ne \omega_{j,0}$ for $i\ne j$) a number of studies
have focused on chains that have {\it rotational symmetry}.
In these systems, each oscillator has the same equation of motion as
the previous oscillator in the chain, but with index shifted by 1. 
For example, 
\citet{Niedermayer_2008,Tilles_2011,Denes_2019} 
studied loops with nearest  neighbor interactions. 
We refer to a chain of $N$ phase oscillators that has a periodic boundary condition, 
$\theta_0 = \theta_N$, $\theta_{N+1} = \theta_1$,  as a {\it loop}. 

For loops with rotationaly symmetric interactions and identical intrinsic frequencies \citep{Denes_2019} 
the linearized system (linearized about a stable synchronous or phase-locked state) has Jacobian that is a circulant matrix. (This is a matrix
where each row is a cyclic permutation of the previous row).
This gives a closed form for the eigenvalues, which can be used to study the stability of synchronous or phase locked states (e.g., \cite{Niedermayer_2008}). 
\citet{Ottino_Loffler_2016} considered chains and loops of nearest neighbor coupled 
oscillators that differ in intrinsic or natural oscillator frequency. 
They found that 
for both topologies, stable phase-locked states exist if and only if the spread or `width' of the natural frequencies is smaller than a critical value called the locking threshold.
By studying a system with the coupling strength of a given link varies from zero (a chain with free ends) to one with a periodic boundary (a ring), \citet{Tilles_2011}  investigated the birth of phase locked solutions. 

\subsection{Loops of identical oscillators -- rotational symmetry}
\label{sec:rot}

We consider 
the class of loop systems that has only nearest neighbor interactions,  
\begin{equation}
\frac{d\theta_i}{dt} = \omega_0 + H_+(\theta_i,\theta_{i+1}) + H_-(\theta_i, \theta_{i-1}) 
\label{eqn:HH}
\end{equation}
which is rotationally symmetric because each oscillator resembles every other oscillator in the loop.  
Here intrinsic oscillator frequencies are the same for each oscillator and
equal to  $\omega_0$. The functions $H_+$ and $H_-$ are periodic in both arguments
so $H_+(\psi_1 + 2 \pi, \psi_2) =  H_+(\psi_1 , \psi_2+ 2 \pi)   = H_+(\psi_1, \psi_2)$ and similarly
for $H_-()$.   Because we don't specify the functions $H_+, H_-$, the model
is more general than the local Kuramoto model (Eqn.~\ref{eqn:local_model}) 
with sinusoidal interactions, and where all oscillators have the same intrinsic frequency ($\omega_{i,0} = \omega_0$ for all $i$). 

The dynamical system of Eqn.~\ref{eqn:HH} 
need not be symmetric to inversion $(j \to N-1-j$ for $j=0,...,N-1$), also known
as {\it mirror symmetry} \citep{Solovev_2022}. 
Equivalently,  
we need not require that the function $H_+()$ be the same as $H_-()$. 
In other words, if $H_+$ differs from $H_-$, then the loop has a directionality. 
If the system is symmetric to inversion we refer to it as {\it bidirectional }
otherwise we refer to it as {\it directional}.  
If one of the functions $H_-$ or $H_+$ is zero, we refer to the model
as {\it unidirectional}. 
For examples of directional models see the coupling called `telescopic coupling' 
by \citet{Ottino_Loffler_2016}, the unidirectional  model by \citet{Quillen_2021}
and interactions with `odd coupling'  by \citet{Solovev_2022}.

For a directional model in the form of Eqn.~\ref{eqn:HH} it is convenient to define two functions 
\begin{align}
H_s(\psi_1,\psi_2) &\equiv H_+(\psi_1, \psi_2)  + H_-(\psi_1, \psi_2) \nonumber \\
H_a(\psi_1,\psi_2) &\equiv H_+(\psi_1, \psi_2)  - H_-(\psi_1, \psi_2) .
\label{eqn:HsHa}
\end{align}
A bidirectional model (with mirror symmetry) has $H_a(\psi_1,\psi_2) = 0$. 

\subsection{Phase differences and the winding number} 
\label{sec:winding}

It is convenient to describe the state of the system with phase shifts or differences
between neighboring oscillator phases. 
 We follow \citet{Denes_2019} and define the phase difference
between two consecutive oscillators with phases $\theta_i$ and $\theta_{i-1}$ 
\begin{equation}
\phi_i \equiv \theta_i - \theta_{i-1} 
    - 2 \pi \ {\rm floor} \left[ \frac{\theta_i - \theta_{i-1} + \pi}{2 \pi} \right]
    \label{eqn:phaseshift}
\end{equation}
where the  function floor$(x)$ gives the largest integer that is less than $x$. 
The phase difference $\phi_i \in [-\pi,\pi]$.
To characterize the slope of a state we define a winding number 
\begin{equation}
w \equiv \frac{1}{2 \pi} \sum_{i=0}^{N-1} \phi_i .  \label{eqn:winding}
\end{equation}
It is convenient to compute a quantitive that is proportional to the cumulative sum
of the phase differences 
\begin{equation}
w_j =\frac{1}{2\pi} \sum_{i=0}^j \phi_i  \label{eqn:w_j}
\end{equation}
where the winding number $w = w_{N-1}$.
The periodic boundary condition and Eqn.~\ref{eqn:phaseshift} 
implies that the sum of the phase
differences must be a multiple of $2 \pi$.  This implies 
that the winding number $w$
must be an integer, with negative integers or zero allowed \citep{Denes_2019}. 
Because the phase shifts are between $-\pi$ and $\pi$, 
the winding number $-N/2 \le w \le N/2$ with $w \in {\mathbb{ Z}}$.

For phase locked or entrained states, phase shifts remain
near a particular mean value and the standard deviation of
the phase shift remains low.  
It is convenient to compute the standard deviation of the phase shift 
\begin{equation}
\sigma_\phi \equiv \sqrt{\langle (\phi-\bar \phi)^2 \rangle}. \label{eqn:sig_phi}
\end{equation}
Here the mean phase shift 
\begin{align}
\bar \phi &= \langle \phi \rangle = \frac{1}{N} \sum_i \phi_i 
= \frac{2 \pi w}{N}  \label{eqn:barphi}
\end{align}
  is proportional to the winding number $w$.

If the system is in a phase locked or entrained state, 
how is winding number related to the wave speed?
The metachronal wave speed 
$v_{\rm MW} \sim  {\tilde \omega}\ dx/{\bar \phi}$ where $\tilde \omega \sim \omega_0$
is the average angular velocity and $dx$ is the separation between oscillators.
This gives $v_{\rm MW} \sim \frac{{ N \tilde \omega}\ dx}{2 \pi w}$,  
thus winding number $w$ and mean phase shift $\bar \phi$ 
are related to wave travel speed $v_{\rm MW}$. 

\section{Associated continuum equations}
\label{sec:cont}

If $N$ is large and there are no large jumps in phase between neighboring oscillators,
the dynamical system of Eqn.~\ref{eqn:HH} can be approximated with a partial
different equation,  (e.g., \cite{Pikovsky_2003,chakrabarti2022multiscale}).

We approximate our describe system of oscillators with 
 a continuous function $\theta(x,t)$ and with coordinate $x \in [0, 2\pi)$ in an interval. The boundary condition is periodic, so $\theta(0,t) = \theta(2 \pi, t)$. 
We associate a position in the interval $x \in [0, 2\pi)$ for each oscillator in the loop
with $x_j = 2 \pi j /N$ giving a separation $dx = 2 \pi/N$ between each oscillator. 
The continuum variable $\theta(x,t)$ is related to oscillator phases with 
$\theta_j(t) \approx \theta(x_j,t)$
 where $x_j$ are the coordinate positions of each oscillator.

To third order in $dx$, where $dx$ is the separation between neighboring oscillators
\begin{align}
\theta_{j+1} &\approx \theta_j + dx \frac{\partial \theta}{\partial x}\Big|_{x_j}
+ \frac{dx^2}{2} \frac{\partial^2 \theta}{\partial x^2} \Big|_{x_j} 
+ \frac{dx^3}{3!} \frac{\partial^3 \theta}{\partial x^3} \Big|_{x_j} 
\nonumber \\
\theta_{j-1} &\approx \theta_j - dx \frac{\partial \theta}{\partial x}\Big|_{x_j}
+ \frac{dx^2}{2} \frac{\partial^2 \theta}{\partial x^2} \Big|_{x_j}
- \frac{dx^3}{3!} \frac{\partial^3 \theta}{\partial x^3} \Big|_{x_j} 
.
\end{align}
We expand the two interaction functions
of Eqn.~\ref{eqn:HH}, keeping only terms to third  order in the phase difference 
\begin{align}
H_+(\theta_j , \theta_{j+1}) =& 
	H_+(\psi_1, \psi_2) \Big|_{\psi_1,  \psi_2 = \theta_j}  \\
	& + \sum_{i=1}^3 \frac{\partial^{(i)} H_+(\psi_1, \psi_2)}{\partial \psi_2^i} \Big|_{\psi_1,  \psi_2 = \theta_j}\! \frac{1}{i!} (\theta_{j+1} - \theta_j)^i \nonumber 
 .
\end{align}
\begin{widetext}
\begin{align}
H_+(\theta_j , \theta_{j+1}) = & 
 H_+(\psi_1, \psi_2) \Big|_{\psi_1,  \psi_2 = \theta_j}  \nonumber \\
	&  + \frac{\partial H_+(\psi_1, \psi_2)}{\partial \psi_2} \Big|_{\psi_1,  \psi_2 = \theta_j}
	  \left(\!dx \frac{\partial \theta }{\partial x}
		+ \frac{dx^2}{2} \frac{\partial^2 \theta}{\partial x^2} 
		+ \frac{dx^3}{3!} \frac{\partial^3 \theta}{\partial x^3}\!\right) \nonumber \\
& +  \frac{\partial^2 H_+(\psi_1, \psi_2)}{\partial \psi_2^2} \Big|_{\psi_1,  \psi_2 = \theta_j}
 \frac{1}{2}\Bigg[\! \left(dx \frac{\partial \theta}{\partial x} \right)^2  + dx^3 \frac{\partial \theta}{\partial x} \frac{\partial^2 \theta}{\partial x^2}\!\Bigg] \nonumber \\
& +   \frac{\partial^3 H_+(\psi_1, \psi_2)}{\partial \psi_2^3} \Big|_{\psi_1,  \psi_2 = \theta_j}
 \frac{1}{3!}\left(dx\frac{\partial \theta}{\partial x} \right)^3. 
 \label{eqn:HPexp}
\end{align}
\end{widetext}
In Eqn.~\ref{eqn:HPexp}, the angle  $\theta$ refers to $\theta(x_j)$ and $x_j$ is the $x$-position of the $j$-th oscillator.  With the same expansion, we derive similar expressions
for $H_-(\theta_j,\theta_{j-1})$. 

It is convenient to compute derivatives  
\begin{align}
c_{0s}(\theta) & = H_s(\theta, \theta) \nonumber\\
c_{is}(\theta) & =  \frac{\partial^{(i)} H_s(\psi_1, \psi_2)}{\partial \psi_2^i} \Big|_{\psi_1,  \psi_2 = \theta} \nonumber\\
c_{ia}(\theta) & =  \frac{\partial^{(i)} H_a(\psi_1, \psi_2)}{\partial \psi_2^i} \Big|_{\psi_1,  \psi_2 = \theta} .
 \label{eqn:cofs}
\end{align}
The index for the coefficient specifies the order of the derivative and the $a$ or $s$ 
specifies which function is used from Eqn.~\ref{eqn:HsHa}. 

We insert the expansions of Eqn.~\ref{eqn:HPexp} and a similar
one for $H_-$ into Eqn.~\ref{eqn:HH} and use short hand $\theta_{xx} = \frac{\partial^2 \theta}{\partial x^2}$
and $\theta_t = \dot \theta$, and similarly for other partial derivatives,  giving 
\begin{align}
\theta_t &= \omega_0 + c_{0s}(\theta) 
+ c_{1a}(\theta) dx\ \theta_x
+ c_{1s}(\theta) \frac{dx^2}{2}  \theta_{xx} 
\nonumber \\ & \qquad 
+ c_{2s}(\theta) \frac{dx^2}{2}  (\theta_{x})^2 
+  c_{1a}(\theta) \frac{dx^3}{3!}  \theta_{xxx}
\nonumber \\ & \qquad 
+ c_{2a}(\theta) \frac{dx^3}{2}  \theta_x \theta_{xx} 
+ c_{3a}(\theta) \frac{dx^3}{3!}  (\theta_x)^3 . \label{eqn:cmodel2}
\end{align}

If the system is bidirectional then the functions $H_+() = H_-()$
and the asymmetric coefficients $c_{1a} = c_{2a} = c_{3a} = 0$. 
The partial differential equation in Eqn.~\ref{eqn:cmodel2} becomes  (expanding to third order in $dx$)
\begin{align}
\theta_t = \omega_0 + c_{0s}(\theta) 
+ c_{1s}(\theta) \frac{dx^2}{2}  \theta_{xx}
+ c_{2s}(\theta) \frac{dx^2}{2}  (\theta_{x})^2.  \label{eqn:cmodel_bi}
\end{align}

Following \citet{Pikovsky_2003} (their chapter 11),
the continuum or large $N$ limit is taken by multiplying the interaction functions with 
a strength $\epsilon$ and then rescaling the strength of the interaction functions
in the continuum equation so that they depend on $dx^2$.  If the interaction
functions depend on phase differences, then the coefficients are independent of angle.
With a bidirectional equation of motion 
\begin{equation}
\frac{d\theta_i}{dt} = \omega_0 + \epsilon \left[ H(\theta_i-\theta_{i+1}) + H(\theta_i- \theta_{i-1}) \right],
\label{eqn:HH3}
\end{equation}
Eqn.~\ref{eqn:cmodel_bi} becomes
\begin{align}
\theta_t = \omega_0' 
	+ \alpha  \theta_{xx}
	+ \beta  (\theta_{x})^2.  \label{eqn:cmodel_bi4}
\end{align}
with $\tilde \epsilon = \epsilon\ dx^2$ (via the continuum limit) and coefficients $\alpha = \tilde \epsilon H'(0)$,  $\beta = \tilde \epsilon H''(0)$, and 
$\omega_0' = \omega_0 + 2H(0)$ \citep{Pikovsky_2003}. 

Eqn.~\ref{eqn:cmodel_bi4} is the one dimensional version of Eqn.~11.4 by \citet{Pikovsky_2003}), 
has been previously discussed in the context of the non-linear phase equation (Eqn.~10.24 by \cite{Cross_2009}, and it is related to the Complex Ginzburg Landau equation). 
With the addition of an additional stochastic term,  this equation becomes 
the Kardar-Parisi-Zhang equation \citep{Pikovsky_2003,Solovev_2022} which is used to describe theory of roughening interfaces \citep{Barabasi_1995}.


Henceforth we don't take the continuum limit, rather we use the associated continuum partial differential equation of Eqn.~\ref{eqn:cmodel2} as an approximation to the more general discrete directional system of Eqn.~\ref{eqn:HH}. 
We discuss each term in the third order (in $dx$) continuum equation of Eqn.~\ref{eqn:cmodel2}.

The term with coefficient $c_{1a}$ is first order in $dx$ and is $\propto \theta_x$ so it is an advective term.
It is only present if the model is directional. 
Its coefficient could be dependent upon $\theta$.  If the time average of $c_{1a}$ is non-zero  then there would be an advection speed associated with perturbations.
The term $\propto  \theta_{xxx}$  is dispersive and only relevant for directional models. 

The term with coefficient $c_{1s}$ that is $\propto \theta_{xx}$ is a diffusive term.
If this is positive then the system should be stable to small perturbations.
Its coefficient could be dependent upon $\theta$ in which case its time average
 would be relevant for stability. 

The term with coefficient $c_{2a}$ that is proportional to $\theta_x \theta_{xx}$ can be considered a diffusive term
with sign that depends on the winding number or local slope.  
This term is only present in directional models.   If this term exceeds the term
proportional to $\theta_{xx}$ then 
only regions where the slope gives a positive term would be stable to growth of small
perturbations.    If  $c_{2a}>0$ is positive, then a monotone continuous solution with negative slope $\theta_x<0$  could be unstable to growth of small perturbations.   The direction of long-lived wave-like states could be set by the sign of this term. 


\subsection{A condition for stability of a smooth initial condition} 

Suppose we have a state described with a smooth function $\theta(x,t)$ 
at time $t$ in a directional model. 
An approximate condition for local stability is that the diffusive terms 
(those $\propto \theta_{xx}$) 
in the continuum equation (Eqn.~\ref{eqn:cmodel2}) 
are positive so that short wavelength 
perturbations are damped diffusively.  
This implies that a local and slope dependent condition for stability 
\begin{align}
c_{1s}(\theta)  + c_{2a}(\theta)  dx\ \theta_x  \gtrsim 0. 
\label{eqn:condition}
\end{align}
As instability might be slow, the above condition should be satisfied 
on average, for example averaged over a few oscillation periods if the state is
approximately periodic or over a few times the period $T_0 = 2\pi/\omega_0$.
We denote the averaged coefficients as $\bar c_{1s}$ and $\bar c_{2a}$.
As the condition for instability is dependent upon slope, if there is a sinusoidal
perturbation, stability would depend upon the product of its amplitude and wavenumber. 

We can relate the stability condition of Eqn.~\ref{eqn:condition} to that of the oscillator chain model by relating the 
phase shift $\phi$ between oscillators  to the slope; 
$\phi \approx \theta_x dx$ where $dx$ is
the separation between oscillators.  Eqn.~\ref{eqn:condition} becomes
\begin{align}
{\bar c_{1s}} + {\bar c_{2a} }\phi  \gtrsim 0.  \label{eqn:condition2}
\end{align}
In a region where phase shifts between oscillators are similar and equal to $\phi$, Eqn.~\ref{eqn:condition2} gives a condition
on the phase shift for stability. 

\subsection{The conserved topological charge}

If the continuum system has periodic boundary conditions, then the 
integrated quantity
\begin{equation}
Q = \frac{1}{2\pi} \int_0^{2 \pi} dx\ \theta_x,
\end{equation}
sometimes called a topological charge,  must be equal to an integer
\citep{Pikovsky_2003}.  The topological charge $Q$ measures the phase shift through the loop.  This charge 
 is analogous to the winding number $w$ that we computed
for the loop of oscillators (Eqn.~\ref{eqn:winding}) and it measures the phase shift across the loop. 
Furthermore, for the continuum model with a periodic boundary, the topological charge is a conserved quantity. 
This follows because
\begin{align}
\dot Q =  \dot \theta(2 \pi ) - \dot \theta(0) = 0.
\end{align}
Because of the periodic boundary condition in the equations of motion, 
the right hand side must vanish. 

Conservation of the topological charge $Q$ in the continuum model 
(Eqn.~\ref{eqn:cmodel2}) 
implies that initial conditions
set the  slope of asymptotic solutions \citep{chakrabarti2022multiscale}.
This means that whether an asymptotic state is synchronous or a wave-like state
would be determined by initial conditions.  
A biological system could still tend to form metachronal waves if it does not
have periodic boundary conditions.
For example, 
\citet{chakrabarti2022multiscale} proposed that gaps in ciliated carpets could
facilitate metachronal wave formation. 
Alternatively,  the continuum approximation may fail if discontinuities or short wavelength perturbations are present or develop in the system.   The continuum approximation
should not hold if there is power at wavevector $k \sim 1/dx$. 
For example,  \citet{Niedermayer_2008} showed
that a rotationally symmetric  bidirectional model similar to the local
Kuramoto model (Eqn.~\ref{eqn:local_model}) was unstable if the phase differences  
between oscillators were large, with $|\phi| > \pi/2$.

In the subsequent section we investigate the possibility that 
jumps in phase (discontinuities) between neighboring oscillators 
in a loop of phase oscillators do not conserve the winding number
and so allow wave-like states to develop, independent of the winding
number of the initial condition.  

\section{Numerical exploration}
\label{sec:num}

We illustrate two models that have been used to describe systems exhibiting
metachronal waves, a bidirectional model by \citet{Niedermayer_2008}, 
and a unidirectional model by \citet{Quillen_2021}.
Numerical integration of the equation, in the form of Eqn.~\ref{eqn:HH}
is done with a fixed timestep 4-th order Runge Kutta integrator where
each step has duration $dt$. 

\subsection{A bidirectional model by \citet{Niedermayer_2008}}
\label{sec:bi}

The model by \citet{Niedermayer_2008},  shown in their Figure 4 
and given by their Eqn.~35) with 
identical intrinsic frequencies and periodic boundary conditions, is described by 
\begin{align}
\frac{d\theta_i}{dt}  &= \omega_0 
+ \mu_c \left( \cos(\theta_i - \theta_{i+1})  +  \cos(\theta_i - \theta_{i-1}) \right) \nonumber \\
& \qquad - K   \left( \sin(\theta_i - \theta_{i+1})  +  \sin(\theta_i - \theta_{i-1}) \right).
\label{eqn:bi}
\end{align}
We relate this model to Eqn.~\ref{eqn:HH} with functions 
\begin{align}
H_+(\psi_1,\psi_2) &= H_-(\psi_1,\psi_2)  \nonumber \\
&= \mu_c \cos(\psi_1-\psi_2) - K \sin (\psi_1-\psi_2).  \label{eqn:HHbi}
\end{align}
This model is bidirectional as it has $H_+ = H_-$ and 
it reduces to the local Kuramoto model of Eqn.~\ref{eqn:local_model} 
with $\mu_c = 0$.  
Computing the coefficients for the continuum model with equations \ref{eqn:cofs}
\begin{align}
c_{0s} & = 2 \mu_c\nonumber \\
c_{1a} &= c_{2a} = c_{3a} = 0 \nonumber \\
c_{1s} &= 2 K  \nonumber \\
c_{2s} &=  -2 \mu_c  .
\end{align}
The related continuum model (using Eqn.~\ref{eqn:cmodel2}, accurate
to third order in $dx$) is 
\begin{align}
\theta_t & = \omega_0 +  2 \mu_c + K dx^2 \theta_{xx} -  \mu_c dx^2 (\theta_x)^2.
\label{eqn:bi_cont}
\end{align}
We note that the continuum model has a diffusive term (that proportional to $\theta_{xx}$)  that causes perturbations to diffusively decay when $K>0$.
Due to the mirror symmetry (bidirectionality) of the model, the continuum equation lacks a term
proportional to $\theta_x \theta_{xx}$ which could cause a slope dependent instability.  

With initial conditions chosen from
a uniform distribution (and containing large phase differences) 
\citet{Niedermayer_2008} showed that the large phase differences decay,
and the system develops a smooth wave-like state.   The model is bidirectional so 
the resulting metachronal waves could be in either direction. 

\subsection{The unidirectional model by \citet{Quillen_2021}}
\label{sec:uni}

We also consider the unidirectional model by \citet{Quillen_2021} which is 
\begin{align}
\frac{d\theta_j}{dt} & = \omega_0  - \frac{\omega_0 K_u}{2} \left[ 
{\rm tanh} \left( \frac{\cos \theta_{j-1} - \cos \theta_j - \beta }{h} \right)  + 1
\right] . \label{eqn:uni}
\end{align}
Here real parameters $\beta,h>0$.  The model was motivated by steric
interactions between nematodes that reduce the phase velocity for parameter $K_u >0$.
In this system, stable long-lived wave-like states are entrained states, 
as they oscillate in
phase velocity \citep{Quillen_2021}.
Because the interaction function was motivated by preventing an overlap between neighboring nematode bodies,
we sometimes refer to this model as the {\it overlap} model.  

We relate the unidirectional model in Eqn.~\ref{eqn:uni} 
to Eqn.~\ref{eqn:HH} with functions 
\begin{align}
H_+(\psi_1, \psi_2) & = 0 \nonumber \\
H_-(\psi_1, \psi_2) &= - \frac{\omega_0 K_u }{2} \left[ 
{\rm tanh} \left( \frac{\cos \psi_2 - \cos \psi_1 - \beta }{h} \right)  + 1
\right] . \label{eqn:HHuni}
\end{align}
The coefficients for the continuum model, computed using Eqn.~\ref{eqn:cofs} are 
\begin{align}
 c_{0s}(\theta) &   = - \frac{ \omega_0 K_u}{2}
\left[ \tanh \left( \frac{\beta}{h}\right) -1 \right]  \nonumber \\
c_{1a}(\theta) & 
= -\frac{\omega_0 K_u}{2} {\rm sech}^2 \left( \frac{\beta}{h} \right) 	 \frac{ \sin \theta}{h} \nonumber \\
c_{1s}(\theta) & 
= -c_{1a}(\theta) \nonumber \\
c_{2a}(\theta)  & 
 =  -\omega_0 K_u {\rm sech^2} \left( \frac{\beta}{h} \right)
 \frac{1}{h^2}\nonumber \\ 
 & \qquad \times \Big[-2 
{\rm tanh} \left( \frac{\beta}{h} \right) \sin^2 \theta + h\cos\theta  \Big] \nonumber \\
c_{2s}(\theta)  & 
= -c_{2a}(\theta) \nonumber \\
c_{3a}(\theta)  & 
= -\omega_0 K_u {\rm sech}^2 \left( \frac{\beta}{h}  \right) \frac{\sin \theta}{h}  \nonumber \\
& \qquad  \times 
\Big[ 2 \tanh^2 \left( \frac{\beta}{h} \right) \frac{\sin^2 \theta}{h^2} - {\rm sech}^2 \left( \frac{\beta}{h} \right) \frac{\sin^2 \theta}{h^2} \nonumber \\ 
& \qquad \qquad
  -2  \tanh \left( \frac{\beta}{h} \right) \frac{ \cos \theta}{h}
- \frac{1}{2} \Big].
 \label{eqn:cofs_tt}
\end{align}
If  $K_u$ is not large then we can assume that 
the oscillator phases advance at a nearly constant rate.  
We can approximately average
over an oscillation cycle by integrating over $\theta$.  
We define an averaged coefficient with
\begin{equation}
\bar c \approx \frac{1}{2\pi} \int_0^{2 \pi} d\theta\ c(\theta) . 
\end{equation} 
Taking the averages of the coefficients 
of Eqn.~\ref{eqn:cofs_tt}, 
\begin{align}
\bar c_{1s}  & =  \bar c_{1a} = c_{3a} =  0 \nonumber \\
\bar c_{2a}  
       & = - \bar c_{2s} = \frac{\omega_0 K_u}{h^2} {\rm sech}^2 \left( \frac{\beta}{h} \right) 
\tanh  \left( \frac{\beta}{h} \right) .
\label{eqn:cofs_uni}
\end{align}
Because the coefficient $c_{0s}(\theta)$ is independent of $\theta$, 
the average $\bar c_{0s} = c_{0s}$. 

Using the averaged coefficients in equations \ref{eqn:cofs_uni}, the continuum equation (Eqn.~\ref{eqn:cmodel2})
for the unidirectional model of Eqn.~\ref{eqn:uni} becomes
\begin{align}
\theta_t  = &\omega_0  - \frac{ \omega_0 K_u}{2}
\left[ \tanh \left( \frac{\beta}{h}\right) -1 \right]   \nonumber \\
& \   + \omega_0 K_u {\rm sech}^2\! \left( \frac{\beta}{h} \right) \!
\tanh \! \left( \frac{\beta}{h} \right)\! \frac{dx^2}{2h^2}
\left[ -(\theta_x)^2 \! + \! dx\ \theta_x \theta_{xx} \right] .
\label{eqn:uni_cont}
\end{align}

As the continuum equation lacks a second order term proportional to $\theta_{xx}$,
equations \ref{eqn:condition} would be violated for 
negative slopes (assuming $K_u>0$).
Thus instability caused by the $\theta_x \theta_{xx}$ term is expected where
the slope or phase shifts are locally negative, even when the
magnitude of the phase shift is small.   
This suggests that the synchronous state itself is unstable.  By
linearizing about the synchronous state and averaging over time,  it is possible
to show that this is true, though the associated Jacobian matrix is degenerate. 

\subsection{Illustrations of numerical integrations}

\begin{table}
\caption{Integration parameters for unidirectional and bidirectional models \label{tab:parms}}
\begin{tabular}{lllllllll}
\hline \hline
\multicolumn{7}{l}{Unidirectional model 
(Eqn.~\ref{eqn:uni}, continuum equation Eqn.~\ref{eqn:uni_cont}) } \\
\hline
Common parameters & $K_u$ & $\beta$ & $h$  & $N$ & $dt$ & $t_{\rm max}$ \\
& 0.7 & 0.1 & 0.05  & 200 & 0.05  & 600 \\
\hline
Integration   names                  & Uni-S1  & Uni-S4  & Uni-U   \\
Initial condition & sine    & sine & \multicolumn{2}{l}{uniform} \\
Amplitude $ A_{\rm init}$      & 0.5  & 0.02 &- \\
Wavelength $n_\lambda$     & 1 &  4 & -\\
Figures  & \ref{fig:ov1}, \ref{fig:sharp_sin}a & \ref{fig:sharp_sin}b & \ref{fig:phase_U}a\\
\hline\hline
\multicolumn{7}{l}{Bidirectional model (Eqn.~\ref{eqn:bi}, continuum equation  Eqn.~\ref{eqn:bi_cont}) } \\
\hline
Common Parameters & $K$ & $\mu_c$ &   $N$ & $dt$ & $t_{\rm  max}$ \\
                                   &   0.03   &  0.05        &   100 & 0.05 & 300 \\
\hline 
Integration name                    & Bi-U \\
Initial condition &  \multicolumn{4}{l}{uniform} \\
Figure   & \ref{fig:phase_U}b \\
\hline \hline
\end{tabular}
{Notes: $n_\lambda$ is the number of wavelengths that fit within the loop
of oscillators for the sinusoidal initial condition. 
When initial conditions are uniform, the initial phases for each oscillator
are independently drawn from a uniform distribution within $[0,2\pi)$.
All models have intrinsic angular frequency $\omega_0=1$ and a periodic
boundary condition. 
}
\end{table}

Parameters for unidirectional and bidirectional model integrations 
are listed in Table \ref{tab:parms}.
We group integrations by the dynamical system integrated, and refer
to the equation describing it in the table. 
The number of oscillators in the loop
is $N$ and $dt$ is the time-step used
for each single 4-th order Runge Kutta integration step.  
All models have intrinsic frequency $\omega_0=1$. 
Total integration time is $t_{\rm max}$.

The initial conditions for the unidirectional integrations, denoted 
Uni-S1 and Uni-S4, are a sine 
\begin{equation}
\theta_j(t=0) = A_{\rm init} \sin( 2 \pi n_\lambda j/N ) 
\end{equation}
with amplitude $A_{\rm init}$ and integer $n_\lambda$ that determines
how many wavelengths fit within the loop. 
For the Uni-U and Bi-U integrations, initial phases are independently drawn from
a uniform probability distribution $\in [0, 2\pi)$.

\begin{figure}[ht]
\includegraphics[width=3.8truein]{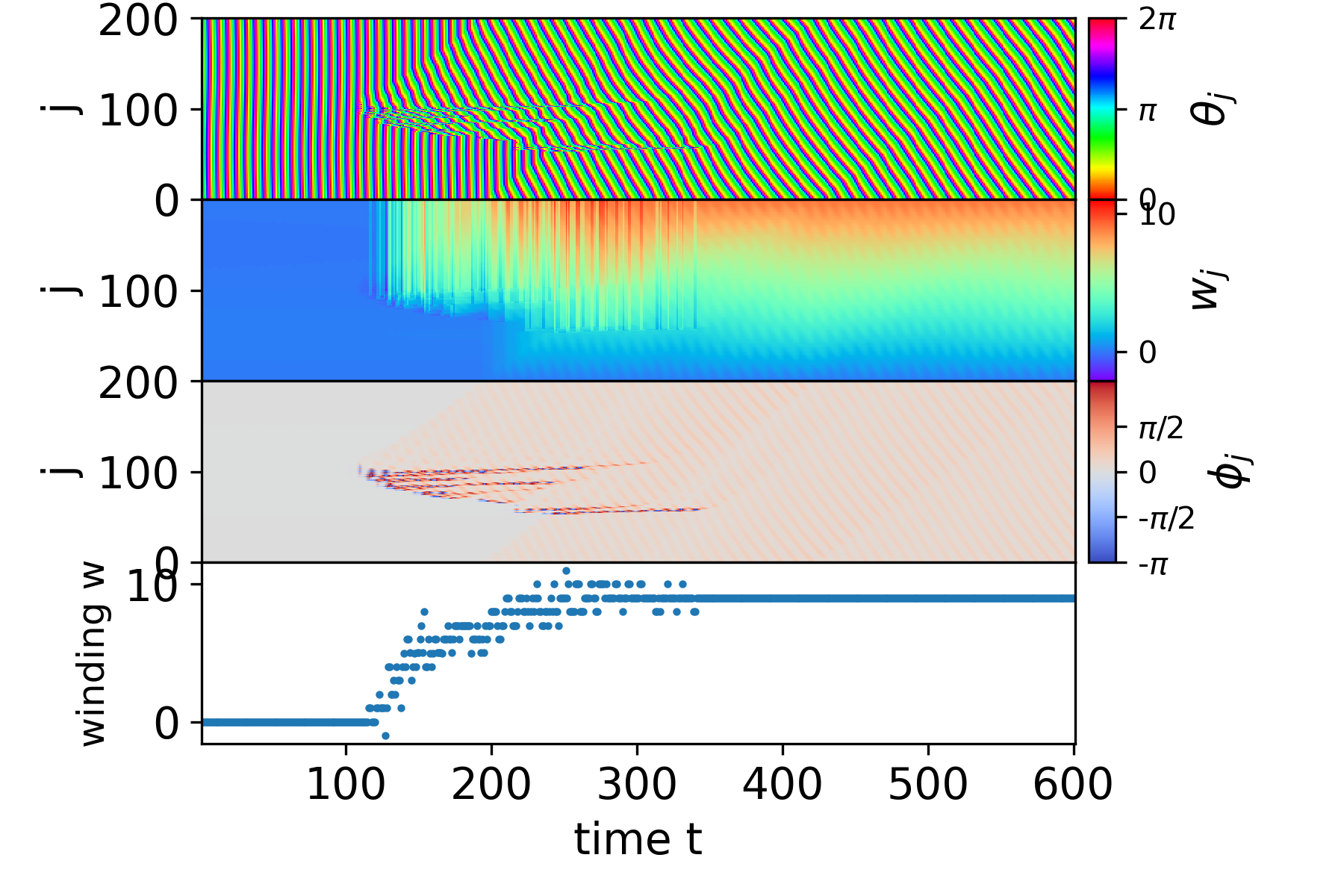}
\caption{An integration, labelled Uni-S1 of the unidirectional model given in 
Eqn.~\ref{eqn:uni}.  The parameters for the model
are listed in Table \ref{tab:parms}. The initial condition is a sine wave
and the boundary condition is periodic. 
The top panel shows phase $\theta_j$ for each oscillator as a function
of index $j$, where index $j$  
increases on the $y$ axis and as a function of time which increases on the $x$ axis. 
The second panel from top shows the cumulative sum  $w_j$ of phase differences 
(defined in Eqn.~\ref{eqn:w_j}).
The phase differences $\phi_j$ (defined in Eqn.~\ref{eqn:phaseshift}) are shown in the third panel.
The winding number $w= w_{N-1}$ (defined in Eqn.~\ref{eqn:winding}) 
is computed from the sum of the phase differences
 and is shown in the bottom panel. 
Groups of oscillators that have phases that differ by about $\pi$
develop, and cause jumps in the cumulative sum of phase differences and these  
give changes in the winding number $w$.   At the end
of the integration, variations in winding number cease and a wave-like state is maintained.
\label{fig:ov1}
}
\end{figure}

In Figure \ref{fig:ov1} we show the Uni-S1 integration of the unidirectional model
with an initial sine perturbation.  
In this figure integration time is along the $x$ axis.  For the top three panels,
the $y$ axis is the index of the oscillator $j$.  
In the top three panels
  we show phase $\theta_j$, phase difference $\phi_j$
(as defined in Eqn.~\ref{eqn:phaseshift}), and the cumulative sum $w_j$ 
of the phase differences, as defined in Eqn.~\ref{eqn:w_j}.  The bottom panel plots  
 the winding number $w$
(defined in Eqn.~\ref{eqn:winding} and equal to $w_{N-1}$).
The cumulative sum of the phase difference shows where differences
in the winding number arise.   The integration shows that a smooth initially smooth
state develops regions where there are jumps in phase between neighboring oscillators. 
We can think of them as discontinuities but they consist of pairs or groups
of oscillators with phase shifts that alternate by approximately $\pi$. 
The changes in the winding number occur where phase differences are near $\pi$. 
When two consecutive oscillators have a phase difference of $\pi$, a small change
in the phase difference can cause a change of $\pm 1$ in the winding number. 
At the end of the integration the winding number is 9
and a metachronal wave has emerged, even though the initial condition 
had a winding number of zero. 

\begin{figure}[ht]
\includegraphics[width=3.5truein]{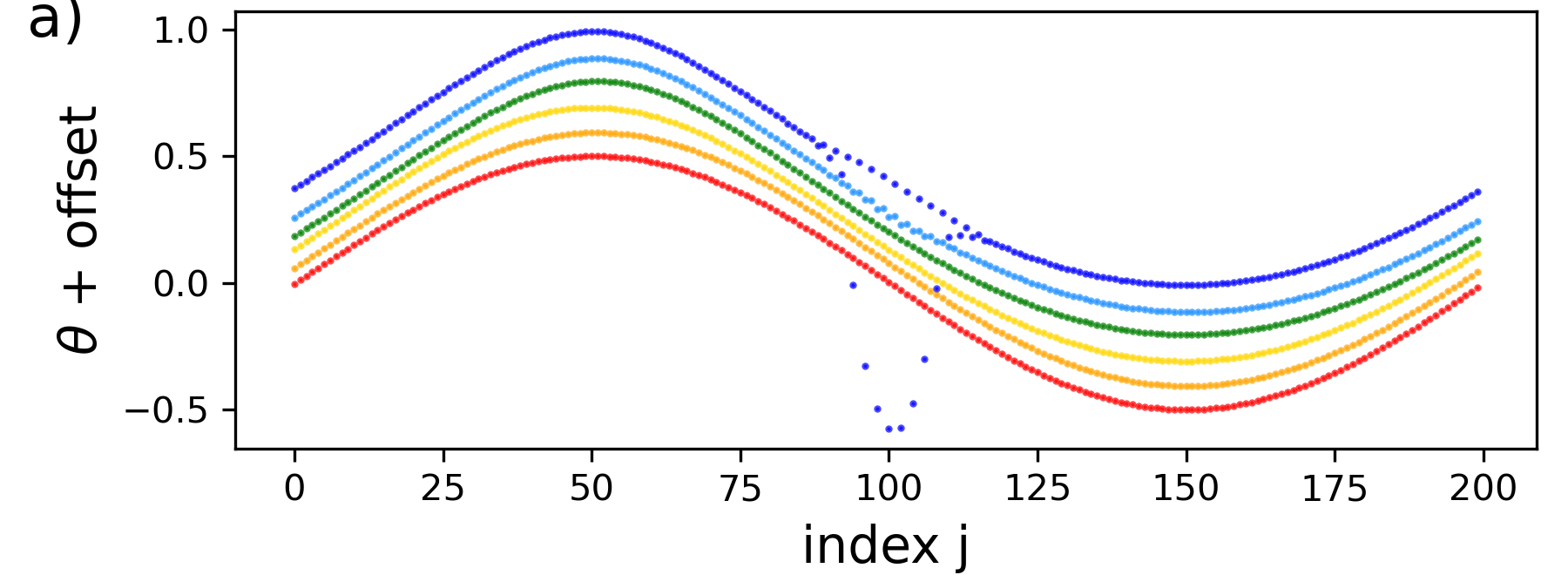}
\includegraphics[width=3.5truein]{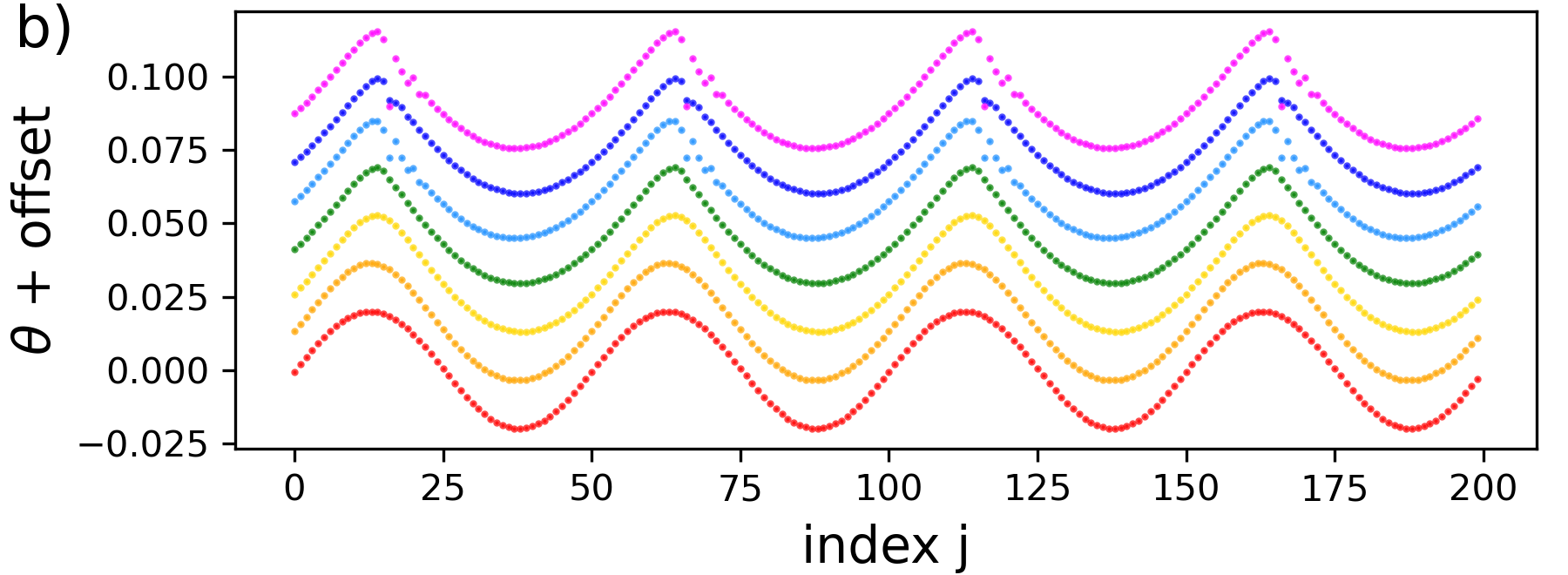}
\caption{Evolution of the unidirectional model (Eqn.~\ref{eqn:uni}) 
with  initial sinusoidal perturbations.
a) We  show the Uni-S1 integration with an initial sine perturbation with wavelength
that exactly fits within the loop. 
b) We show the Uni-S4 integration where the initial sine perturbation is small and 
has wavelength 1/4 of the length of the loop.  
The phases $\theta_j$ of each oscillator  are plotted
at different times as a function of index $j$ (on the $x$ axis) labeling the oscillator. 
The oscillator phases at the different times have been offset so that the curves are plotted
in order of time, with the later times on the top. 
While the initial conditions (shown red as the bottom curves) are smooth,
non-linearity in the model causes an increase in the height of
the peaks.  Regions with negative phase difference (negative slope) are 
unstable to the growth of short wavelength perturbations. 
}
\label{fig:sharp_sin}
\end{figure} 

\begin{figure}[ht]
\includegraphics[width=3.5truein]{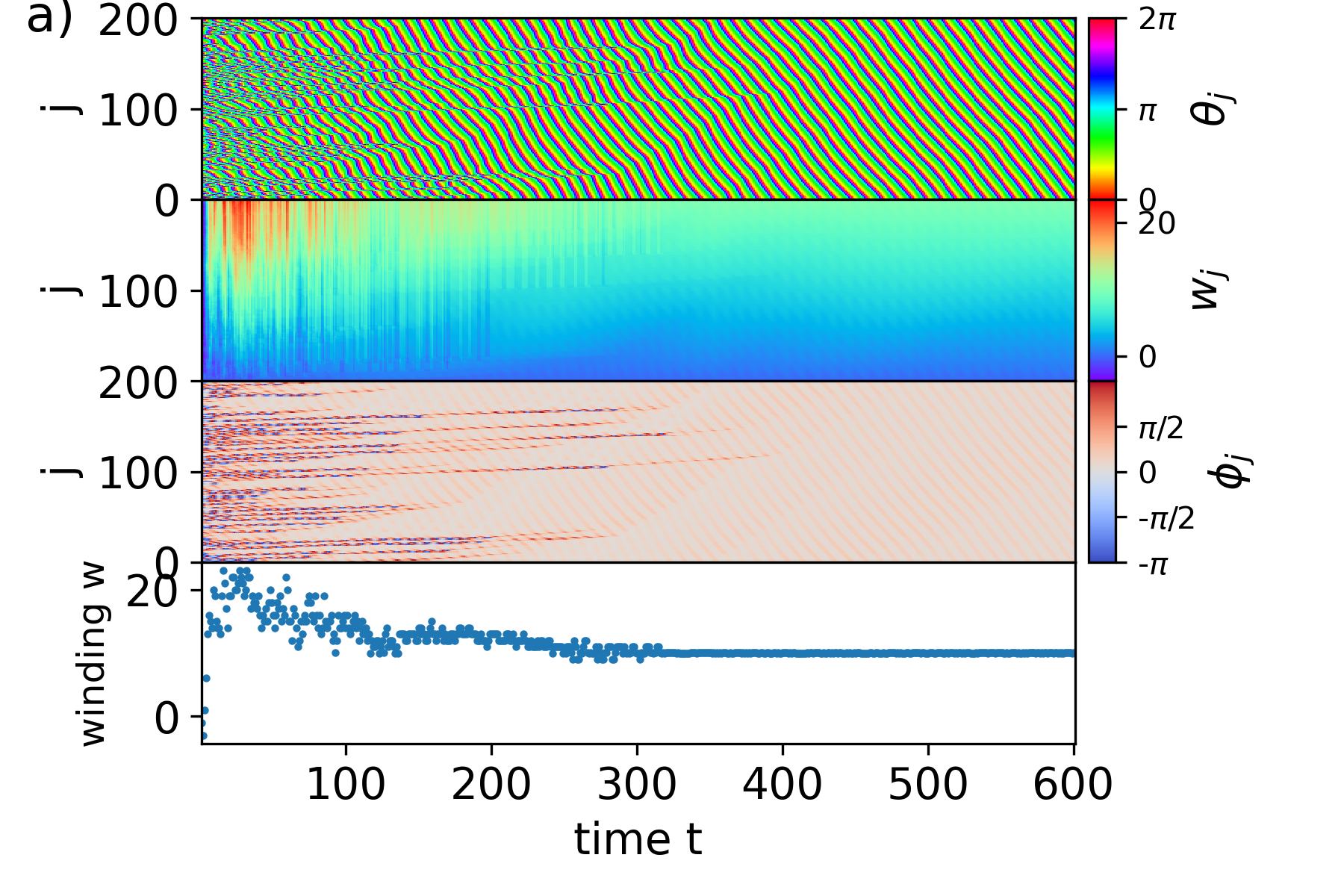}
\includegraphics[width=3.5truein]{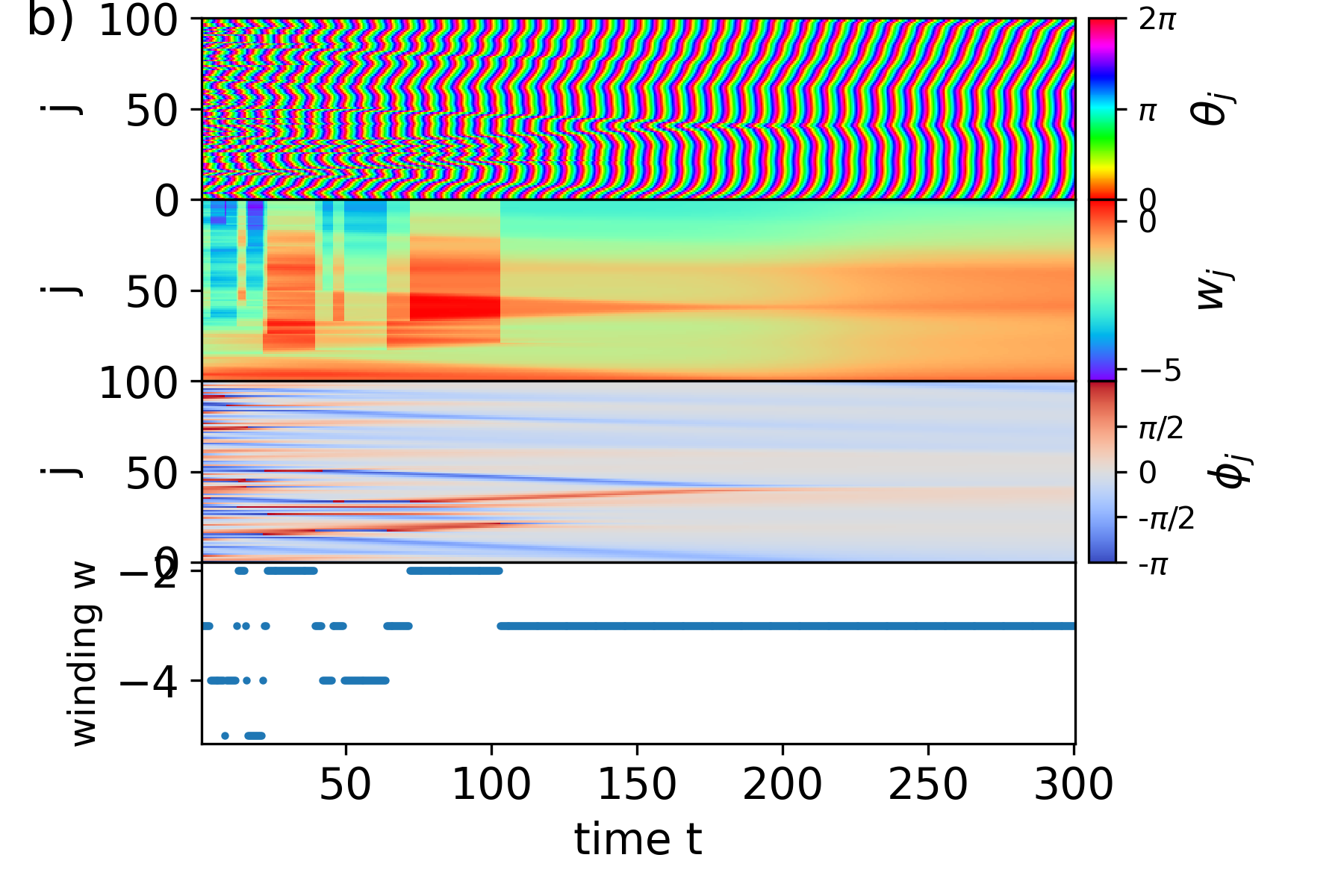}
\caption{Similar to Figure \ref{fig:ov1} except the initial conditions are drawn from a uniform distribution $\in [0,2 \pi)$.  
a) We show the unidirectional Uni-U integration.  
b) We show  the bidirectional Bi-U integration.  
Groups of oscillators that have phase differences of about $\pi$ cause
jumps in the cumulative sum of phase differences. 
The initial conditions have large phase differences,
and while these persist,  the winding number is not conserved.  
Clusters of oscillators form in wave-states with waves going in either direction in the
bidirectional model but only moving in a single direction in the unidirectional model.
After large phase differences
decay, the winding number ceases to vary in both models. 
}
\label{fig:phase_U}
\end{figure} 

In Figure \ref{fig:sharp_sin} we show phases as a function of oscillator index
at different times in two integrations of the unidirectional model
of Eqn.~\ref{eqn:uni}, the Uni-S1 and Uni-S4 integrations.
The Uni-S4 integration also has a small sinusoidal initial perturbation but it is 
shorter wavelength and lower amplitude than that in the Uni-S1 integration. 
Figure \ref{fig:sharp_sin} shows that the short wavelength perturbations
only grow where the phase difference (or slope) is negative.
The continuum approximation for this model (Eqn.~\ref{eqn:uni_cont})
contains a term proportional to $\theta_x \theta_{xx}$ which 
causes instability depending upon the sign of the slope.  For $K_u>0$
the sign of this term is only positive if $\theta_x>0$.  This means
that instability is expected if the slope or phase difference is negative. 
This expectation is consistent with what is seen in Figure \ref{fig:sharp_sin}.
The times of the plotted curves are 
$t=1,20,50,91,106,108$ in Figure \ref{fig:sharp_sin}a and 
$t=1,100,200,300,330,340,350$ in Figure \ref{fig:sharp_sin}b.

The bidirectional model (Eqn.~\ref{eqn:bi}) exhibits some differences
when compared to the unidirectional model. 
For the bidirectional model, when the initial conditions are smooth,
 and phase differences between neighboring oscillators are small, 
the winding number is conserved.   
This is consistent with the stability
limit computed by \citet{Niedermayer_2008} who found that instability
arises only if phase differences exceed $\pm \pi/2$. 
In the unidirectional model, even smooth initial conditions can lead to growth
of large phase differences (depending upon the sign of the slope). 

If initial conditions contain large jumps in phase, then discontinuities can persist
that cause variations in winding number in both bidirectional and unidirectional models. 
We show two integrations, one for the unidirectional model (denoted Uni-U)
and one for the bidirectional model (denoted Bi-U).  The phases for these integrations
are independently  initialized with random
angles drawn from uniform probability distributions in $[0,2\pi)$. 
These integrations are shown in Figure \ref{fig:phase_U} and the parameters
of the models are listed in Table \ref{tab:parms}.
In both models, jumps in phase cause changes in winding number. 
However, over long periods of time
the high frequency power decays and both system approach a long
lived solution with an approximately constant
slope.  After the decay of the large jumps in phase, variations in winding number cease. 

In the unidirectional model, 
 both continuous and random initial conditions generate a wave-like state
with a preferred direction. 
However, in the bidirectional model, only initial conditions that include jumps
in phase allow variations in winding number.    
In the unidirectional model, jumps in phase resolve into waves traveling in 
a single direction, whereas in the bidirectional model jumps in phase 
resolve
into clusters of oscillators exhibiting waves that travel in either direction. 
In the bidirectional model, and with smooth initial conditions, 
the direction of an emergent wave is set by the initial winding number. 
Because regions of negative slope can result in instability in the unidirectional model,
a smooth initial condition with an initial winding number of zero can still 
lead to an emergent wave. 
With random initial conditions, emergent 
waves in either direction are equally likely for the bidirectional model. 
In the unidirectional model, emergent waves only travel in one direction. 
In both models, there are multiple stable long live entrained states, that 
are characterized by different winding numbers. 

\begin{table}
\caption{Integration parameters for directional models \label{tab:parmsmod}}
\begin{tabular}{lllllllllll}
\hline \hline
\multicolumn{6}{l}{Directional sinusoidal model (Eqn.~\ref{eqn:bimod}) } \\
\multicolumn{6}{l}{Associated continuum equation (Eqn.~\ref{eqn:bimod_cont}) } \\
\hline
Common Parameters & $K_+$ & $K_-$ & $\mu_{c+}$ &   $N$ & $dt$ & $t_{\rm max}$ \\
                                   &  0.01  &0.01      &     0              &     64 & 0.05 & 450\\
\hline
Integration series name & \multicolumn{4}{l}{Di-Series-A} \\  
Type of initial condition & \multicolumn{4}{l}{sine, $n_\lambda=4$}     \\ 
\multicolumn{1}{l}{Amplitude $A_{\rm init} $}      &   \multicolumn{2}{l}{[0,0.5]}       \\
\multicolumn{1}{l}{Parameter $\mu_{c-}$}            &  \multicolumn{2}{l}{[0,0.9] }\\
Figure                         & \ref{fig:crit} \\
\hline \hline
\multicolumn{6}{l}{Directional overlap model (Eqn.~\ref{eqn:unimod})} \\
\multicolumn{6}{l}{Associated continuum equation (Eqn.~\ref{eqn:unimod_cont}) } \\
\hline
Common Parameters & $\beta$ & $h$ & $K$ &   $N$ & $dt$ & $t_{\rm max}$ \\
                                   &  0.1  &0.05     &   0.01    &     64 & 0.05  & 600 \\
\hline
Integration series name & \multicolumn{4}{l}{Di-Series-B} \\  
Type of initial condition & \multicolumn{4}{l}{sine, $n_\lambda=4$}     \\ 
 \multicolumn{1}{l}{ Amplitude $A_{\rm init}$}    &  \multicolumn{2}{l}{[0.01,0.4]}   \\
 \multicolumn{1}{l}{Parameter $K_u$}     & \multicolumn{2}{l}{[0.01,0.4]} \\
Figure                         & \ref{fig:crit_ov} \\
\hline\hline
\end{tabular}
{\\ Notes: 
All models have intrinsic angular frequency $\omega_0=1$ and a periodic
boundary condition.   For integration series we show ranges for varied parameters. 
}

\end{table}

\subsection{Causing instability in a bidirectional model to make an adjustable directional model}
\label{sec:bimod}

To explore how directionality affects the behavior of oscillator chain models, 
we desire  simple
models with sufficient numbers of parameters that
we can smoothly adjust whether it is directional or bidirectional. 
We modify the sinusoidal bidirectional model in section \ref{sec:bi}, Eqn.~\ref{eqn:bi}, so that it can be directional
\begin{align}
\frac{d\theta_i}{dt}  &= \omega_0 
+ \mu_{c+}  \cos(\theta_i - \theta_{i+1})  +  \mu_{c-} \cos(\theta_i - \theta_{i-1})  \nonumber \\
& \qquad - K_+ \sin(\theta_i - \theta_{i+1})  - K_- \sin(\theta_i - \theta_{i-1}) .
\label{eqn:bimod}
\end{align}
With $K_+ = K_-$ and $\mu_{c+} = \mu_{c-}$ we recover the bidirectional 
model of Eqn.~\ref{eqn:bi}. 
The interaction functions are 
\begin{align}
H_+(\theta_i,\theta_{i+1}) &= \mu_{c+}  \cos(\theta_i - \theta_{i+1})  - K_+ \sin(\theta_i - \theta_{i+1}) \nonumber \\
H_-(\theta_i,\theta_{i-1}) &= \mu_{c-}  \cos(\theta_i - \theta_{i-1})  - K_- \sin(\theta_i - \theta_{i-1}) . 
\end{align}
The coefficients computed via Eqn.~\ref{eqn:cofs} become 
\begin{align}
c_{0s} & =  \mu_{c+} + \mu_{c-} \nonumber \\
c_{1a} &= K_+ - K_- \nonumber \\
c_{1s} & = K_+ + K_-\nonumber \\
c_{2a} & = -\mu_{c+} + \mu_{c-}\nonumber \\
c_{2s} & = -c_{0s} \nonumber \\
c_{3a} & = -c_{1a},  
\label{eqn:cofs_bimod}
\end{align}
and are independent of $\theta$.

\begin{figure}[ht]
\includegraphics[width=3truein]{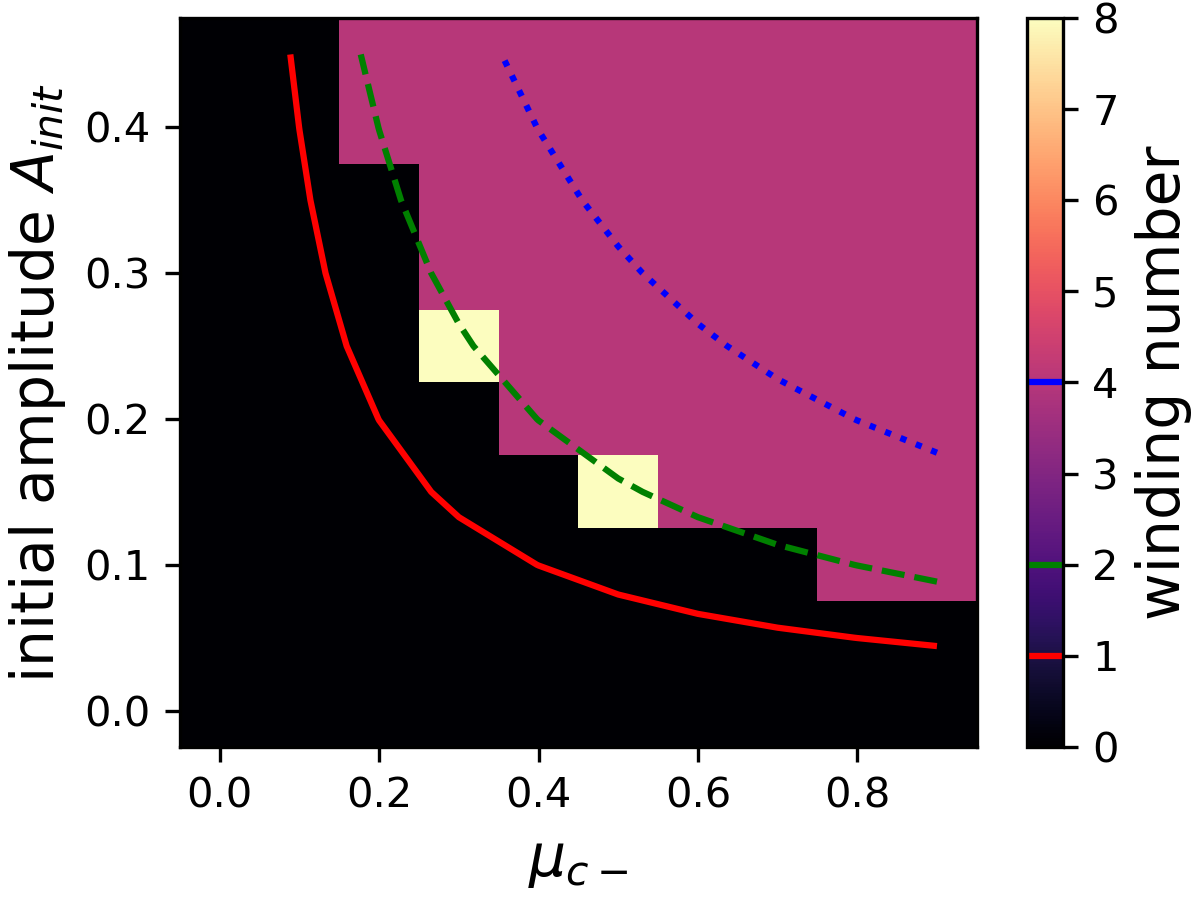}
\caption{We integrate the directional sinusoidal model of Eqn.~\ref{eqn:bimod}
with a sinusoidal initial condition with a range of initial amplitudes $A_{\rm init}$, on the $y$ axis, and a range for the parameter $\mu_{c-}$, on the $x$ axis.  
Remaining parameters for the series of integrations are listed in Table \ref{tab:parmsmod}
with the name Di-Series-A.
The initial winding number $w=0$. Plotted as an image is the final 
 winding number at the end of each integration. 
 A change in winding number 
implies that an instability occurred during the integration. 
The contours show the stability parameter $y_{\rm crit}$ of Eqn.~\ref{eqn:ycrit}
which is derived by comparing the strength of diffusive terms in the continuum equation
(following Eqn.~\ref{eqn:crit}).   We plot $y_{\rm crit} = 1$ (red solid line), 2 (green dashed
line) and 4 (blue dotted line). 
When the stability parameter $y_{\rm crit}$ is greater than 1, and to the right of the red solid line, instability occurs giving clusters of oscillators with larger phase differences.  These resolve by increasing the winding number.  
}
\label{fig:crit}
\end{figure} 

The related continuum model (using Eqn.~\ref{eqn:cmodel2}, accurate
to third order in $dx$) is 
\begin{align}
\theta_t & = \omega_0 +   \mu_{c+}  + \mu_{c-} + (K_+ + K_-) \frac{dx^2}{2} \theta_{xx} 
 \nonumber \\ & \qquad 
-  (\mu_{c+} + \mu_{c-}) \frac{dx^2}{2} (\theta_x)^2  + (K_+ - K_-) \frac{dx^2}{2}  \theta_{xxx}  \nonumber \\
& \qquad + (-\mu_{c+} + \mu_{c-}) \frac{dx^3}{2} 
\theta_x \theta_{xx} + (K_+ - K_-) \frac{dx^3}{3!} (\theta_x)^3
.
\label{eqn:bimod_cont}
\end{align}

For instability caused by the $\theta_x \theta_{xx}$ term that is sensitive to the sign of the slope,  Eqn.~\ref{eqn:condition}  approximately gives 
\begin{equation}
(\mu_{c+} - \mu_{c-}) (dx \ \theta_x) \gtrsim (K_+ + K_-).
\end{equation}
For phase shift $\phi$ between oscillators, this condition for instability (following Eqn.~\ref{eqn:condition2}) becomes 
\begin{equation}
(\mu_{c+} - \mu_{c-}) \phi \gtrsim (K_+ + K_-). \label{eqn:crit}
\end{equation}
The sign of $\mu_{c+} - \mu_{c-}$ determines the sign of unstable slopes.

To check to see if we can predict when a system develops instability
we run a series of integrations, denoted Di-Series-A in Table \ref{tab:parmsmod}, 
that begin with a small sinusoidal variation
and a winding number of zero.   
We measure the change in winding number after integrating a specific period of time.
The model has common parameters
$N=64$, $K_+ = K_-  = 0.01$, $\omega_0=1$, and $\mu_{c+}=0$.
The sinusoidal initial condition has $n_\lambda=4$ wavelengths within the loop of oscillators. 
We do integrations with a range of amplitudes $A_{\rm init}$ for the initial condition
and a range of parameter $\mu_{c-}$.   In each integration, we measure the winding number
at the end of the integration.   The final winding number is plotted
as an image in Figure \ref{fig:crit}.  
Integrations in which perturbations grow exhibit
changes in winding number. 

For the integrations shown in Figure \ref{fig:crit}, 
the maximum phase shift in 
the initial condition depends on the amplitude and wavelength of the sine 
$|\phi_{max}| = A_{\rm init} n_\lambda dx$ with $dx = 2 \pi/N$.
The contours in Figure \ref{fig:crit} show the value of 
\begin{align}
y_{\rm crit}(A_{\rm init}, \mu_{c-})  & =| \phi_{max}| \frac{|\mu_{c+} - \mu_{c-}|}{(K_+ + K_-)}
\nonumber  \\
& = A_{\rm init}n_\lambda \frac{2 \pi}{N}\frac{|\mu_{c+} - \mu_{c-}|}{(K_+ + K_-)}, 
\label{eqn:ycrit}
\end{align} 
which is derived from the stability limit estimate 
of Eqn.~\ref{eqn:crit}.     
Near where this function is above 1, we expect instability.
This is indeed seen in these numerical integrations as 
 changes in winding number are only seen to the right of 
 the solid red contour which has $y_{\rm crit}=1$.
 We numerically confirm that Eqn.~\ref{eqn:condition2} can give
 a slope dependent estimate for the local stability of smooth initial conditions
 in a directional model.  
 
\subsection{Stabilizing a unidirectional model to make an adjustable directional model}
\label{sec:unimod}

The unidirectional model discussed in section \ref{sec:uni} (Eqn.~\ref{eqn:uni}),
when averaged has coefficient $\bar c_{1s} = 0$, so its associated continuum 
equation (Eqn.~\ref{eqn:uni_cont}) lacks a stabilizing term proportional to $\theta_{xx}$.
To this unidirectional model, 
we add an additional term, that with coefficient $K$ from the bidirectional model of Eqn.~\ref{eqn:bi}, 
that gives a non-zero 
coefficient $c_{1s}$ (see  section \ref{sec:bi});  
\begin{align}
\frac{d\theta_j}{dt}  = & \omega_0  - \frac{\omega_0 K_u}{2} \left[ 
{\rm tanh} \left( \frac{\cos \theta_{j-1} - \cos \theta_j - \beta }{h} \right)  + 1
\right]  \nonumber \\
&- K [\sin(\theta_j - \theta_{j+1}) + \sin(\theta_j - \theta_{j-1}) ].
 \label{eqn:unimod}
\end{align}
With the addition of the term with coefficient $K$, the model is no longer unidirectional, rather
it is directional and we can adjust the relative strengths of the symmetric and
antisymmetric interactions by varying $K$. 

The averaged coefficients present in the continuum equation 
for this model are the same as in 
equations \ref{eqn:cofs_uni}, except the coefficient $c_{1s} = 2K$. 
The associated continuum equation is similar to Eqn.~\ref{eqn:uni_cont}
but with an additional term, 
\begin{align}
\theta_t  = &\omega_0  - \frac{ \omega_0 K_u}{2}
\left[ \tanh \left( \frac{\beta}{h}\right) -1 \right]   + K dx^2 \theta_{xx} \nonumber \\
&  + \omega_0 K_u {\rm sech}^2\! \left( \frac{\beta}{h} \right) \!
\tanh \! \left( \frac{\beta}{h} \right)\! \frac{dx^2}{2h^2}
\left[ -(\theta_x)^2 \! + \! dx\ \theta_x \theta_{xx} \right] .
\label{eqn:unimod_cont}
\end{align}

\begin{figure}[ht]
\includegraphics[width=3.4truein, trim = 10 0 10 0, clip]{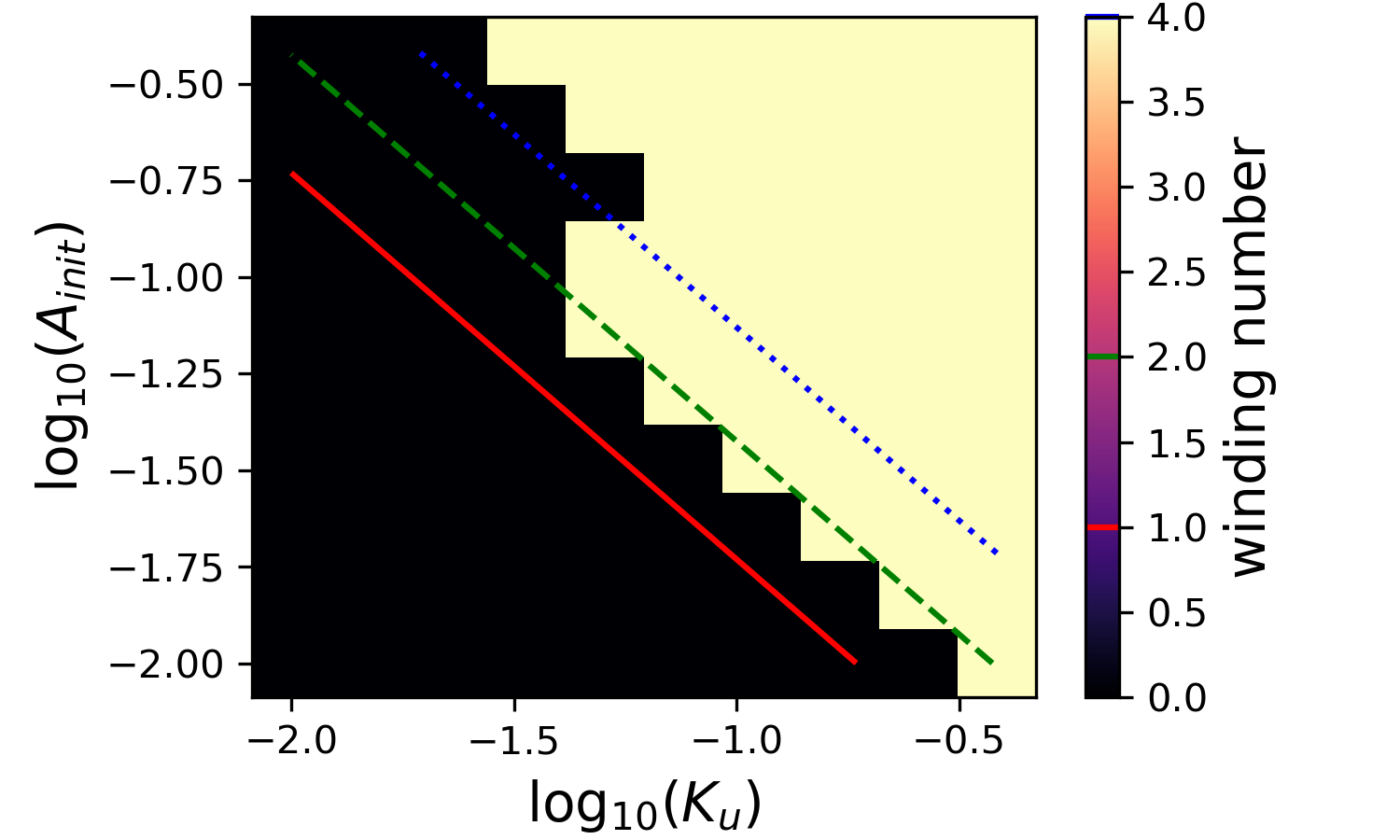}
\caption{Similar to Figure \ref{fig:crit} except we integrate the directional overlap model of Eqn.~\ref{eqn:unimod}
with a sinusoidal initial condition.  Integrations have  
 a range of initial amplitudes $A_{\rm init}$, shown with 
 a log-scale on the $y$ axis, and a range for the parameter $K_u$, shown with
a log-scale on the $x$ axis.  
Remaining parameters for the series of integrations are listed in Table \ref{tab:parmsmod}
with the name Di-Series-B.
Plotted as an image is the final 
 winding number at the end of each integration.  A change in winding number
implies that an instability occurred during the integration. 
The contours show the stability parameter $y_{\rm crit}$ of Eqn.~\ref{eqn:ycrit2}, 
which is derived by comparing the strength of diffusive terms in the continuum equation.
We plot $y_{\rm crit} = 1$ (red solid line), 2 (green dashed
line) and 4 (blue dotted line). 
When the stability parameter $y_{\rm crit}$ is greater than 1, and to the right of the red solid line, instability occurs giving perturbations
in regions where the slope (or equivalently the phase difference) is negative.  These resolve by increasing the winding number.  
}
\label{fig:crit_ov}
\end{figure}

As in section \ref{sec:bimod}, we run a series of integrations, denoted Di-Series-B 
and with parameters listed 
in Table \ref{tab:parmsmod}, have 
 sinusoidal initial conditions, and cover a range of amplitudes and
parameters $K_u$ to see which ones develop instabilities that cause variations
in winding number.   The final winding numbers are plotted in figure \ref{fig:crit_ov}.
These integrations have 
 parameters $h, \beta, \omega_0$  giving 
   coefficient 
 $\bar c_{2a} \approx 27 K_u $ (evaluated using Eqn.~\ref{eqn:cofs_uni}).
 The coefficient $c_{1s} = 2K$ for this dynamical system. 
The estimate for instability of Eqn.~\ref{eqn:condition2} depends on 
 \begin{align}
y_{\rm crit}(A_{\rm init}, K_u)  & \approx 13 A_{\rm init}n_\lambda \frac{2 \pi}{N} \frac{K_u}{K},   
\label{eqn:ycrit2}
\end{align} 
with unstable phase shifts for a sinusoidal initial condition likely for
$y_{\rm crit} \gtrsim 1$. 
Contours with $y_{\rm crit} = 1, 2, 4$ are shown on Figure \ref{fig:crit_ov}.
The $y_{\rm crit} =1$ curve delineates the region where winding number
remains fixed.  Thus Figure \ref{fig:crit_ov}
illustrates that the condition (Eqn.~\ref{eqn:condition2}) 
based on coefficients of diffusive terms in the associated 
continuum equation is consistent with the development of short wavelength
instabilities in the dynamical system of Eqn.~\ref{eqn:unimod}. 

\section{Stochastic directional phase oscillator models}
\label{sec:noise}

In the previous sections we found that the initial condition
can affect the properties of the system after integration.  
How is it possible for a biological system to ensure
that a metachronal wave is robustly generated?
As fluctuations are likely to be present in ciliated systems (e.g., \cite{Ma_2014})
and following \citet{Solovev_2022}, we consider the role of white noise
in influencing the properties of long-lived states. 

To the each oscillator in the 
direction models of Eqns.~\ref{eqn:bimod} and \ref{eqn:unimod} 
we add a continuous random variable
that is Gaussian white noise, $\xi(t)$.
We characterize the strength of the noise with parameter $\eta$ where  
 the probability distribution of
the integral $W(\Delta t) = \int_0^{\Delta t} \xi(t) dt$  is a normal distribution
 with zero mean and  with variance $\eta \Delta t$.   
Equivalently $\langle \xi(t) \xi(t') \rangle = \eta \delta(t-t')$.
In our numerical integrations,
at each time step of duration $dt$ we add an independent random variable
to each oscillator phase that is drawn from a normal distribution with zero mean
and variance $\eta dt$.   

\subsection{A sinusoidal directional model with white noise}
\label{sec:bi_noise}

We modify the directional model of Eqn.~\ref{eqn:bimod} discussed
in section \ref{sec:bimod} with the addition of a stochastic term 
\begin{align}
\frac{d\theta_i}{dt}  &= \omega_0 
+ \mu_{c+}  \cos(\theta_i - \theta_{i+1})  +  \mu_{c-} \cos(\theta_i - \theta_{i-1})  \nonumber \\
& \qquad - K_+ \sin(\theta_i - \theta_{i+1})  - K_- \sin(\theta_i - \theta_{i-1}) 
\nonumber \\
& \qquad + \xi_i(t).
\label{eqn:noise}
\end{align}
Here each $\xi_i(t)$ is an independent continuous random variable that is Gaussian white noise with strength $\eta$, as discussed at the beginning of section \ref{sec:noise}.

The associated continuum equation for the model of Eqn.~\ref{eqn:noise}
is the same as Eqn.~\ref{eqn:bimod_cont} with the addition of white noise
that depends on both space and time;
\begin{align}
\theta_t & = \omega_0 +   \mu_{c+}  + \mu_{c-} + (K_+ + K_-) \frac{dx^2}{2} \theta_{xx} 
 \nonumber \\ & \qquad 
-  (\mu_{c+} + \mu_{c-}) \frac{dx^2}{2} (\theta_x)^2  + (K_+ - K_-) \frac{dx^2}{2}  \theta_{xxx}  \nonumber \\
& \qquad + (-\mu_{c+} + \mu_{c-}) \frac{dx^3}{2} 
\theta_x \theta_{xx} + (K_+ - K_-) \frac{dx^3}{3!} (\theta_x)^3 \nonumber \\
&\qquad +  \xi(x,t)
.
\label{eqn:bimod_noise_cont}
\end{align}
Here $\xi(x,t)$ denotes uncorrelated Gaussian white noise with
$\langle \xi(x,t) \xi(x',t') \rangle =  D \delta(x-x') \delta(t-t')$.
We relate $D$ to the noise strength $\eta$ for the discrete system  
via $D = \eta dx $
where $dx$ is the distance between neighboring oscillators.  

\begin{table*}
\caption{Parameters for integrations of directional models with noise \label{tab:noise}}
\begin{tabular}{lllllllllllllll}
\hline \hline
\multicolumn{8}{l}{Directional sinusoidal model with white noise (Eqn.~\ref{eqn:noise}) } \\
\multicolumn{8}{l}{Associated stochastic continuum equation 
	(Eqn.~\ref{eqn:bimod_noise_cont}) } \\
\hline
Common parameters & $K_+$ & $K_-$ & $\mu_{c+}$     & $dt$ & $t_{\rm max}$ \\
                                   &  0.01  &0.01      &     0       &     0.05  & 600 \\
\hline
Integrations  & \multicolumn{2}{l}{DWN1} 
			& \multicolumn{2}{l}{DWN2} 
			& \multicolumn{2}{l}{DWN-SerA} 
			& \multicolumn{2}{l}{DWN-SerW} 
			& \multicolumn{2}{l}{DWN-SerN} 
			& \multicolumn{2}{l}{BWN-SerN} 
			\\
Noise strength  $\eta$  &   \multicolumn{2}{l}{0.005}   
	&  \multicolumn{2}{l}{0.02}    
	& \multicolumn{2}{l}{[$10^{-3}$,0.09]}   
	&  \multicolumn{2}{l}{[$10^{-3}$,0.04]}  
	&  \multicolumn{2}{l}{[$10^{-3}$,0.03]} 
	&  \multicolumn{2}{l}{[$10^{-3}$,0.03]}  
	\\
Parameter $\mu_{c-}$  & \multicolumn{2}{l}{-0.006}  
	& \multicolumn{2}{l}{0.03}     
	 & \multicolumn{2}{l}{[-0.07,0.07]} 
	 &  \multicolumn{2}{l}{0.06} 
	 &  \multicolumn{2}{l}{0.06} 
	 &  \multicolumn{2}{l}{0.0} 
	 \\
Initial winding number $w_0$         &  \multicolumn{2}{l}{0} 
					&   \multicolumn{2}{l}{0} 
					&  \multicolumn{2}{l}{0} 
					&  \multicolumn{2}{l}{[-15,15]} 
					& \multicolumn{2}{l}{0} 
					& \multicolumn{2}{l}{0} 
					\\
Number of oscillators $N$     & \multicolumn{2}{l}{50} 
	& \multicolumn{2}{l}{50} 
	& \multicolumn{2}{l}{50} 
	& \multicolumn{2}{l}{50} 
	&  \multicolumn{2}{l}{[32,4096]} 
	&  \multicolumn{2}{l}{[32,4096]} 
	\\
Figures                        & \multicolumn{2}{l}{\ref{fig:phase_eta}a} 
	& \multicolumn{2}{l}{ \ref{fig:phase_eta}b} 
	& \multicolumn{2}{l}{\ref{fig:red_blue} } 
	& \multicolumn{2}{l}{\ref{fig:w0}a,b}
	& \multicolumn{2}{l}{\ref{fig:NN}a,b} 
	& \multicolumn{2}{l}{\ref{fig:NN}e,f} 
	\\
\hline \hline
\multicolumn{8}{l}{Directional overlap model with white noise 
(Eqn.~\ref{eqn:uni_noise}) } \\
\multicolumn{8}{l}{Associated stochastic continuum equation (Eqn.~\ref{eqn:unimod_noise_cont}) } \\
\hline
Common parameters &  $\beta$ & $h$ &    $dt$  & $t_{\rm max}$ \\
                                   &   0.1      &     0.05    & 0.05  & 600 \\
\hline
Integrations  &  \multicolumn{2}{l}{OWN-SerA} & \multicolumn{2}{l}{OWN-SerB} & \multicolumn{2}{l}{OWN-SerC} & \multicolumn{2}{l}{OWN-SerW} &\multicolumn{2}{l}{OWN-SerN} \\
Noise strength  $\eta$  &    \multicolumn{2}{l}{$10^{-3}$}  &  
                                            \multicolumn{2}{l}{[$10^{-4}$,0.04]}  & 
                                            \multicolumn{2}{l}{[$10^{-4}$,0.04]}  & 
                                            \multicolumn{2}{l}{[$10^{-3}$,0.04] } &
                                            \multicolumn{2}{l}{[$10^{-3}$,0.03]} \\
Parameter $K$              & \multicolumn{2}{l}{[0.0003,0.13]\!} 
                                      & \multicolumn{2}{l}{0.002} 
                                      &  \multicolumn{2}{l}{[0.0006,0.26]\!} 
                                      & \multicolumn{2}{l}{0.01} 
                                      & \multicolumn{2}{l}{0.01} 
                                      \\
Parameter $K_u$          & \multicolumn{2}{l}{[0.01,0.41]} 
                                      &  \multicolumn{2}{l}{[0.01,0.28]} 
                                      &  \multicolumn{2}{l}{0.2} 
                                      & \multicolumn{2}{l}{0.2} 
                                      & \multicolumn{2}{l}{0.2} 
                                      \\
Initial winding number $w_0$    &  \multicolumn{2}{l}{0} 
	&   \multicolumn{2}{l}{0} &  \multicolumn{2}{l}{0} 
	&  \multicolumn{2}{l}{[-15,15]} 
	&  \multicolumn{2}{l}{0} \\
Number of oscillators $N$     & \multicolumn{2}{l}{100} 
	& \multicolumn{2}{l}{100} 
	& \multicolumn{2}{l}{100} 
	& \multicolumn{2}{l}{50} 
	&  \multicolumn{2}{l}{[32,4096]} \\
Figures           & \multicolumn{2}{l}{\ref{fig:ov_noise}a,b} 
	& \multicolumn{2}{l}{\ref{fig:ov_noise}c,d}  
	& \multicolumn{2}{l}{\ref{fig:ov_noise}e,f} 
	& \multicolumn{2}{l}{\ref{fig:w0}c,d} 
	& \multicolumn{2}{l}{\ref{fig:NN}c,d} 
	\\
\hline \hline
\end{tabular}
{\\ Notes: 
All models have intrinsic angular frequency $\omega_0=1$ and a periodic boundary condition. Initial conditions have a constant slope.  The initial phase differences are determined by the initial winding number $w_0$.   For integration series we show ranges for the varied parameters.
}
\end{table*}

\begin{figure}[ht]
\includegraphics[width=3.6truein]{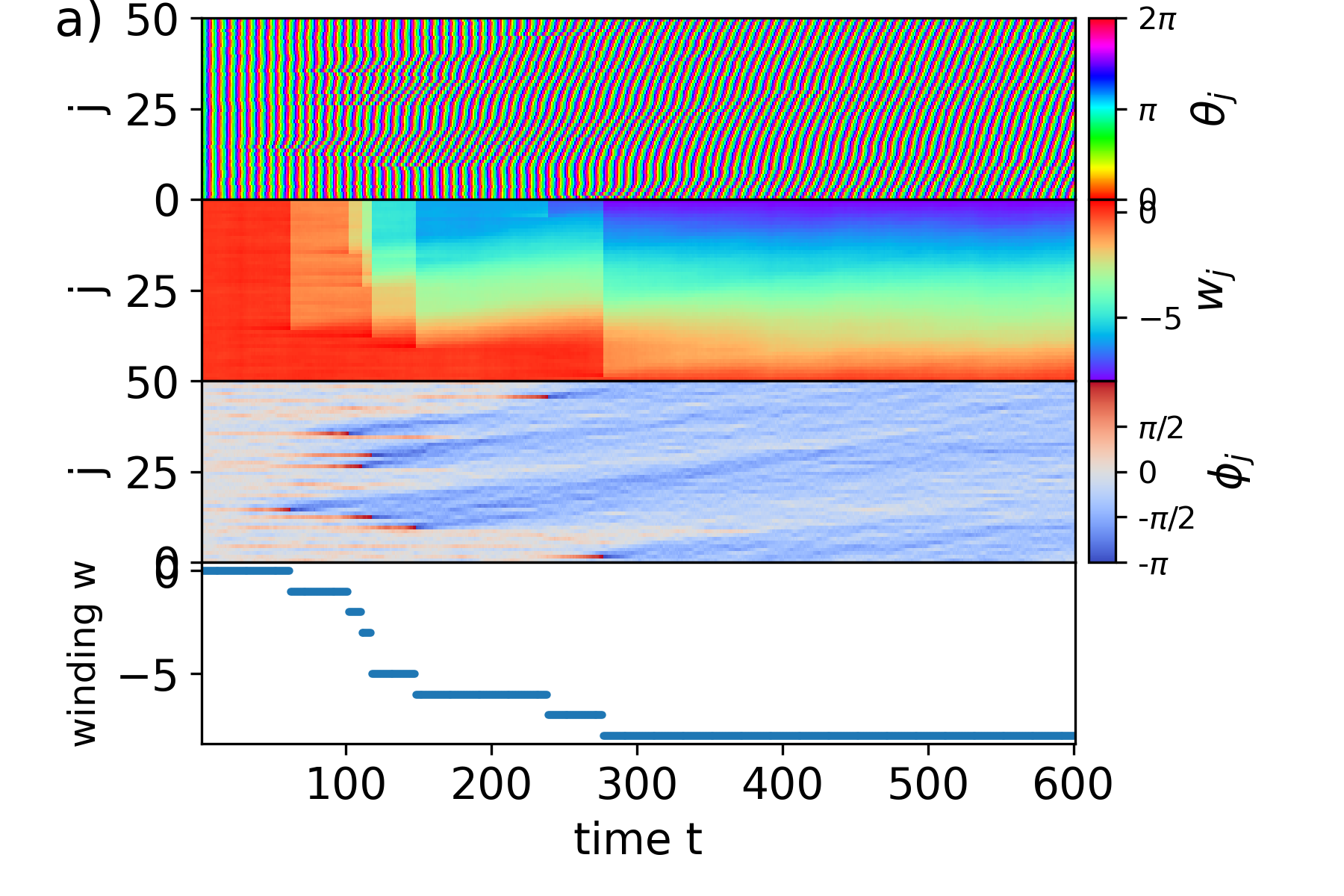}
\includegraphics[width=3.6truein]{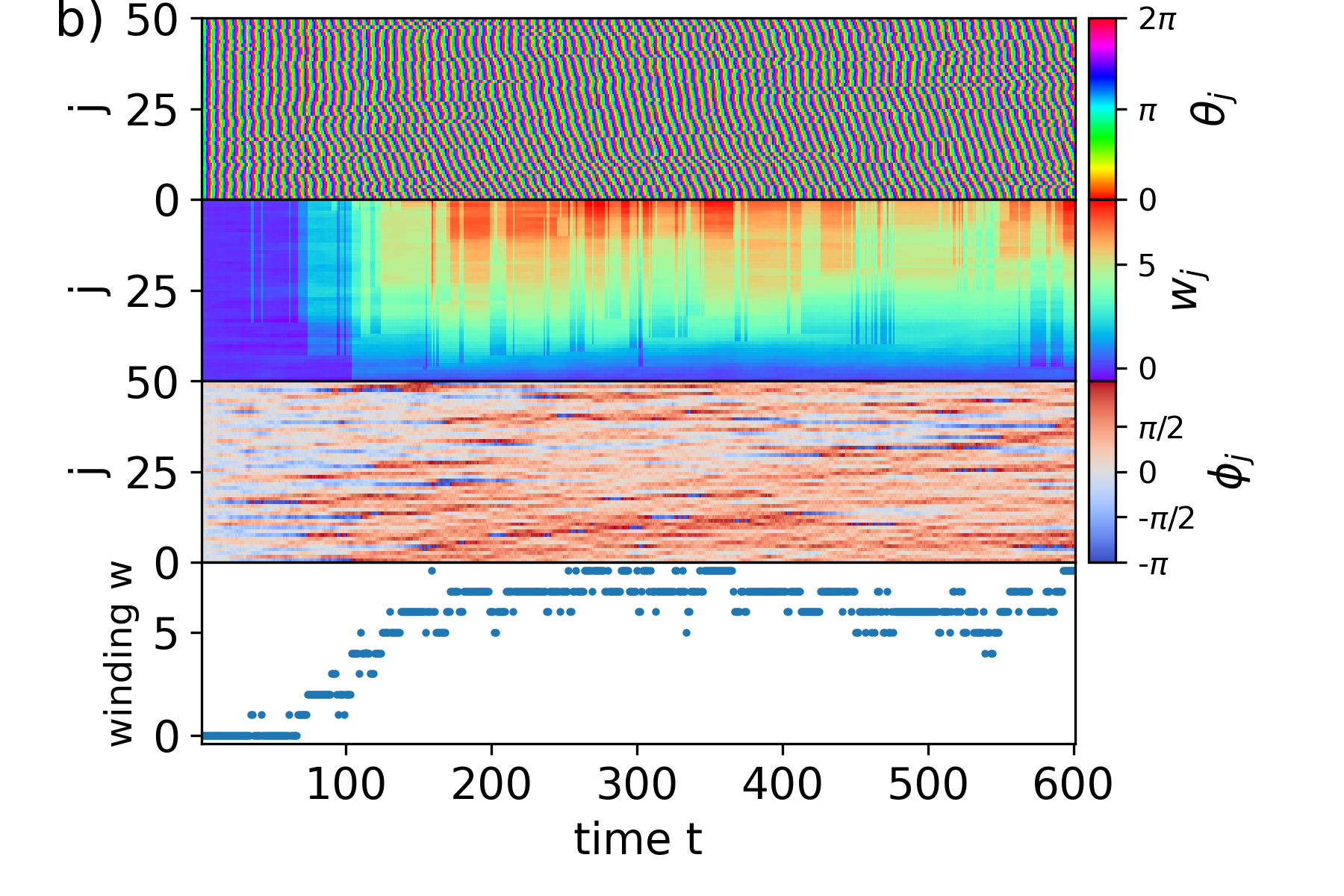}
\caption{Similar to Figure \ref{fig:ov1} except we show
two integrations of the stochastic directional sinusoidal model of Eqn.~\ref{eqn:noise}, with parameters listed in Table \ref{tab:noise}. 
Initially all oscillator phases are set to zero.    
a) We show the DWN1 integration.  This model has enough noise to seed 
perturbations that grow.
Groups of oscillators that have phase differences of about $\pi$ cause
jumps in the cumulative sum of phase differences, and these 
  cause variations in the winding number.  
These groups resolve into negative phase differences and the system enters a long-lived wave-like state. 
We attribute the later stability of the resulting wave to the slope dependent
diffusive term in the continuum equation (that $\propto \theta_x \theta_{xx}$).
b) We show the DWN2 integration. This model has $\mu_{c-}$ with the opposite
sign as the DWN1 model, so noise induced perturbations resolve
into a wave that travels in the opposite direction.
The noise strength is higher in this integration so only clusters of oscillators 
maintain a constant phase delay and the winding number continues
to vary.  
}
\label{fig:phase_eta}
\end{figure}

We run a series of integrations of Eqn.~\ref{eqn:noise}, 
where initial phases are all set to zero,
so the system begins in the synchronous state. 
In these models we set $K_+ = K_-$ and $\mu_{c+}=0$.
We vary the strength of the noise $\eta$ and the parameter
$\mu_{c-}$ which makes the model directional. 
The integrations denoted DWN1 and DWN2, have parameters listed in Table \ref{tab:noise} and 
 are shown in Figure \ref{fig:phase_eta}
In both integrations the noise seeds perturbations that grow sufficiently
large that they cause variations in winding number.
As the phase jumps resolve, the system enters a coherent wave state that persists. 

The DWN1 integration, shown in Figure  
\ref{fig:phase_eta}a has parameter $\mu_{c-}$ with opposite
sign to that of the DWN2 model, which is shown in Figure  
\ref{fig:phase_eta}b.   The sign difference causes the resultant waves
to be in opposite directions.   The DWN2 integration has stronger noise
than the DWN1 integration.  While the winding number remains 
constant at the end of the DWN1 integration, it continues to vary in the DWN2
integration.  The phase shift is fairly smooth in the DWN1 integration,
indicating that the phase shift is sufficiently high that perturbations caused
by the noise are damped diffusively.  We attribute the increased stability
to the strength of the slope dependent diffusion term, 
proportional to $\theta_x \theta_{xx}$, in the associated
continuum equation, 
is Eqn.~\ref{eqn:bimod_noise_cont}. 

The higher level of
noise in the DWN2 integration,  shown in Figure  
\ref{fig:phase_eta}b,  causes changes in the winding number to persist
throughout the integration.   While the winding number never drops to zero, 
variations in slope or phase shift persist and 
only clusters of oscillators 
maintain a constant phase delay.    This integration has a higher value
of the standard deviation of the phase shift than the DWN1 integration,
indicating that the wave is not entirely coherent.  There are regions or clusters
of oscillators in wave-like states with jumps in phase between them.  
The sensitivity of the collective motion to the strength of the noise
is consistent with the study by \citet{Solovev_2022} who found that
white noise could suppress synchronization in two-dimensional models of interacting phase oscillators. 

\begin{figure}[ht]
\includegraphics[width=3.7truein,trim = 10 0 0 0,clip]{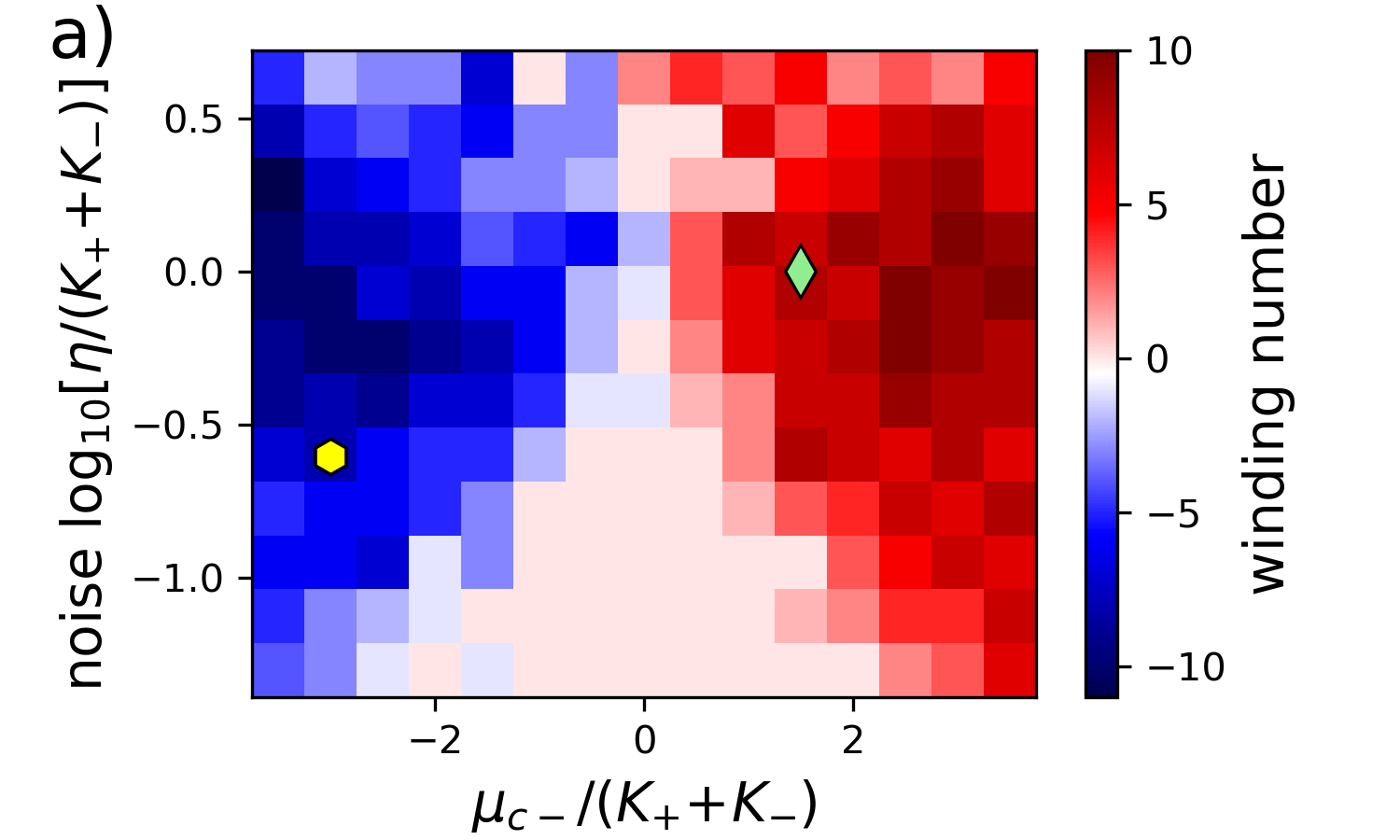}
\includegraphics[width=3.7truein,trim = 10 0 0 0,clip]{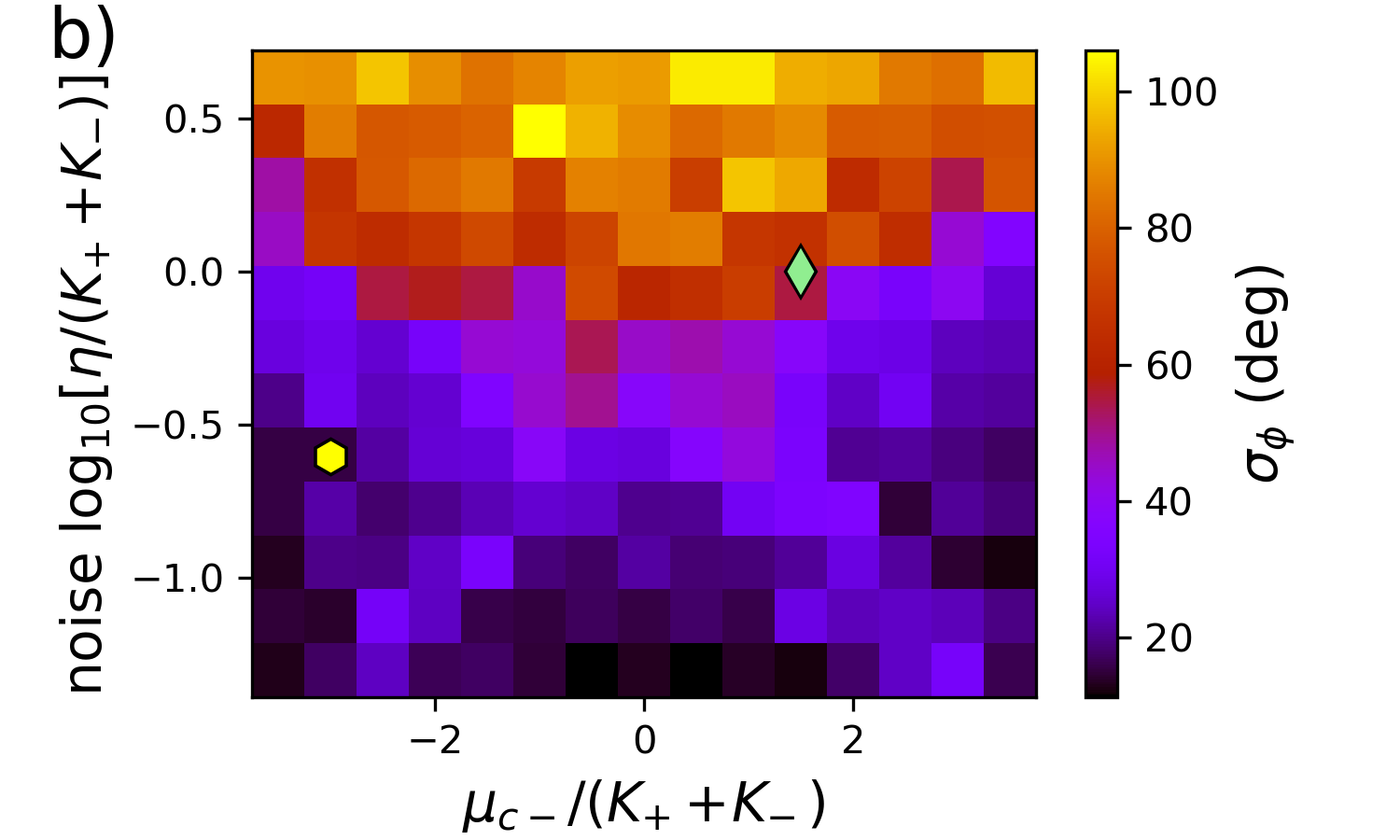}
\caption{We show integration series  denoted DWN-SerA with
different levels of white noise strength $\eta$ and parameter $\mu_{c-}$ for
the directional sinusoidal model of Eqn.~\ref{eqn:noise}.  
Parameters for the integrations are given in Table \ref{tab:noise}. 
a) We show final winding number as an image.
b) We show the standard deviation of the average phase difference $\sigma_\phi$
(defined in Eqn.~\ref{eqn:sig_phi}).
The blue diamond shows the DWN1 integration of Figure \ref{fig:phase_eta}a
and the yellow hexagon shows the DWN2 integration of Figure  \ref{fig:phase_eta}b.
The direction of the waves is set by the parameter $\mu_{c-}$.  
Simulations with sufficient noise and asymmetry in nearest neighbor interaction functions 
enter wave-like states, but if the noise is too strong, coherence is reduced or lost.  
\label{fig:red_blue}}
\end{figure} 

We run a series of integrations varying the strength
of the noise $\eta$ and the $\mu_{c-}$ parameter setting
the asymmetry in the interactions.  The series
is denoted DWN-SerA in Table \ref{tab:noise}.
At the end of each integration we record the winding number $w$ 
and the standard deviation of the phase difference $\sigma_\phi$.  Both quantities are 
plotted as images in Figure \ref{fig:red_blue}.  We use $\sigma_\phi$
to characterize the coherence of wave-like states at the end of the integrations. 

Figure \ref{fig:red_blue}a shows that wave-like states are long lived in the presence
of noise and it is possible to chose the direction of the waves by adjusting
the sign of the parameter $\mu_{c-}$.   In these integrations $\mu_{c+} = 0$.  More
generally the sign of $\mu_{c+} - \mu_{c-}$ would determine 
the direction of the waves.  This follows as this difference 
sets the sign of the $c_{2a}$ coefficient
which in turn determines the sign
of the $\theta_x \theta_{xx}$ term in the associated continuum equation 
(Eqn.~\ref{eqn:bimod_cont}).

The size of term that is proportional to $\theta_{xx}$ in the continuum equation 
(Eqn.~\ref{eqn:bimod_cont}) depends on $c_{1s} = (K_+ + K_-)$ (Eqn.~\ref{eqn:cofs_bimod}).   As the coefficient
is positive, this diffusive term damps short wavelength perturbations. 
The linearized stochastic continuum equation would resemble the Edwards-Wilkinson
equation, where the variance of the phase is sensitive to the ratio 
$\eta/(K_+ + K_-)$, which is why we use $\eta/(K_+ + K_-)$ on the $y$ axis in Figure \ref{fig:red_blue}.

Slope dependent instability depends on the size of the term that
is proportional to $\theta_x \theta_{xx}$  in the continuum equation  (Eqn.~\ref{eqn:bimod_noise_cont}).  This term depends on the coefficient $c_{2a} = -\mu_{c+} + \mu_{c-}$  
(Eqn.~\ref{eqn:cofs_bimod}).  The slope dependent stability condition
depends on the ratio of $c_{2a}$ and $c_{1s} = K_+ + K_-$ (Eqn.~\ref{eqn:condition})
which is why we use $\mu_{c-} / (K_+ + K_-)$ on the $x$ axis.
With noise able to cause jumps in phase
($\eta/(K_+ + K_-)$ not too small) and jumps in phase able
to cause changes in winding number $|\mu_{c-} |/ (K_+ + K_-) \gtrsim 1$, 
the system maintains a wave-like state.   With larger $\eta/(K_+ + K_-)$
the noise dominates over local synchronization causing the system
to loose coherence.  The system breaks up into clusters
of oscillators that are moving together.  This is evident in Figure \ref{fig:red_blue}b
showing the standard deviation of the phase shift  $\sigma_\phi$ (computed 
with Eqn.~\ref{eqn:sig_phi}).   When the wave
is coherent across the system, the standard deviation $\sigma_\phi$
is lower ($\lesssim 20^\circ$).  
When the system loses coherence and breaks up into small 
clusters, $\sigma_\phi$ is higher. 

Two integrations with the same 
value of $\eta$ and $\mu_{c-}$ can give different final winding numbers,
but the scatter in final winding numbers is not large. 
This can be seen from the differences in final winding between neighboring pixels
 in Figure \ref{fig:red_blue}a, as each pixel represents a single numerical integration. 
 
\begin{figure*}[ht]
\includegraphics[width=3.3truein,trim = 10 0 0 0,clip]{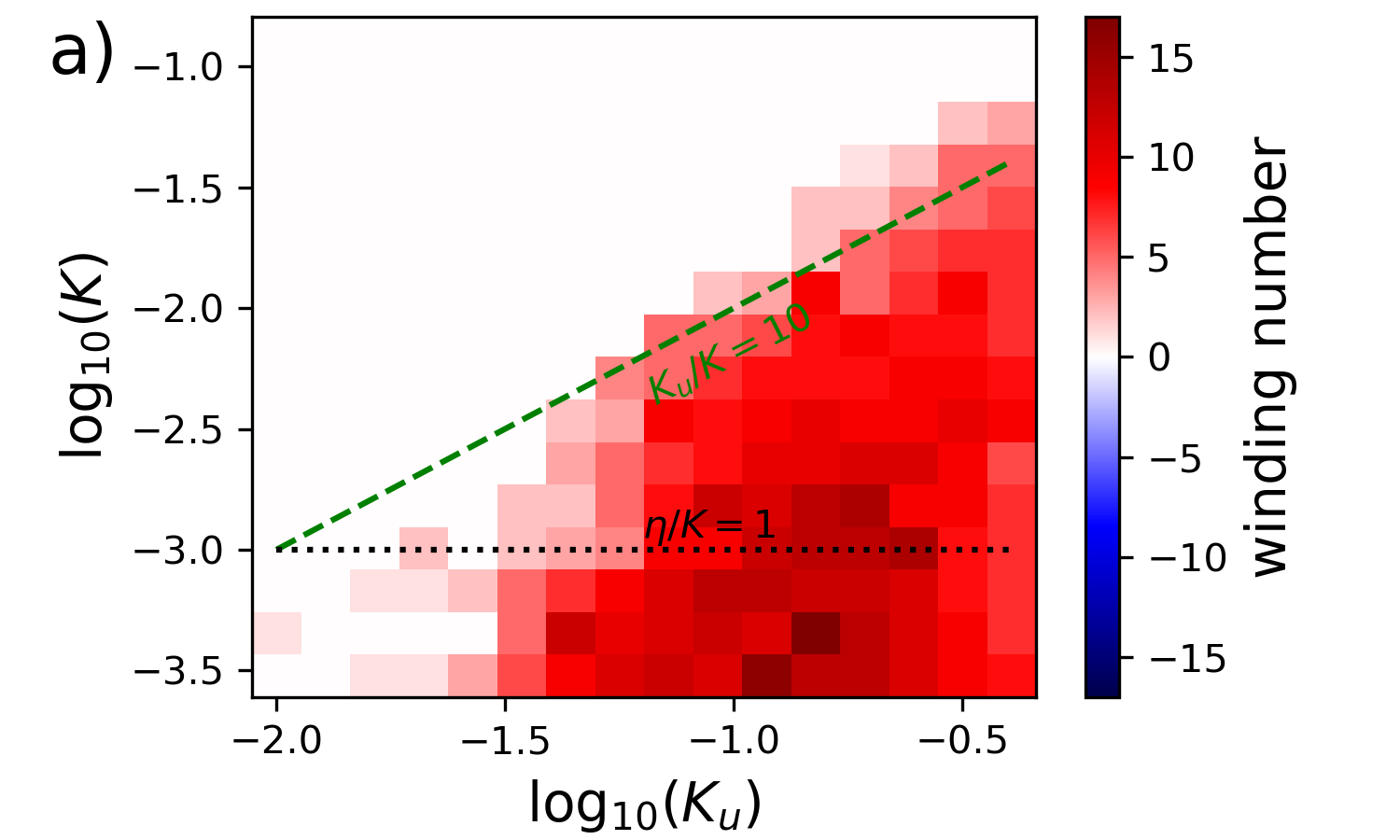}
\includegraphics[width=3.3truein,trim = 10 0 0 0,clip]{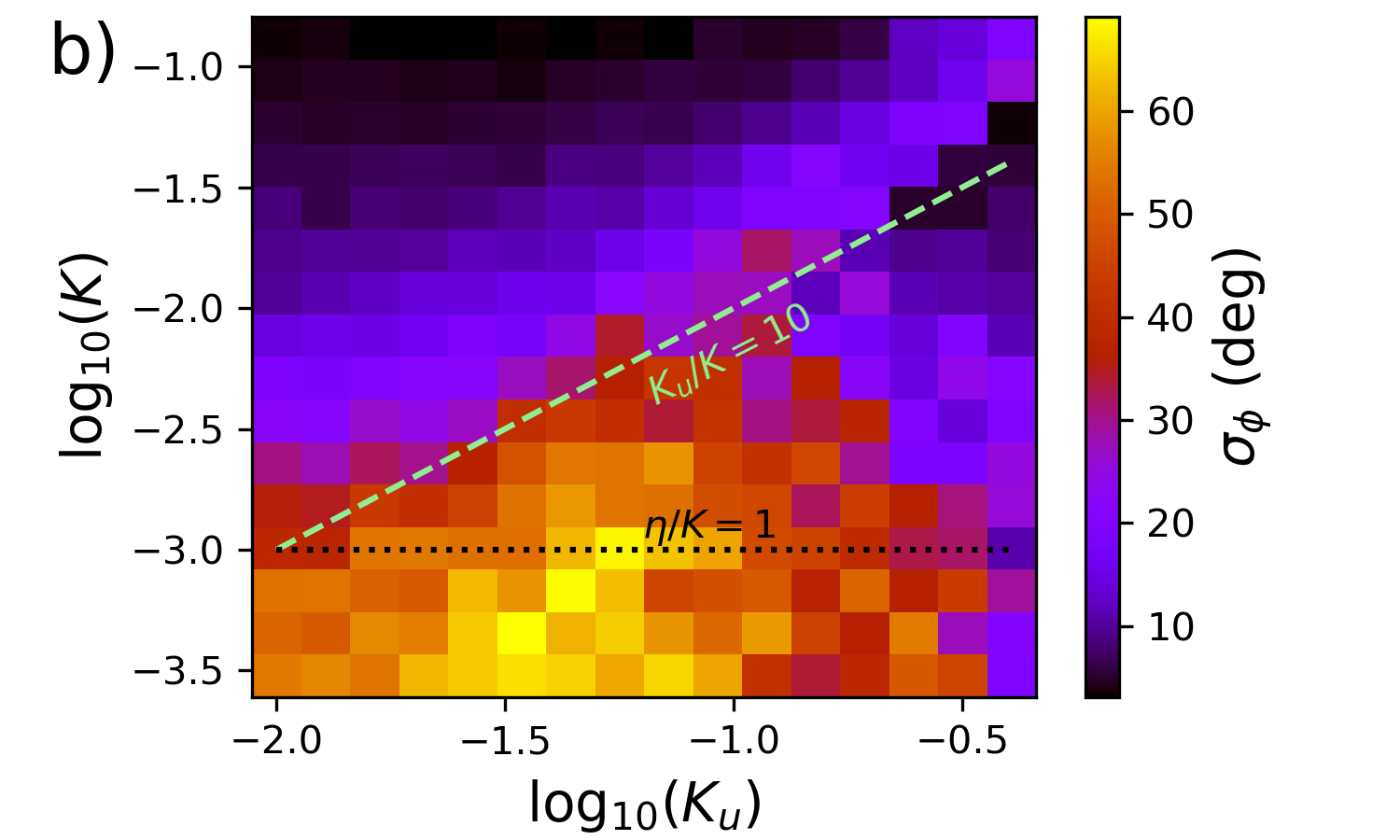}
\includegraphics[width=3.3truein,trim = 10 0 0 0,clip]{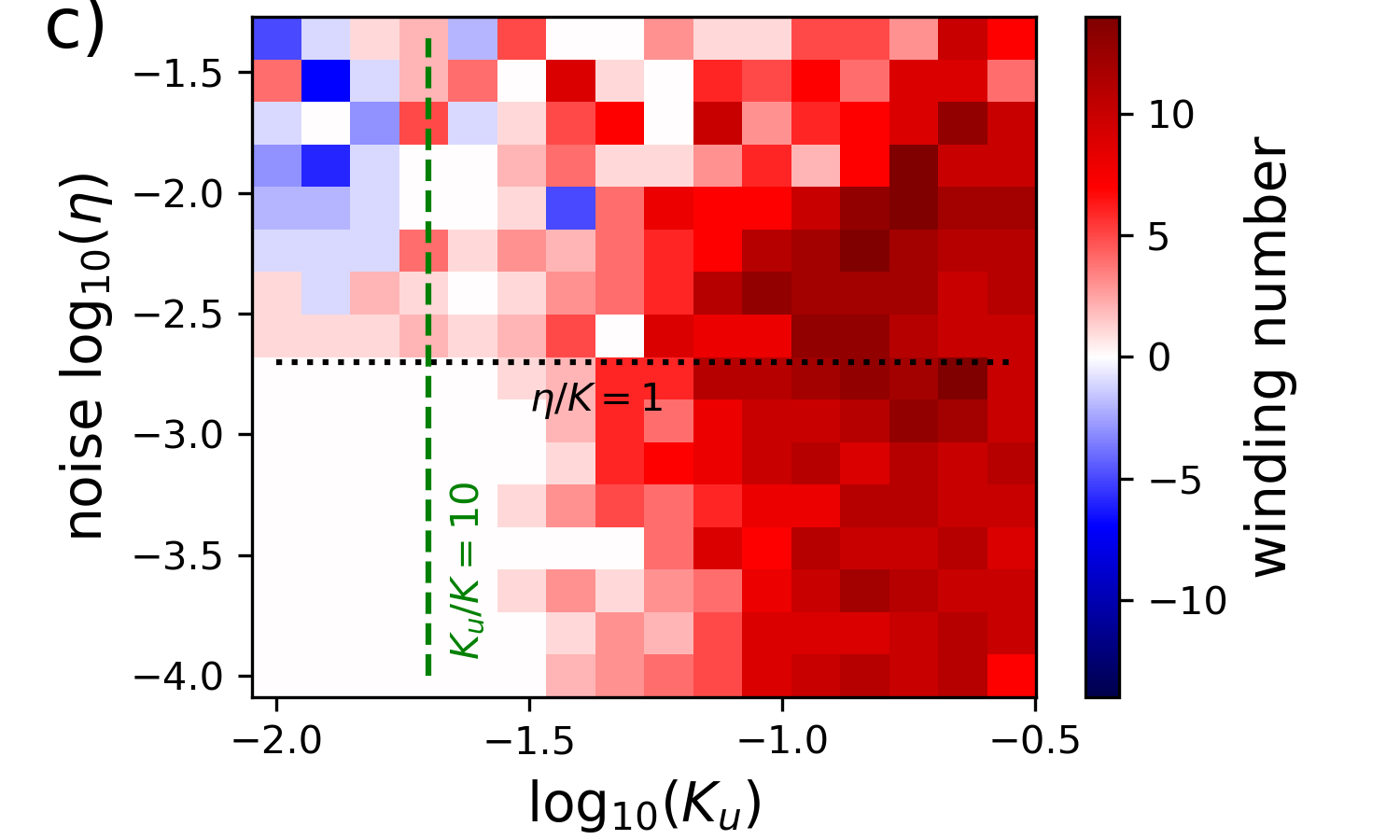}
\includegraphics[width=3.3truein,trim = 10 0 0 0,clip]{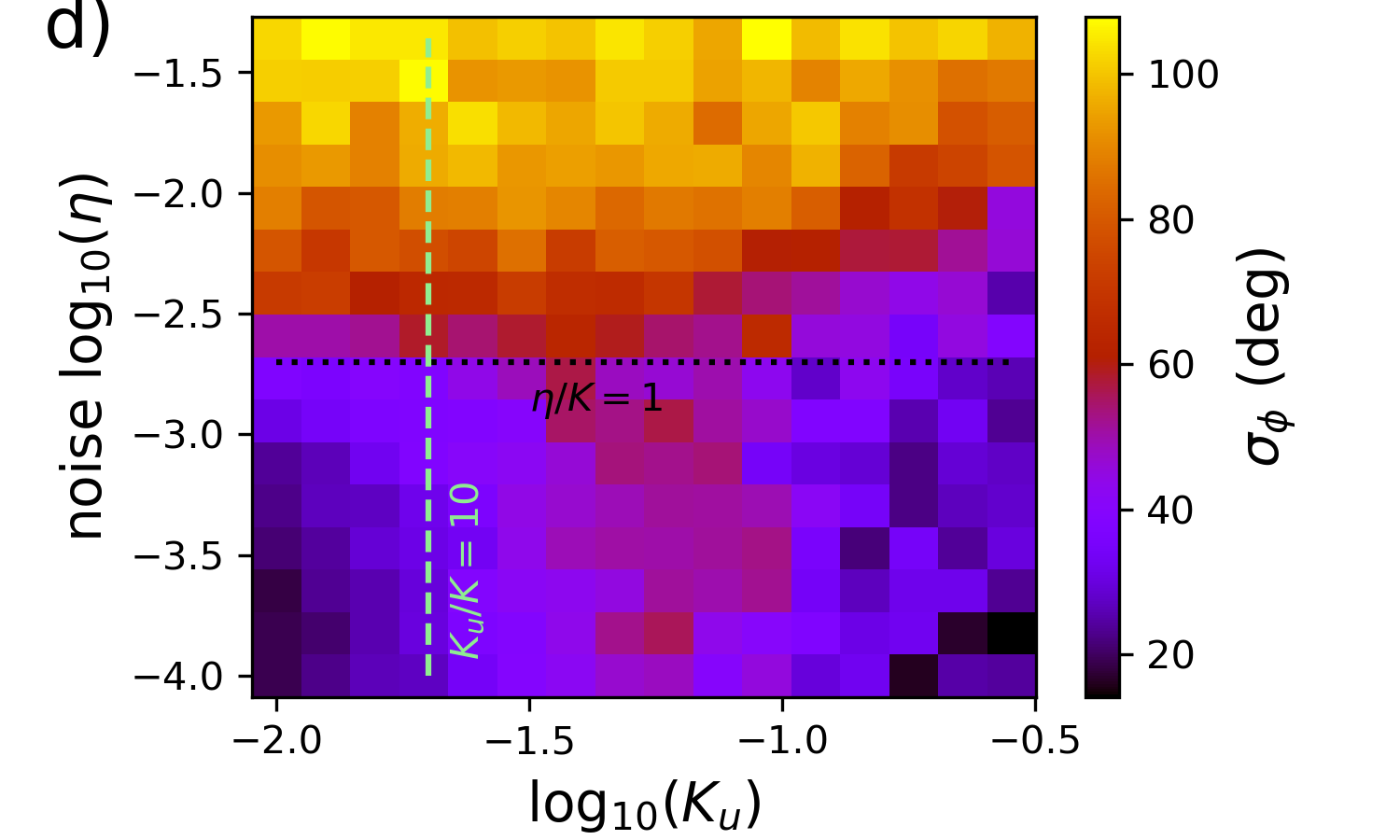}
\includegraphics[width=3.3truein,trim = 10 0 0 0,clip]{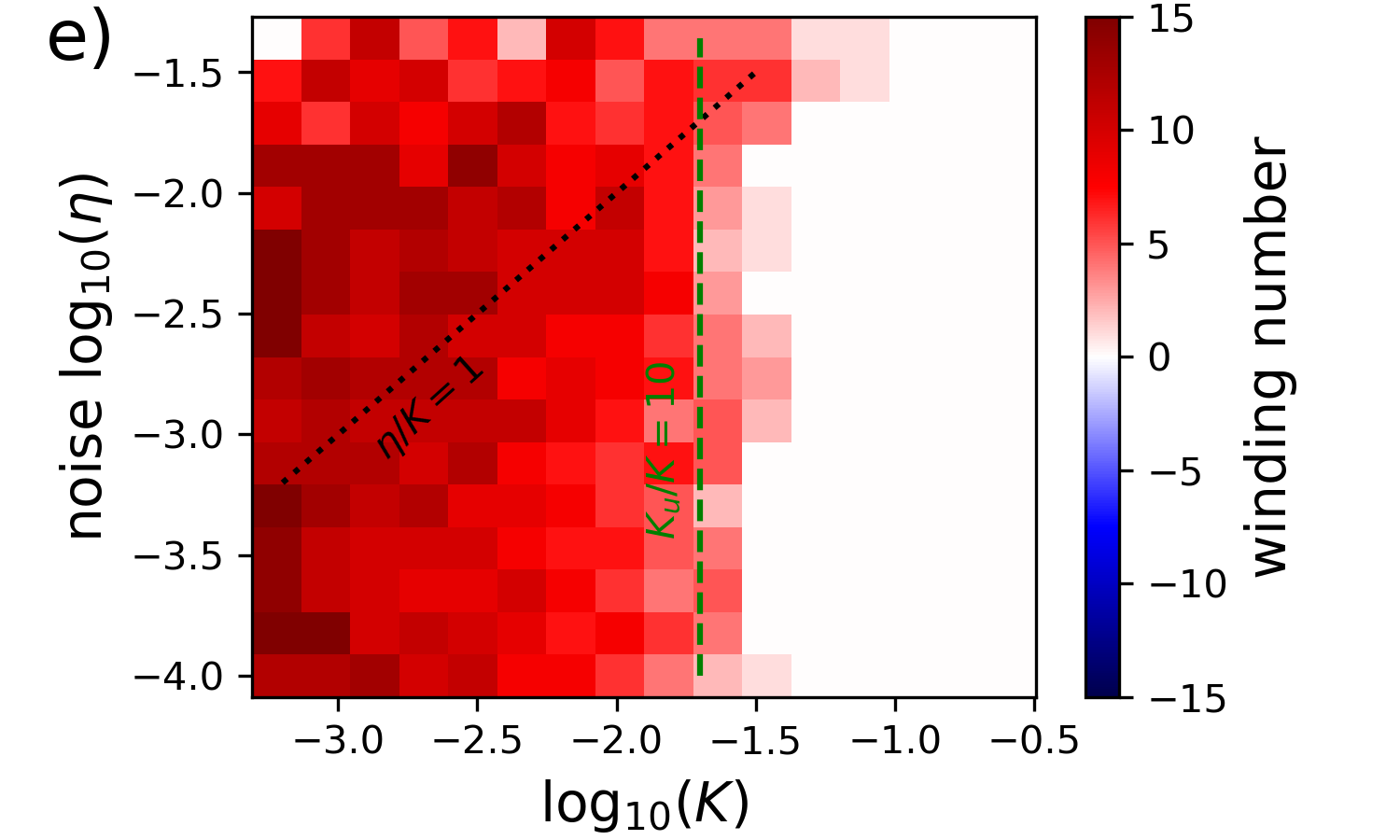}
\includegraphics[width=3.3truein,trim = 10 0 0 0,clip]{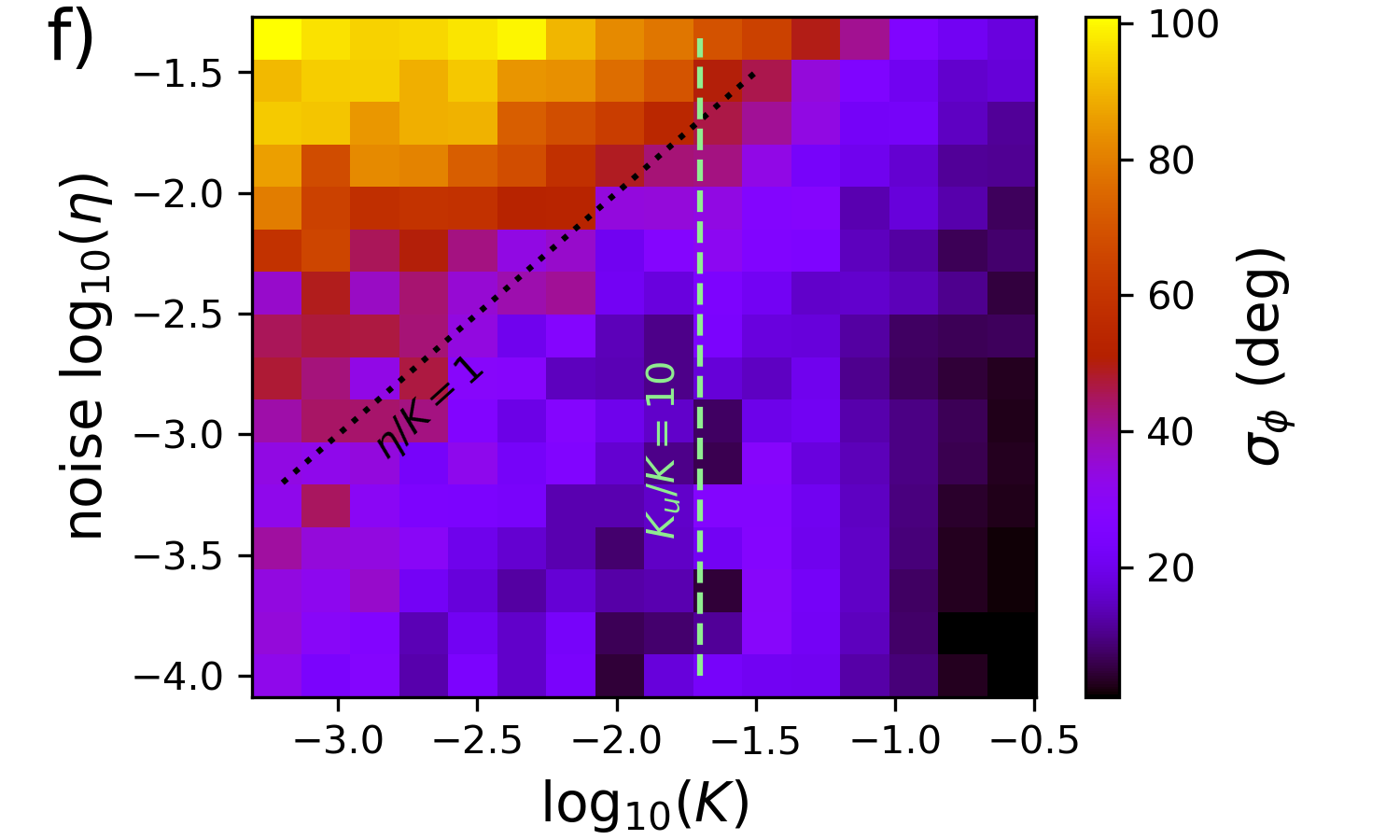}
\caption{a) 
We show final winding number as a image for the OWN-SerA integrations with
different  parameters $K$ and $K_u$ for
the model of Eqn.~\ref{eqn:uni_noise}.  
b) Similar to a) except we show the final value of the standard deviation 
of the phase shift $\sigma_\phi$ for the OWN-SerA integrations.
c) Similar to b) but showing the winding number for 
the OWN-SerB integrations with
different noise strength $\eta$ and parameter $K_u$. 
d) Similar to c) but showing the final value of the standard deviation 
of the phase shift $\sigma_\phi$ for 
the OWN-SerB integrations. 
e) Similar to a) except we show the final value of the standard deviation 
of the phase shift $\sigma_\phi$ for the OWN-SerC integrations
with different noise strength $\eta$ and parameter $K$.
f) Similar to c) but showing the final value of the standard deviation 
of the phase shift $\sigma_\phi$ for 
the OWN-SerC integrations. 
Parameters for the integrations are given in Table \ref{tab:noise}. 
The standard deviation  $\sigma_\phi$ of the phase shift is high if coherence is low.
The winding number remains zero if waves do not form as the integrations
begin in a synchronous state. 
Non-zero winding number and low $\sigma_\phi$ are typical of a coherent 
wave-state.   Sufficiently high noise strength 
(that with $\eta/K \sim 1$, with division shown with dotted black lines)
can cause the system to loose coherence by breaking into clusters of oscillators
that move together.  
Asymmetric interactions ($K_u/K \gtrsim 10$, with division shown with dashed green lines) facilitate wave formation.  
Text labeling the lines are on the side of the line where 
 where coherent metachronal waves can exist. 
}
\label{fig:ov_noise}
\end{figure*} 

\subsection{A modified overlap model with noise}

To the directional model discussed in section \ref{sec:unimod} (Eqn.~\ref{eqn:unimod}),
we add a Gaussian white noise term $\xi_i(t)$ with strength $\eta$ 
(with properties
as discussed at the beginning of  section \ref{sec:noise}).
The oscillator model is 
\begin{align}
\frac{d\theta_i}{dt}  = & \omega_0  - \frac{\omega_0 K_u}{2} \left[ 
{\rm tanh} \left( \frac{\cos \theta_{i-1} - \cos \theta_i - \beta }{h} \right)  + 1
\right]  \nonumber \\
&- K [\sin(\theta_i - \theta_{i+1}) + \sin(\theta_i - \theta_{i-1}) ]\nonumber \\
&+ \xi_i(t).
 \label{eqn:uni_noise}
\end{align}
The associated continuum equation is that of Eqn.~\ref{eqn:unimod_cont}
but with an additional white noise term
\begin{align}
\theta_t  = &\omega_0  - \frac{ \omega_0 K_u}{2}
\left[ \tanh \left( \frac{\beta}{h}\right) -1 \right]   + K dx^2 \theta_{xx} \nonumber \\
&  + \omega_0 K_u {\rm sech}^2\! \left( \frac{\beta}{h} \right) \!
\tanh \! \left( \frac{\beta}{h} \right)\! \frac{dx^2}{2h^2}
\left[ -(\theta_x)^2 \! + \! dx\ \theta_x \theta_{xx} \right] .
 \nonumber \\
 & + \xi(x,t) .
\label{eqn:unimod_noise_cont}
\end{align}

In Figures \ref{fig:ov_noise}a and b we show the final winding number 
and standard deviation $\sigma_\phi$ of the series OWN-SerA 
of integrations where we vary the parameter $K_u$ and the parameter $K$ that diffusively stabilizes the model.  
Figures \ref{fig:ov_noise}c and d are similar except they show  
the series OWN-SerB where we vary $K_u$ and the noise strength $\eta$.
Figures \ref{fig:ov_noise}e and f are similar, except they show 
the series OWN-SerC where we vary $K$ and the noise strength $\eta$.
The integrations
 have parameters, including those held fixed, listed in Table \ref{tab:noise}.
 On all panels in Figure \ref{fig:ov_noise} we show a dotted black line
 corresponding to $\eta/K = 1$ and a dashed green or light green line
 showing $K_u/K = 10$.   
 
 Wave generation, giving positive winding
 number at the end of the integrations, is seen to the right of
 the dashed green line in Figure \ref{fig:ov_noise}a and c
 and to the left of it in Figure \ref{fig:ov_noise}e.
 Low values of $K$ (diffusively preventing perturbations from growing) 
 and high values of $K_u$, giving strongly asymmetric interaction,     
 are required for wave generation, as seen in Figures 
 \ref{fig:ov_noise}a, c and e.
 With noise strength above $\eta > K $ and if waves are generated,  
 coherence is lost. This is seen in Figures \ref{fig:ov_noise}b 
 where the standard deviation of the phase shift $\sigma_\phi$ 
 is high below the $\eta/K=1$ line and in 
 Figures \ref{fig:ov_noise}d and f where
 $\sigma_\phi$ 
 is high above the $\eta/K=1$ line.  
 
Figure \ref{fig:ov_noise} illustrates that given a particular
 noise strength, the parameters $K$ and $K_u$  of equation 
 \ref{eqn:uni_noise} can be adjusted to put the system in a region
 of parameter space that allows waves to form and remain coherent. 
 
We find that clusters waves in a single direction tend to 
be generated for $K_u/K > 10$, independent of noise strength, with more 
coherent waves requiring larger values of $K_u$. 
These integrations have 
 parameters $h, \beta, \omega_0$  giving coefficient 
 $\bar c_{2a} \approx 27 K_u $ (evaluated using Eqn.~\ref{eqn:cofs_uni}).
The coefficient $c_{1s} = 2K$ for this dynamical system. 
The $K_u/K=10$, gives ratio $\bar c_{2a}/c_{1s} =  135$.
Using the stability criterion of Eqn.~\ref{eqn:condition2}, 
this gives a remarkably small phase shift of $\phi \sim 0.4^\circ$.
So even though we found that Eqn~\ref{eqn:condition2} could predict
the level of sinusoidal perturbations that cause instability (as discussed
in section \ref{sec:unimod}), if applied with a critical phase shift of order 1 radian, 
this criterion can underestimate the regime where noise 
can help drive clusters of waves in a single direction. 
In this respect, the stochastic directional model of Eqn.~\ref{eqn:uni_noise},
considered here, is more sensitive to noise than 
the sinusoidal directional model of Eqn.~\ref{eqn:noise},
discussed in the previous section, Sec.~\ref{sec:bi_noise}. 


\begin{figure*}[ht]
\includegraphics[width=3.3truein,trim = 0 0 0 0,clip]{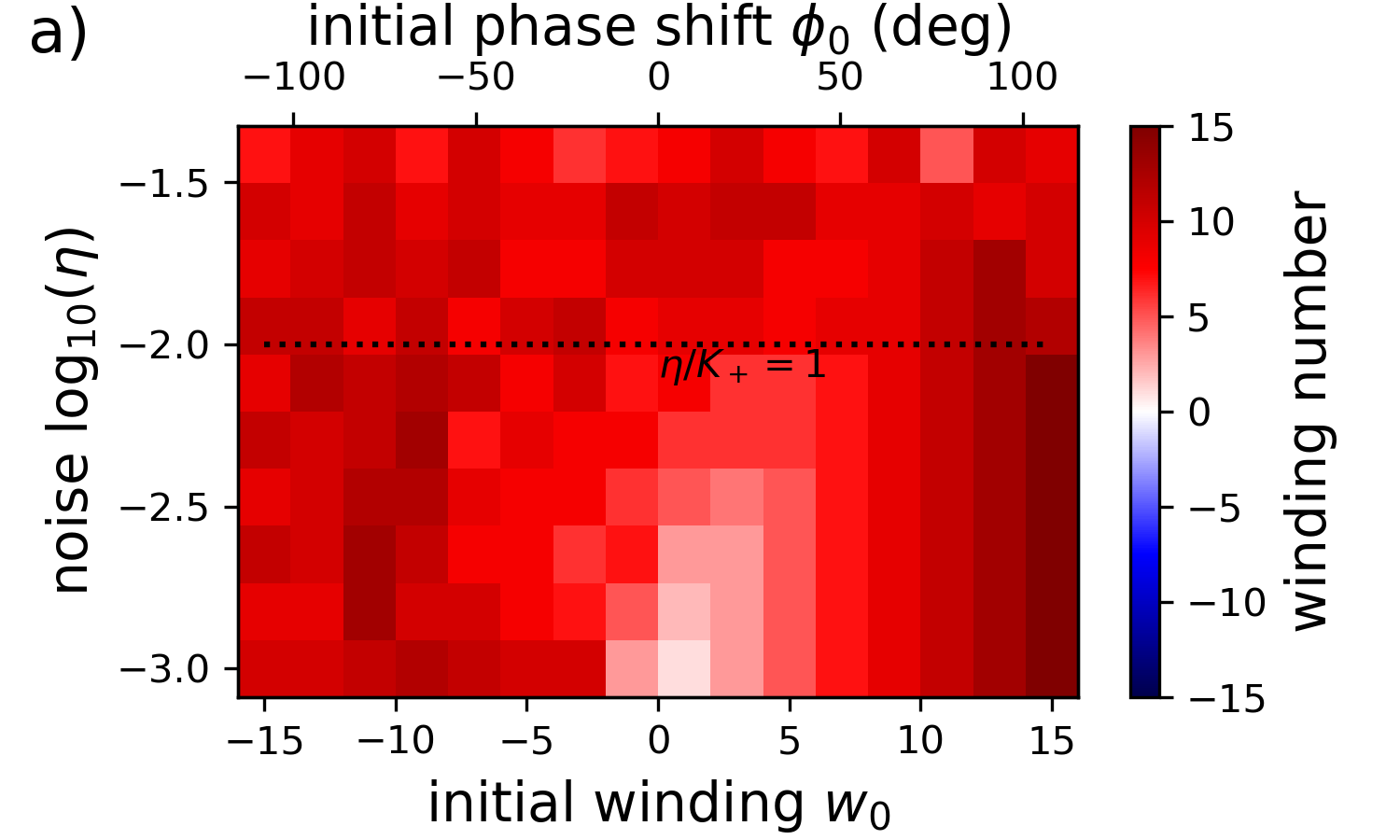}
\includegraphics[width=3.3truein,trim = 0 0 0 0,clip]{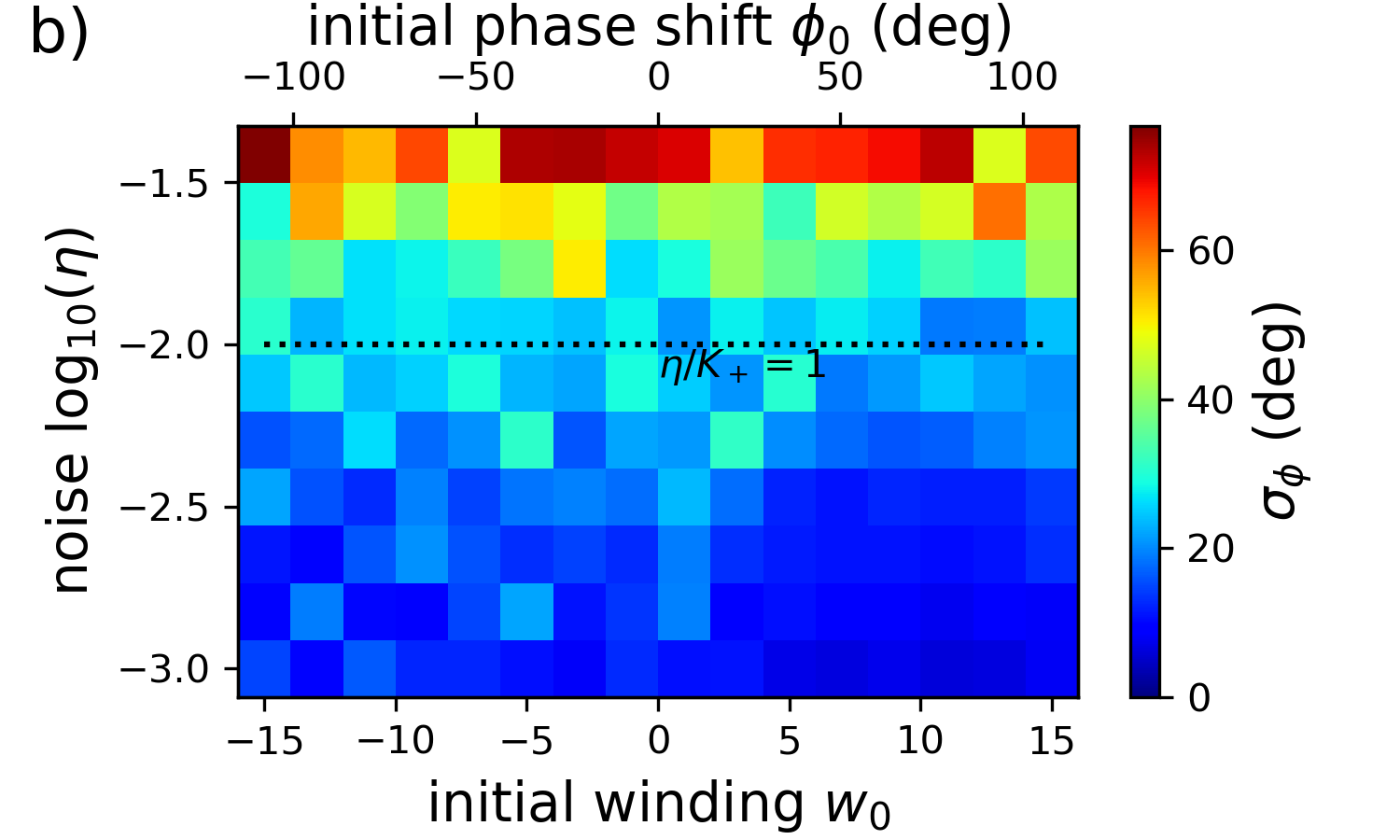}
\includegraphics[width=3.3truein,trim = 0 0 0 -10,clip]{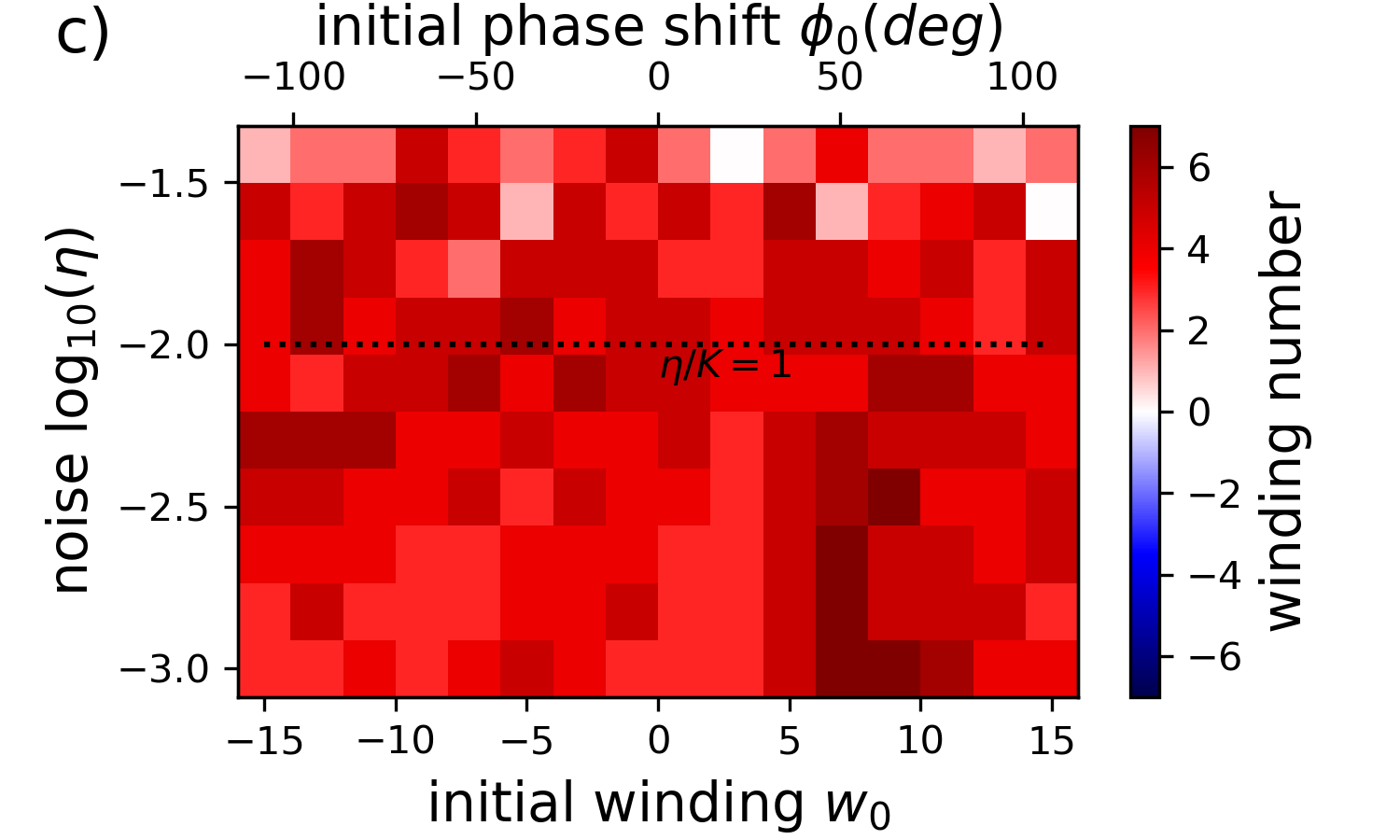}
\includegraphics[width=3.3truein,trim = 0 0 0 -10,clip]{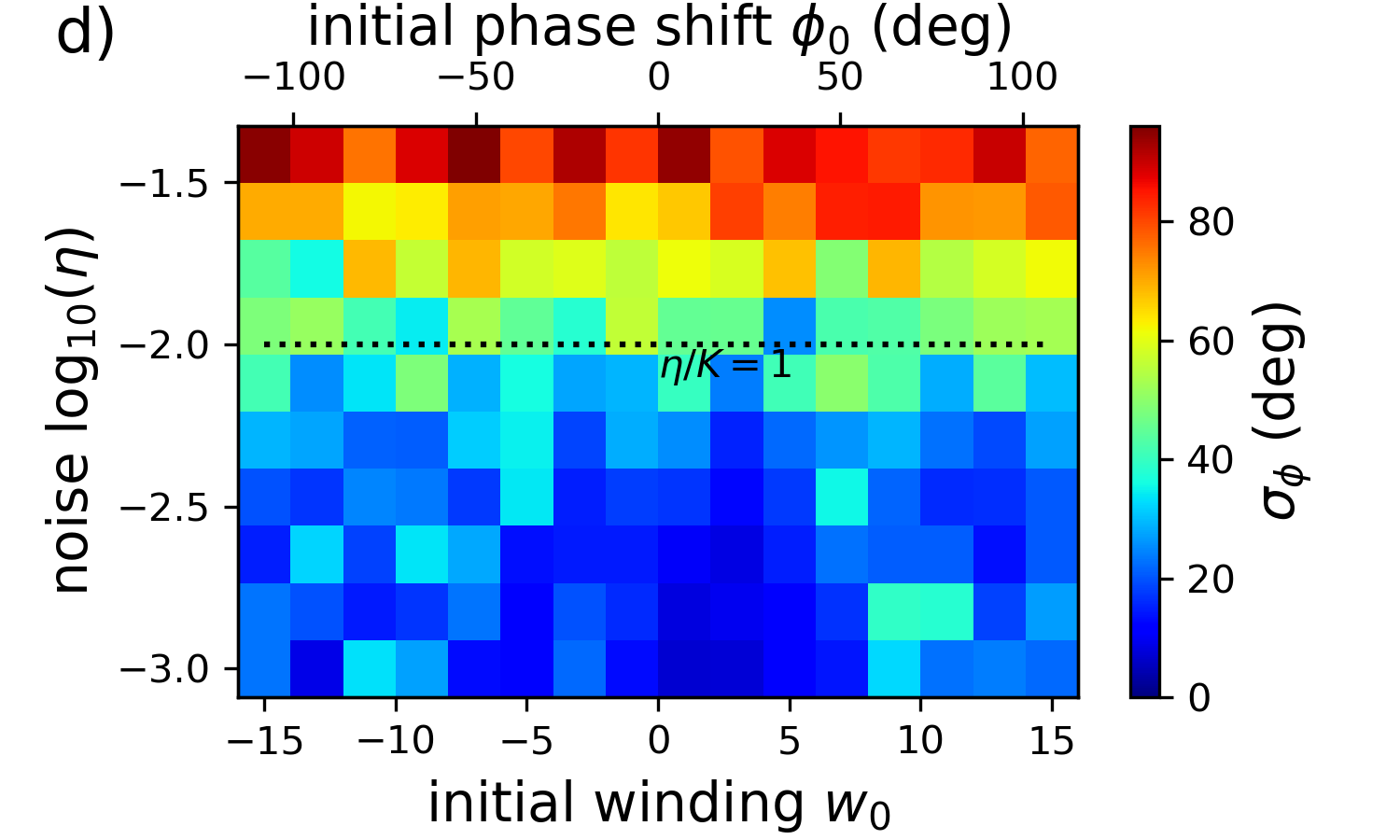}
\caption{Sensitivity of final winding number and standard deviation 
of phase difference to initial winding number $w_0$ and noise strength $\eta$.  
Initial conditions have a constant slope with phase difference determined by
the initial winding number.  Parameters
for each integration series are given in Table \ref{tab:noise}.
a) Final winding number is shown as a function of initial winding number $w_0$
(on the $x$ axis) and noise strength (on the $y$ axis).  
Initial phase differences are shown in degrees on the top $x$ axis.  
We show the DWN-SerW integration series which are for the stochastic sinusoidal directional model of Eqn.~\ref{eqn:noise}.   
b)  The standard deviation of the phase difference at the end of the integrations
also for the DWN-SerW integrations. 
c) Similar to a) but showing winding number for the OWN-SerW integrations.  
These are for
the stochastic directional overlap model of Eqn.~\ref{eqn:uni_noise}. 
d) Similar to b) but showing $\sigma_\phi$ for the OWN-SerW integrations. 
For both models a positive initial slope can be stable, and 
the system would not exhibit variations in 
winding number, giving sensitivity of the final state to initial conditions. 
The regions where there are no changes in winding number
have constant color in the vertical direction in panels a and c. 
For the stochastic directional overlap model (shown in panel c), the region where initial winding number is equal
to the final one, on the lower middle right, is much smaller than for the stochastic sinusoidal directional 
model (shown in panel a).  
The stochastic overlap model is less sensitive to initial conditions
and so more robustly gives metachronal wave states. 
\label{fig:w0} }
\end{figure*}

\subsection{Sensitivity of stochastic directional models to initial mean phase shift or slope}
\label{sec:w0}

The integrations shown in Figures \ref{fig:phase_eta},  \ref{fig:red_blue},
and \ref{fig:ov_noise}
began with all oscillators set to zero, so the initial winding number, slope  and mean
phase shift are all zero.  Because of the slope dependent diffusive term in the associated
continuum equation, perturbations caused by noise can grow.   The system
increases or decreases in slope, depending upon the sign of $\mu_{c+} - \mu_{c-}$
in the stochastic sinusoidal directional model of Eqn.~\ref{eqn:noise}, or the sign
of $K_u$ in the stochastic overlap model of Eqn.~\ref{eqn:uni_noise}.
What if the initial condition was a smooth ramp, so that the initial winding 
number and slope is non-zero?   If the slope's sign allows perturbations
to grow,  then the integrations
evolve, as shown in Figure \ref{fig:phase_eta}, \ref{fig:red_blue}, and \ref{fig:ov_noise}, 
until
the system reaches a winding number and associated slope that is stable.
However if the slope's sign is in the opposite direction, the system could remain 
sufficiently stable that the winding number would remain fixed. 
This would imply that the long-lived states of the stochastic models 
can be sensitive to initial conditions.   To investigate this possibility 
we explore simulations with initial conditions that are linear ramps,
with a single phase shift between neighboring oscillators. 

In Figure \ref{fig:w0} we show series of integrations for
both stochastic models, denoted the DWN-SerW and OWN-SerW simulations,
where we vary initial phase shift and noise strength. 
Initial conditions are ramped so that the phase shift between neighboring oscillators
is fixed and determined from the initial winding number via
Eqn.~\ref{eqn:barphi}.
 In all panels we show on the top $x$ axis the initial phase difference
$\phi_0$ in degrees. 
Parameters of the simulations are listed in Table \ref{tab:noise}.
In Figures \ref{fig:w0}a and c we show final winding number at the end of the integrations 
and in Figures \ref{fig:w0}a and c we show the standard deviation of the phase shifts, 
$\sigma_\phi$. 

Positive slopes (corresponding to positive $w_0$) are more stable 
for both sinusoidal stochastic model and stochastic overlap models
shown in Figure \ref{fig:w0}.
Figure \ref{fig:w0}a, showing the stochastic sinusoidal directional model, has a
 region on the lower right, with $w_0$ ranging from 1 to 15,
  giving 
final winding number that is equal to the initial one. 
The region appears to have vertical bars with the same color. 
The region contains 
 integrations that did not vary 
in winding number.   Thus the integrations began in a stable 
state.  With noise sufficiently strong (on the top right), variations
in winding number occur, but at the expense of coherence in
the resulting wave-like states, as seen in Figure \ref{fig:w0}b.

The stochastic overlap directional model has a similar region
on the lower right in Figure \ref{fig:w0}c (with $w_0$ ranging from 1 to 5)
but it is much
smaller than the stable region in Figure \ref{fig:w0}a.
The stochastic overlap model is more sensitive
to the growth of instabilities from noise than the sinusoidal 
stochastic directional model.   
For the sinusoidal model, the ratio $c_{1s}/c_{2a}$ 
(derived from the stability criterion of Eqn.~\ref{eqn:condition2}) 
corresponds to an
unstable angle of $19^\circ$, whereas for the stochastic overlap
model $c_{1s}/c_{2a}$ is only $0.2^circ$. The difference
between these ratios 
could in part account for the different behavior of the two models. 
We increased the $\mu_{c-}$ parameter in
the sinusoidal stochastic model but we did not see the 
stable region in $w_0$ significantly decrease in size. 
We suspect that 
the shape of the interaction functions influences 
their behavior and the criterion of Eqn.~\ref{eqn:condition2}
 is not sufficient to fully characterize the behavior
 of the stochastic models.  
 
\begin{figure*}[ht]
\includegraphics[width = 3truein]{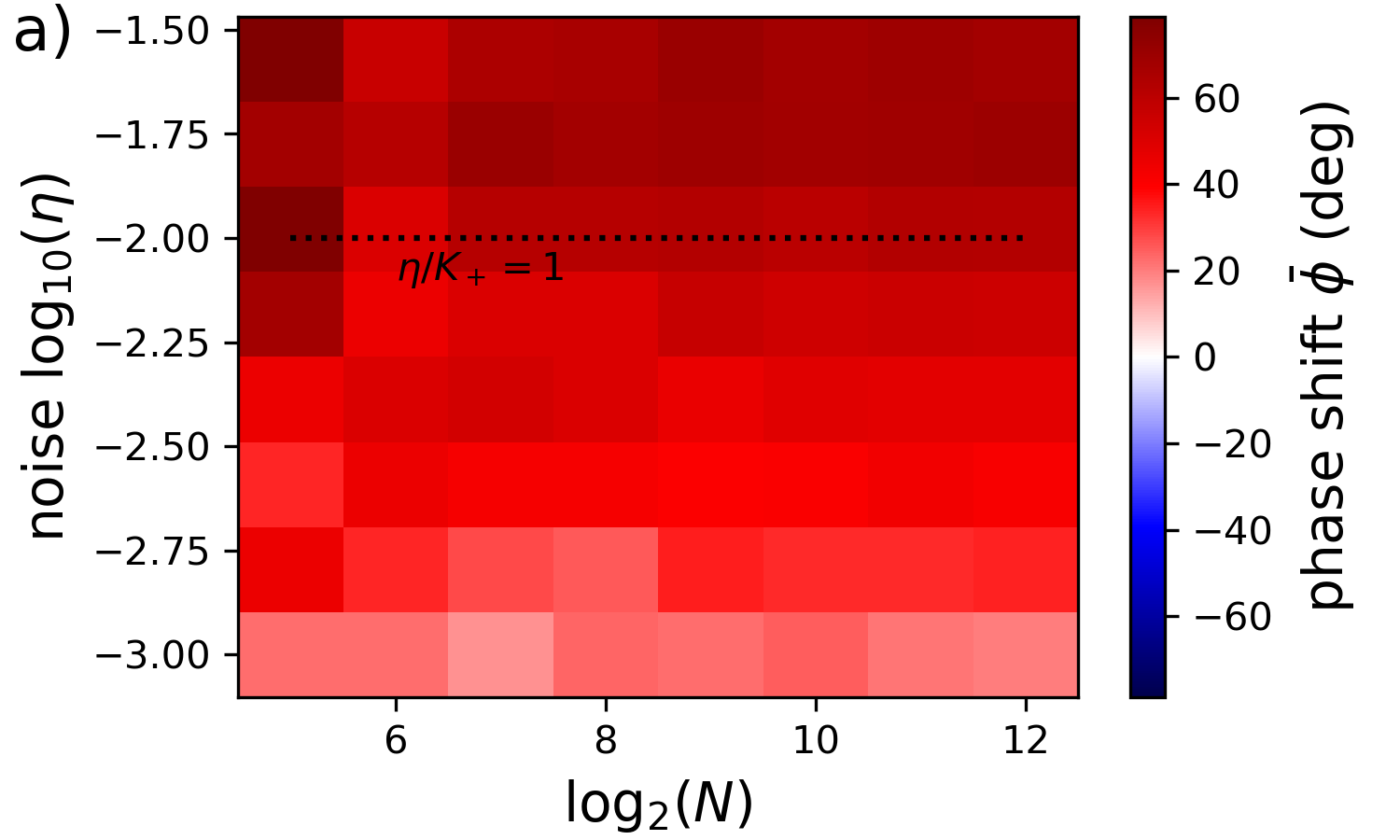}
\includegraphics[width =3truein]{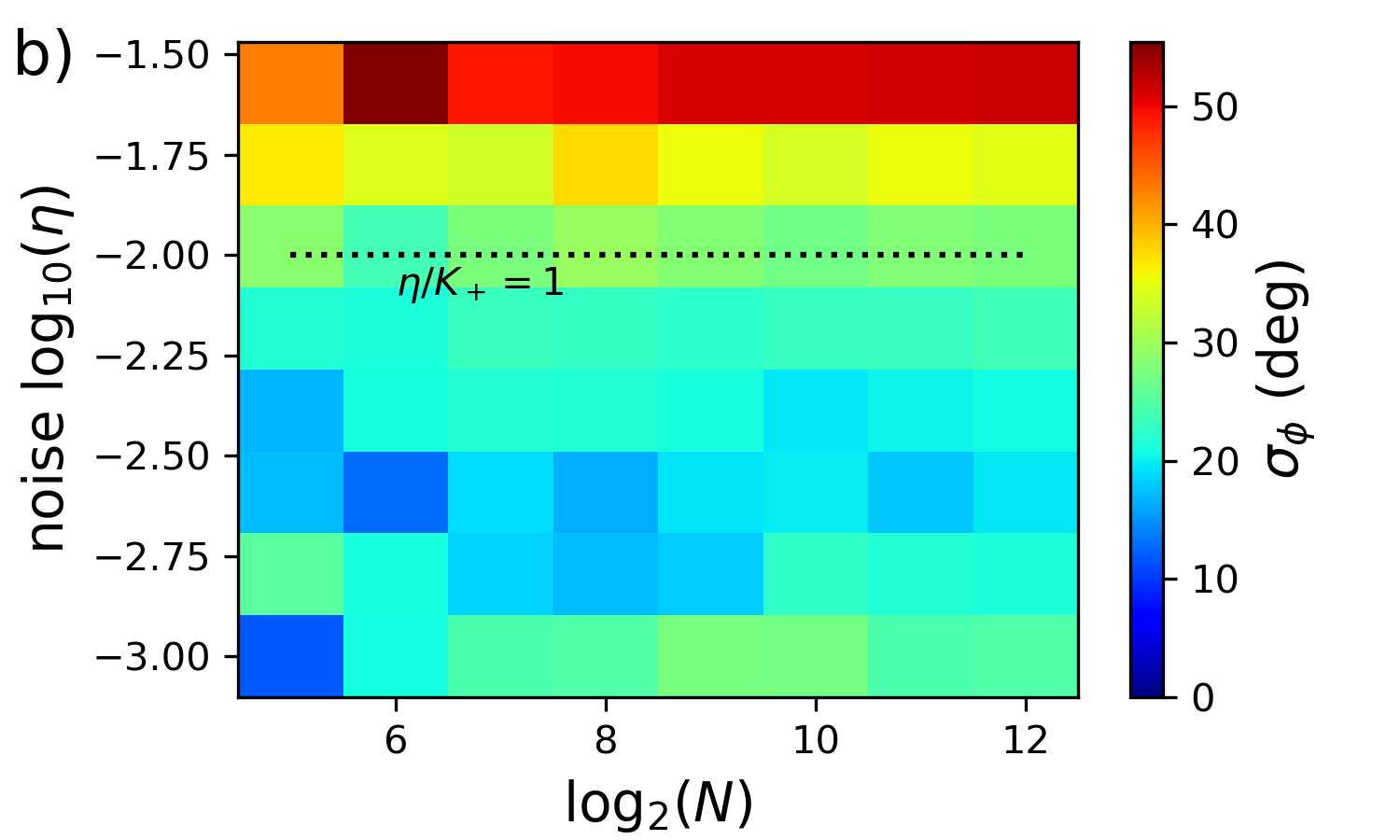}
\includegraphics[width =3truein]{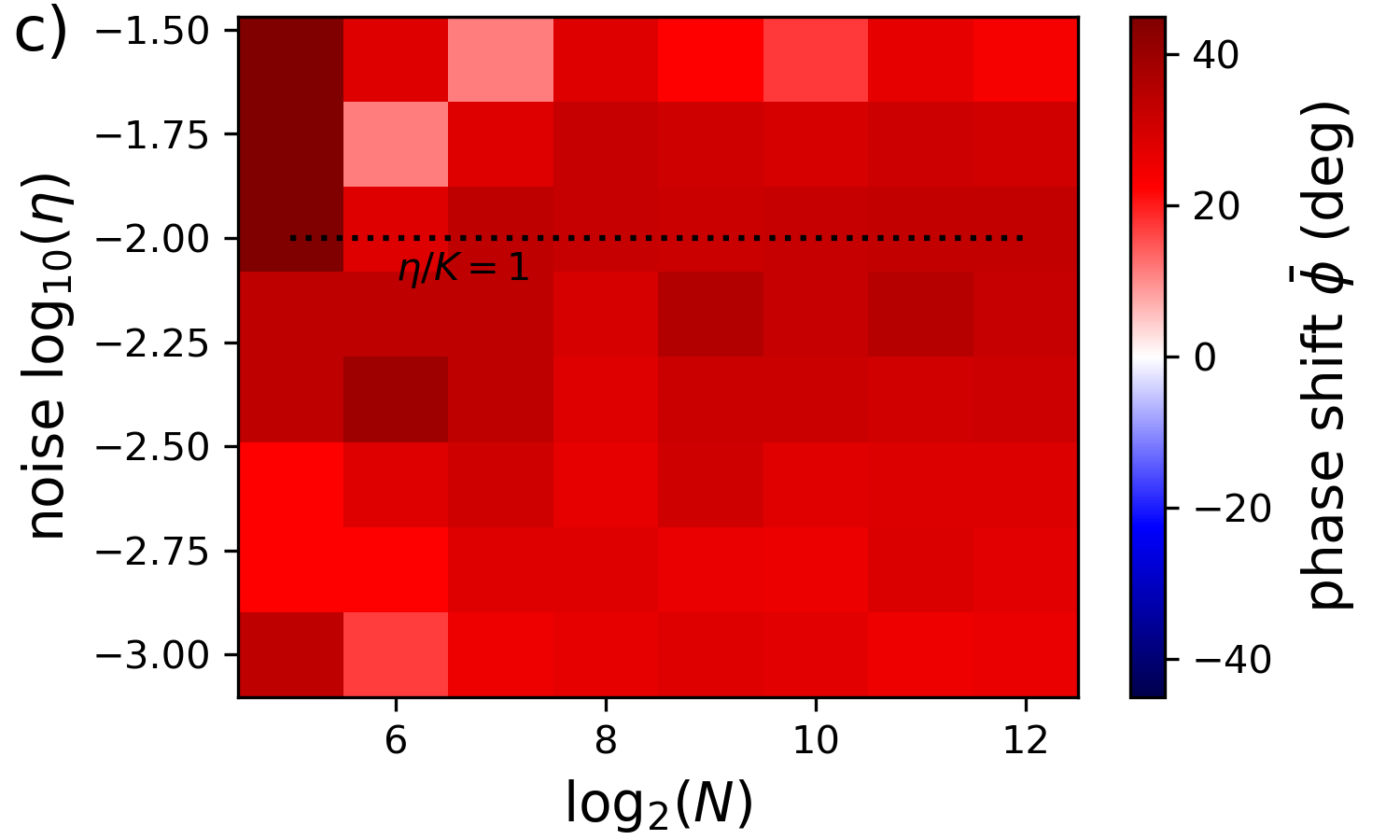}
\includegraphics[width =3truein]{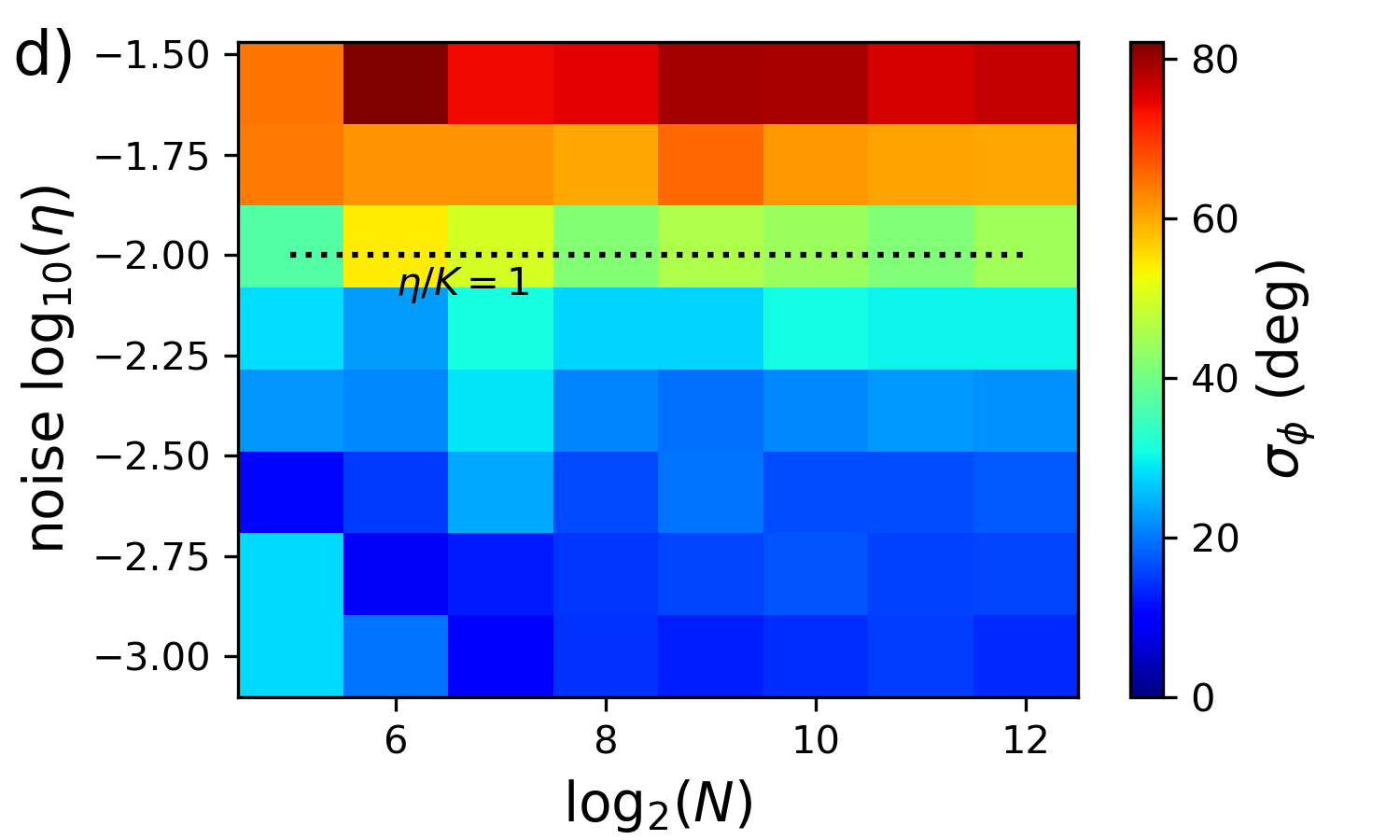}
\includegraphics[width =3truein]{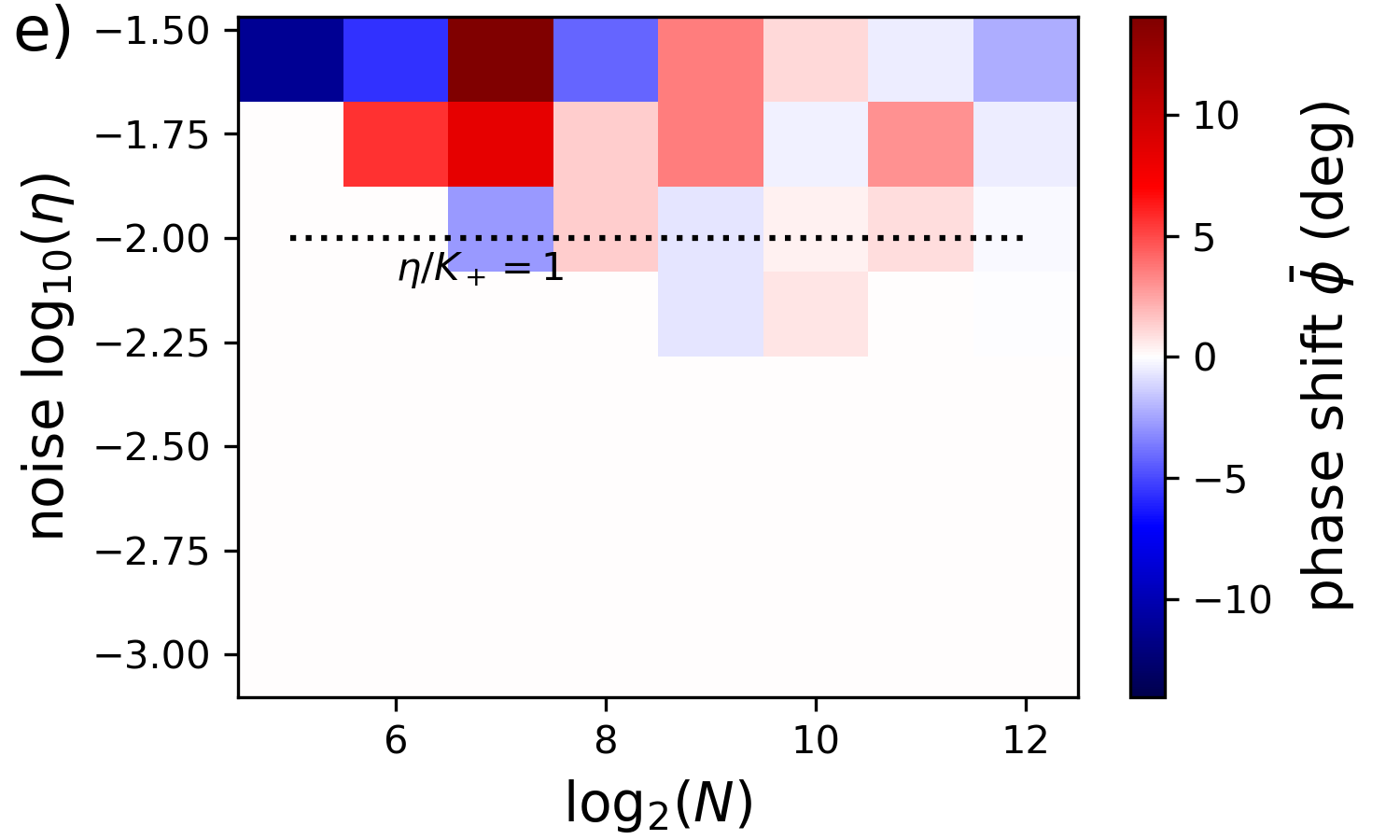}
\includegraphics[width =3truein]{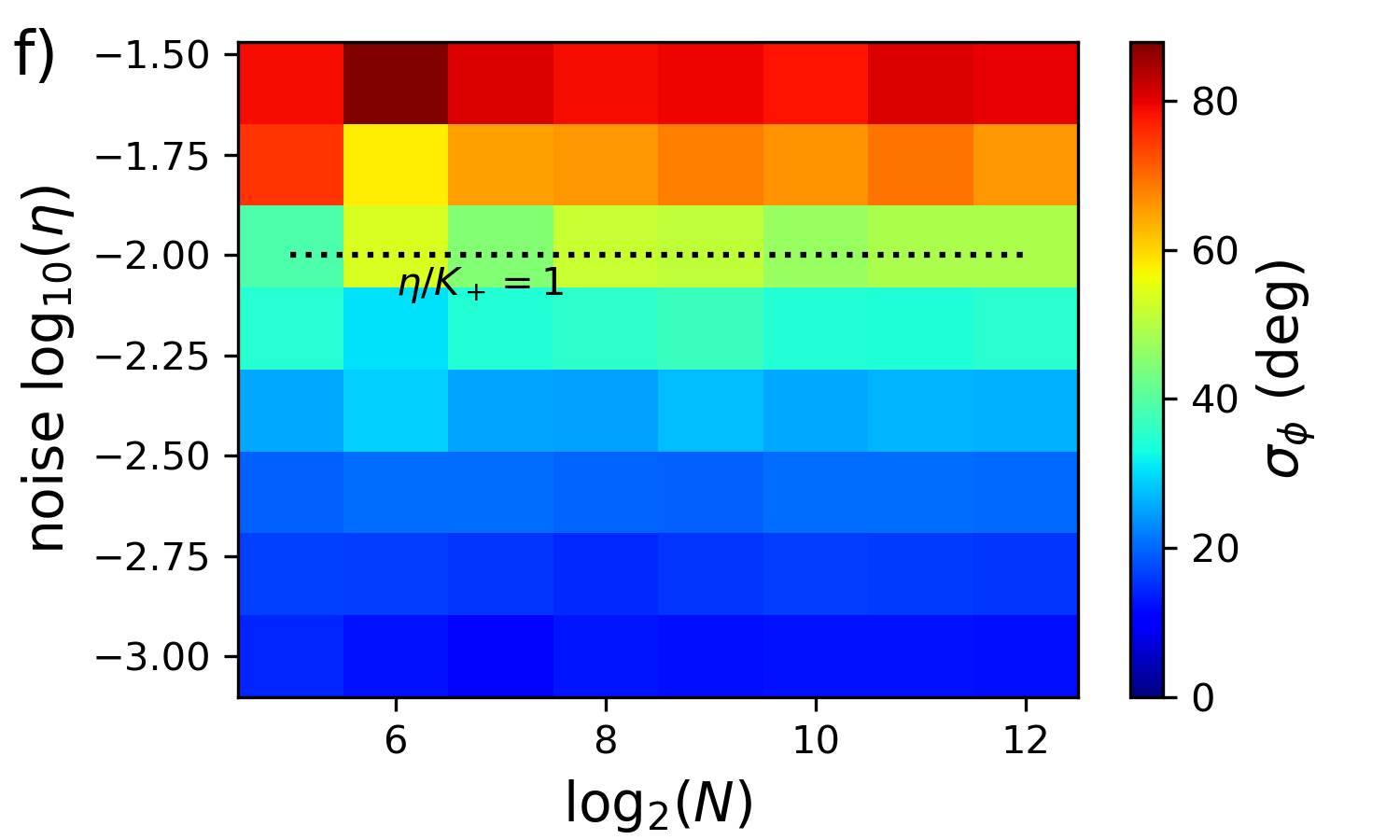}
\caption{
Sensitivity of final mean phase shift and standard deviation 
of the phase shift to noise strength $\eta$ and number of oscillators $N$.  
All phases are initially set to zero.  Parameters
for each integration series are given in Table \ref{tab:noise}.
a) Final mean phase shift $\bar \phi$ is shown as a function of $N$ (on the $x$ axis) and noise strength (on the $y$ axis).  
We show the DWN-SerN integration series which are for the stochastic sinusoidal directional model of Eqn.~\ref{eqn:noise}.  
b) The standard deviation of the phase difference at the end of the integrations
also for the DWN-SerN integrations. 
c) Similar to a) but showing winding number for the OWN-SerN integrations.  
These are for
the stochastic directional overlap model of Eqn.~\ref{eqn:uni_noise}. 
d) Similar to b) but showing $\sigma_\phi$ for the OWN-SerN integrations. 
e) Similar to a) but showing the BWN-SerN integrations which are of a bidirectional model.
The parameters and model are the same as for the DWN-SerN integration, (shown in a)
except the parameter giving asymmetry $\mu_{c-}=0$. 
f) similar to b) except showing $\sigma_\phi$ for the BWN-SerN integrations. 
In the directional models, we find that the mean and standard deviation of the
phase shift, $\bar \phi$ and $\sigma_\phi$, are insensitive to the number of oscillators
in the loop.   
  \label{fig:NN}}
\end{figure*}

\subsection{Sensitivity to the number of oscillators}

Following \citep{Pikovsky_2003,Solovev_2022}, 
the variance of the phases in a chain of interacting stochastic phase oscillators
is predicted to depend on the number of oscillators
in the system.  The argument is based on the stochastic differential
equation in Fourier space, that arises through linearizing the Pardari-Parisi-Zhang
equation (e.g., \citep{Barabasi_1995}), which arises in the continuum limit for the bidirectional model
(as in Eqn.~\ref{eqn:cmodel_bi4} or Eqn.~\ref{eqn:bi_cont} with the addition of noise).  
A long-wavelength cut-off arises from the number of oscillators $N$ in a chain 
and this is predicted to cause the variance of the phases to depend
upon the size of the system. 

Is the mean phase shift in the directional models sensitive
to the number of oscillators in a loop?
To answer this question we ran three integration series where we 
vary the number of oscillators $N$ and the strength of the noise $\eta$.   
The integrations parameters are listed in Table \ref{tab:noise} 
and are shown in Figure \ref{fig:NN}.  The number of oscillators integrated
are powers of 2 ranging from 32 to 4096. 
The series are denoted DWN-SerN, for the directional sinusoidal model,
the OWN-SerN, for the directional overlap model, and BWN-SerN
for a bidrectional model that is same as the DWN-SerN, except the parameter
setting asymmetry in the interactions $\mu_{c-}=0$.
Phases in these integrations are initialized to zero. 
Instead of computing the winding number, which depends on $N$,
we compute the mean phase shift $\bar \phi$ (via Eqn.~\ref{eqn:barphi}) at the end of each integration.   The mean phase shifts $\bar \phi$ are shown as images in 
Figures \ref{fig:NN}a,c,e and the 
and standard deviations of the phase shifts  $\sigma_\phi$
are shown as images in Figures \ref{fig:NN}b,d,f. 

Figures \ref{fig:NN}a,b,c,d shows that neither mean phase shift or standard deviation 
of the phase shift is sensitive to the number of oscillators in the loop for the directional 
stochastic models. 
We were curious whether this insensitivity is only a property of
the directional models. Figure \ref{fig:NN}e and f shows a bidirectional 
model.  The mean phase shift at the end of these integrations
decreases with increasing $N$, which is opposite to what is expected
if the mean phase shift scales with the phase variance which is predicted 
 via Fourier analysis to be larger in a larger system in one-dimension. 
We consider explanations for this discrepancy. 
Our numerical investigations of section \ref{sec:num} found that 
variations in mean phase shift and winding number only 
 occur when there are larger phase differences between 
neighboring oscillators.  However, when the phase differences are large
we do not expect 
the associated continuum equations to be good approximations to
the discrete models. 
Predictions based on the Kardar-Parisi-Zhang equation 
may be only be accurate in the discrete model before phase differences between
oscillators become large.   Possibly in addition, 
the term proportional to $\theta_x \theta_{xx}$
in the associated stochastic continuum equation that is only present for directional models
could give different behavior than
predicted for the Kardar-Parisi-Zhang equation which lacks this term.

\begin{figure}[ht]\centering
\includegraphics[width=3truein]{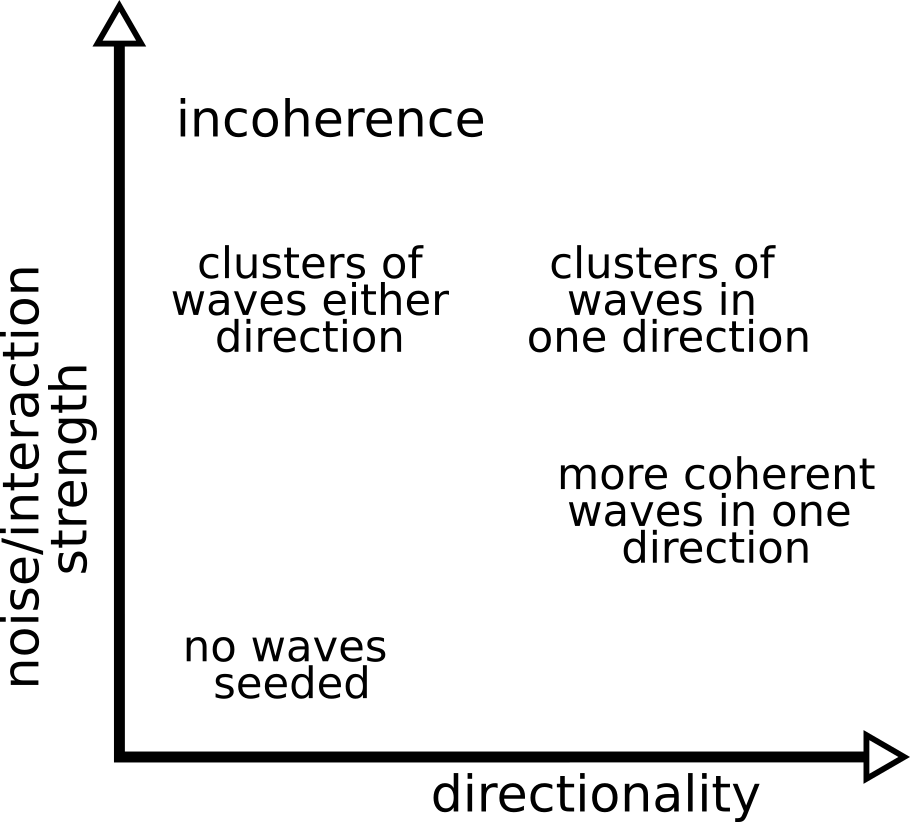}
\caption{Illustration of regimes for stochastic directional phase oscillator models.
\label{fig:regime}
}
\end{figure}

\section{Summary and Discussion}
\label{sec:sum}

We have explored dynamical systems of chains of identical phase oscillators with nearest neighbor interactions that are arranged in a loop. 
We derive a continuum partial differential equation, accurate to third order
in the separation between oscillators,  that 
 is a good approximation if phase differences are small. 
Numerical integrations and related continuum equations illustrate that directional models differ in some respects from bidirectional models (those with mirror reversal symmetry). 
We show a unidirectional model \citep{Quillen_2021} that exhibits
instability to small perturbations even for smooth initial conditions.  The instability depends on the sign of the local slope and there is a preferred direction for emergent waves.  
The instability causes growth of short wavelength perturbations
that grow to $\pm \pi$ phase differences between neighboring oscillators. 
We attribute
the instability to a third order diffusive term in the associated continuum partial
differential equation that has sign that depends on the local slope. 
We also explore a bidirectional model \citep{Niedermayer_2008}
that only exhibits instabilities 
with initial conditions that contain large phase 
differences between neighboring oscillators. 

In the continuum limit, and with a periodic boundary condition,
winding number is like a topological charge and is a conserved quantity
\citep{chakrabarti2022multiscale}.  We use 
numerical integrations of unidirectional and bidirectional discrete phase oscillator 
models with oscillators in a loop to find out whether and how winding number varies. 
We find that variations in winding number occur when there are groups
of neighboring oscillators with phase differences near $\pi$.  
Variations in winding number cease after short wavelength perturbations
decay.   The resulting long-lived state is a synchronous or wave-like 
phase locked state
with waves in either direction for the bidirectional model, but  is 
a wave-like entrained state with a preferred direction in the unidirectional model. 

The two lowest order diffusive terms ($\propto \theta_{xx}$)  in the associated continuum equations give a criterion for slope or phase shift dependent
 instability to the growth of small perturbations. 
With two phase oscillator models that let us adjust the directionality, we show that this
criterion approximately predicts when small sinusoidal perturbations can grow and
cause changes in winding number, resolving into metachronal waves. 

With adjustable directional models we explored the role of white noise
in influencing the states of these phase oscillator dynamical systems.  
An advantage of studying stochastic systems, is that the properties of 
long lived states could be insensitive to initial conditions. 
We find that as long as the strength of
the noise does not destroy the coherence of the system (as previously noted by
\citep{Solovev_2022}),  
noise helps in developing and maintaining a wave-like state through
seeding instabilities.  
The direction of the resulting waves 
 is set by the asymmetry in the oscillator interaction functions. 
We support prior studies \citep{Solovev_2022a,Solovev_2022} finding
that directional models (lacking mirror symmetry in the oscillator interaction
functions) are preferable for modeling phase oscillator systems that
robustly enter and maintain a metachronal wave collective state.
 
We find that wave generation, as seen from the winding number during
 integrations that are initialized with zero phases, is sensitive to  
the strength of coefficients in the associated continuum equation, that depend
upon derivatives of the oscillator interaction functions.  
However, in terms of ratios of these coefficients, 
the regions in parameter space where waves are found differed in the two directional 
stochastic models we explored. 

We explored sensitivity of the stochastic directional models
to initial conditions with a set slope or phase shift, corresponding to 
different initial winding numbers.  
Due to the directionality of these models, only smooth initial conditions
with either positive or negative winding number can be stable.  In the stable case, 
the final winding number can be set by the initial condition. 
The range of possible winding numbers (or initial phase shifts)
where long-lived wave-like states depend upon the initial phase shift 
 depends upon the oscillator interaction functions. 
With the stochastic bidirectional overlap model we explored, the region where
winding number is set by the initial slope is much smaller
than for the stochastic sinusoidal directional model. 
This suggests that stochastic models with strong directionality would more robustly 
enter metachronal wave states and would be less sensitive to initial conditions. 

We explored sensitivity of the stochastic directional models to the number
of oscillators in the loop.  Contrary to expectations based on Fourier analysis
of  stochastic continuum equations 
\citep{Pikovsky_2003,Solovev_2022}, 
we find that mean phase shift and the standard deviation of the phase shift,
after integration, are insensitive to the number of oscillators.  
The scaling estimated via Fourier analysis may fail because 
the continuum equation is a poor approximation to the discrete interacting
oscillator chain models when phase differences between neighboring oscillators
are large.

Given a particular level of noise, is it possible to choose phase oscillator
interaction functions that would robustly give long-lived metachronal wave states?
Based on our exploration of two bidirectional models, we roughly
illustrate regimes of collective behavior in Figure \ref{fig:regime}.
Because noise seeds perturbations that can cause variations
in winding number,  the strength of the symmetric interactions must not be so strong 
that perturbations are damped rapidly. 
For waves to be formed, the strength of the directionality, 
set by asymmetry in the interactions,  
should be sufficiently strong that small perturbations are unstable and can
grow to large enough values to change the winding number. 
The interaction strengths cannot be so weak that noise causes
generated waves to completely loose coherence. 
With sufficiently strong asymmetric interactions, we suspect that a stochastic
 model is relatively insensitive to initial conditions, in the sense
 that only for a small range of smooth and sloped initial conditions would
 the system's long-lived states depend upon the initial slope. 
 Most of our integrations of stochastic directional models 
exhibited clusters of oscillators in wave-like states,
but with waves in a particular direction, rather than a coherent wave that spanned
the entire system. 
If metachronal waves in biological systems rely on noise to seed waves,  
then there might be a trade-off between wave 
coherence and sensitivity to initial conditions.  
Robustly generated states consisting of clusters of oscillators driving waves,
may be functionally preferable to unreliably generated but coherent wave states. 

Future study could improve upon our understanding of
how the characteristics of the interaction functions and
the nature of stochastic perturbations affect wave-generation
(as seen from statistics of the winding number or mean phase shift),  
 the coherence of the generated waves and the sensitivity to initial conditions. 
In biological systems, statistics of wave speeds, variations in wave speed
and coherence of clusters of oscillators showing coherent phase shifts,  
might pin down the role of noise in seeding 
and maintaining metachronal wave states and better constrain the
nature of interactions between the oscillators. 

For the models we explored, large phases differences led to instability
which resolved with changes in winding number.
Using a perturbative analysis and by computing eigenvalues of a circulant matrix,
\citet{Niedermayer_2008} shows explicitly for their bidirectional model (Eqn.~\ref{eqn:bi})  that a phase locked state with a phase shift
above $\pi/2$ between each oscillators would be unstable. 
It is more difficult to similarly assess (via a perturbative linear analysis) 
 the stability of an entrained state with a large
phase shift in the unidirectional model (Eqn.~\ref{eqn:uni}) because this requires averaging over the oscillation period of the entrained state. 
Instability when the phase differences are large is not necessarily sufficient
for robust formation of waves.  To form waves in particular direction,
 jumps in phase should resolve in a particular direction.  In other words, the phase
difference should preferentially cross $\pi$ in either the clockwise or counter
clockwise direction (as is true for our unidirectional model but not
the bidirectional model) to ensure that waves form moving in a specific direction.
Perhaps insight can be sought by studying propagation of phase kinks in other settings
(e.g., \citep{Chate_1999}).
 
We gained intuition by looking at the partial differential equation
that approximates an oscillator chain model. 
However the continuum equations, which are derived in
the limit of small phase shift,  do not help us predict how jumps in phase 
evolve.  We have noticed that a single large jump in phase is not sufficient
to cause a change in winding number, rather at least two large phase jumps
in sequence are required.   If the collective behavior of the model
is sensitive to the dynamics of strong short wavelength 
perturbations, then 
 desirable models for actual biological systems should be 
 good approximations in both short and long wavelength limits.  
The interaction functions for the models we have explored do not contain more than one minimum or maximum.  There may be constraints on the shapes of the periodic functions that permit changes in winding number in the associated phase oscillator
dynamical systems. 

The partial differential equation
that approximates a directional oscillator chain model in the continuum limit
resembles the Kardar-Parisi-Zhang equation but with the addition of 
a third order non-linear term that is proportional to $\theta_x \theta_{xx}$. 
The slope dependent instability we see in the discrete models 
suggests that the stochastic version of this 
differential equation may exhibit novel phenomena 
that is not present with the Kardar-Parisi-Zhang equation.  

Hydrodynamic interaction models for cilia can be directional 
(e.g., \cite{Solovev_2022,chakrabarti2022multiscale}).  
We have shown here that there are directional models that 
exhibit changes in winding number, but that does not necessarily mean
that this class of models is appropriate for real biological systems. 
Quantitative measurements describing  
the coherence of generated waves may help differentiate between
stochastic models. 
By evaluating the strength and shape 
of the symmetric and antisymmetric hydrodynamic interaction functions 
for cilia it may be possible 
to determine if instabilities mediate changes in the winding number 
so that these systems can preferentially enter and maintain sufficiently  
coherent wave-like states.   
If this is not the case, then alternate physical
mechanisms are required to account for the formation of
metachronal waves.    For example,  
physical gaps in ciliated systems, which relax
the constraint of a periodic boundary condition,  
could facilitate metachronal wave formation, as proposed by  
\citet{chakrabarti2022multiscale}.
Additional physical mechanisms for oscillator interaction \citep{Narematsu_2015}, 
 variations in properties and additional degrees of freedom describing the 
 individual oscillators could influence 
 the collective behavior of these interacting systems. 

\begin{acknowledgments}
We thank Nathan Skerrett and Brato Chakrabarti for helpful discussions. 
\end{acknowledgments}

\bibliography{chains}

\begin{thebibliography}{33}
\expandafter\ifx\csname natexlab\endcsname\relax\def\natexlab#1{#1}\fi
\expandafter\ifx\csname bibnamefont\endcsname\relax
  \def\bibnamefont#1{#1}\fi
\expandafter\ifx\csname bibfnamefont\endcsname\relax
  \def\bibfnamefont#1{#1}\fi
\expandafter\ifx\csname citenamefont\endcsname\relax
  \def\citenamefont#1{#1}\fi
\expandafter\ifx\csname url\endcsname\relax
  \def\url#1{\texttt{#1}}\fi
\expandafter\ifx\csname urlprefix\endcsname\relax\def\urlprefix{URL }\fi
\providecommand{\bibinfo}[2]{#2}
\providecommand{\eprint}[2][]{\url{#2}}

\bibitem[{\citenamefont{Wiener}(1958)}]{Wiener_1958}
\bibinfo{author}{\bibfnamefont{N.}~\bibnamefont{Wiener}},
  \emph{\bibinfo{title}{Nonlinear Problems in Random Theory}}
  (\bibinfo{publisher}{MIT Press, Cambridge, MA}, \bibinfo{year}{1958}).

\bibitem[{\citenamefont{Kuramoto}(1975)}]{Kuramoto_1975}
\bibinfo{author}{\bibfnamefont{Y.}~\bibnamefont{Kuramoto}}, in
  \emph{\bibinfo{booktitle}{Int. Symposium on Mathematical Problems in
  Theoretical Physics}}, edited by
  \bibinfo{editor}{\bibfnamefont{H.}~\bibnamefont{Araki}}
  (\bibinfo{publisher}{Springer}, \bibinfo{year}{1975}),
  vol.~\bibinfo{volume}{39} of \emph{\bibinfo{series}{Lecture Notes in
  Physics}}, pp. \bibinfo{pages}{420--422}.

\bibitem[{\citenamefont{Kuramoto and Nishikawa}(1987)}]{Kuramoto_1987}
\bibinfo{author}{\bibfnamefont{Y.}~\bibnamefont{Kuramoto}} \bibnamefont{and}
  \bibinfo{author}{\bibfnamefont{I.}~\bibnamefont{Nishikawa}},
  \bibinfo{journal}{Journal of Statistical Physics}
  \textbf{\bibinfo{volume}{49}}, \bibinfo{pages}{569} (\bibinfo{year}{1987}),
  \urlprefix\url{https://doi.org/10.1007%2Fbf01009349}.

\bibitem[{\citenamefont{Pikovsky et~al.}(2003)\citenamefont{Pikovsky, Kurths,
  Rosenblum, and Kurths}}]{Pikovsky_2003}
\bibinfo{author}{\bibfnamefont{A.}~\bibnamefont{Pikovsky}},
  \bibinfo{author}{\bibfnamefont{J.}~\bibnamefont{Kurths}},
  \bibinfo{author}{\bibfnamefont{M.}~\bibnamefont{Rosenblum}},
  \bibnamefont{and} \bibinfo{author}{\bibfnamefont{J.}~\bibnamefont{Kurths}},
  \emph{\bibinfo{title}{Synchronization: a universal concept in non-linear
  sciences, 12}}, no.~\bibinfo{number}{12} in \bibinfo{series}{Cambridge
  Nonlinear Science Series} (\bibinfo{publisher}{Cambridge university press},
  \bibinfo{year}{2003}).

\bibitem[{\citenamefont{Strogatz}(2012)}]{Strogatz_2012}
\bibinfo{author}{\bibfnamefont{S.}~\bibnamefont{Strogatz}},
  \emph{\bibinfo{title}{Sync: How Order Emerges From Chaos In the Universe,
  Nature, and Daily Life}} (\bibinfo{publisher}{Hachette Books},
  \bibinfo{year}{2012}), ISBN \bibinfo{isbn}{9781401304461},
  \urlprefix\url{https://books.google.com/books?id=vHw44RSiOCwC}.

\bibitem[{\citenamefont{Chakrabarti et~al.}(2022)\citenamefont{Chakrabarti,
  Furthauer, and Shelley}}]{chakrabarti2022multiscale}
\bibinfo{author}{\bibfnamefont{B.}~\bibnamefont{Chakrabarti}},
  \bibinfo{author}{\bibfnamefont{S.}~\bibnamefont{Furthauer}},
  \bibnamefont{and} \bibinfo{author}{\bibfnamefont{M.~J.}
  \bibnamefont{Shelley}}, \bibinfo{journal}{Proceedings of the National Academy
  of Sciences} \textbf{\bibinfo{volume}{119}}, \bibinfo{pages}{e2113539119}
  (\bibinfo{year}{2022}).

\bibitem[{\citenamefont{Tamm}(1972)}]{Tamm_1972}
\bibinfo{author}{\bibfnamefont{S.~L.} \bibnamefont{Tamm}},
  \bibinfo{journal}{The Journal of Cell Biology} \textbf{\bibinfo{volume}{55}},
  \bibinfo{pages}{250} (\bibinfo{year}{1972}).

\bibitem[{\citenamefont{Sleigh et~al.}(1988)\citenamefont{Sleigh, Blake, and
  Liron}}]{Sleigh_1988}
\bibinfo{author}{\bibfnamefont{M.~A.} \bibnamefont{Sleigh}},
  \bibinfo{author}{\bibfnamefont{J.~R.} \bibnamefont{Blake}}, \bibnamefont{and}
  \bibinfo{author}{\bibfnamefont{N.}~\bibnamefont{Liron}},
  \bibinfo{journal}{American Review of Respiratory Disease}
  \textbf{\bibinfo{volume}{137}}, \bibinfo{pages}{726} (\bibinfo{year}{1988}).

\bibitem[{\citenamefont{Afzelius}(2004)}]{Afzelius_2004}
\bibinfo{author}{\bibfnamefont{B.~A.} \bibnamefont{Afzelius}},
  \bibinfo{journal}{Journal of Pathology} \textbf{\bibinfo{volume}{204}},
  \bibinfo{pages}{470} (\bibinfo{year}{2004}).

\bibitem[{\citenamefont{Faubel et~al.}(2016)\citenamefont{Faubel, Westendorf,
  Bodenschatz, and Eichele}}]{faubel2016cilia}
\bibinfo{author}{\bibfnamefont{R.}~\bibnamefont{Faubel}},
  \bibinfo{author}{\bibfnamefont{C.}~\bibnamefont{Westendorf}},
  \bibinfo{author}{\bibfnamefont{E.}~\bibnamefont{Bodenschatz}},
  \bibnamefont{and} \bibinfo{author}{\bibfnamefont{G.}~\bibnamefont{Eichele}},
  \bibinfo{journal}{Science} \textbf{\bibinfo{volume}{353}},
  \bibinfo{pages}{176} (\bibinfo{year}{2016}).

\bibitem[{\citenamefont{Peshkov et~al.}(2022)\citenamefont{Peshkov, McGaffigan,
  and Quillen}}]{Peshkov_2022}
\bibinfo{author}{\bibfnamefont{A.}~\bibnamefont{Peshkov}},
  \bibinfo{author}{\bibfnamefont{S.}~\bibnamefont{McGaffigan}},
  \bibnamefont{and} \bibinfo{author}{\bibfnamefont{A.~C.}
  \bibnamefont{Quillen}}, \bibinfo{journal}{Soft Matter}
  \textbf{\bibinfo{volume}{18}}, \bibinfo{pages}{1174} (\bibinfo{year}{2022}).

\bibitem[{\citenamefont{Quillen et~al.}(2021)\citenamefont{Quillen, Peshkov,
  Wright, and McGaffigan}}]{Quillen_2021}
\bibinfo{author}{\bibfnamefont{A.~C.} \bibnamefont{Quillen}},
  \bibinfo{author}{\bibfnamefont{A.}~\bibnamefont{Peshkov}},
  \bibinfo{author}{\bibfnamefont{E.}~\bibnamefont{Wright}}, \bibnamefont{and}
  \bibinfo{author}{\bibfnamefont{S.}~\bibnamefont{McGaffigan}},
  \bibinfo{journal}{Phys. Rev. E} \textbf{\bibinfo{volume}{104}},
  \bibinfo{pages}{014412} (\bibinfo{year}{2021}).

\bibitem[{\citenamefont{Wiley et~al.}(2006)\citenamefont{Wiley, Strogatz, and
  Girvan}}]{Wiley_2006}
\bibinfo{author}{\bibfnamefont{D.~A.} \bibnamefont{Wiley}},
  \bibinfo{author}{\bibfnamefont{S.~H.} \bibnamefont{Strogatz}},
  \bibnamefont{and} \bibinfo{author}{\bibfnamefont{M.}~\bibnamefont{Girvan}},
  \bibinfo{journal}{Chaos} \textbf{\bibinfo{volume}{16}},
  \bibinfo{pages}{015103} (\bibinfo{year}{2006}).

\bibitem[{\citenamefont{Tilles et~al.}(2011)\citenamefont{Tilles, Ferreira, and
  Cerdeira}}]{Tilles_2011}
\bibinfo{author}{\bibfnamefont{P.~F.~C.} \bibnamefont{Tilles}},
  \bibinfo{author}{\bibfnamefont{F.~F.} \bibnamefont{Ferreira}},
  \bibnamefont{and} \bibinfo{author}{\bibfnamefont{H.~A.}
  \bibnamefont{Cerdeira}}, \bibinfo{journal}{Physical Review E}
  \textbf{\bibinfo{volume}{83}} (\bibinfo{year}{2011}),
  \urlprefix\url{https://doi.org/10.1103%2Fphysreve.83.066206}.

\bibitem[{\citenamefont{D{\'e}nes et~al.}(2019)\citenamefont{D{\'e}nes,
  S{\'a}ndor, and N{\'e}da}}]{Denes_2019}
\bibinfo{author}{\bibfnamefont{K.}~\bibnamefont{D{\'e}nes}},
  \bibinfo{author}{\bibfnamefont{B.}~\bibnamefont{S{\'a}ndor}},
  \bibnamefont{and} \bibinfo{author}{\bibfnamefont{Z.}~\bibnamefont{N{\'e}da}},
  \bibinfo{journal}{Communications in Nonlinear Science and Numerical
  Simulation} \textbf{\bibinfo{volume}{78}}, \bibinfo{pages}{104868}
  (\bibinfo{year}{2019}), ISSN \bibinfo{issn}{1007-5704},
  \urlprefix\url{http://www.sciencedirect.com/science/article/pii/S1007570419301881}.

\bibitem[{\citenamefont{Niedermayer et~al.}(2008)\citenamefont{Niedermayer,
  Eckhardt, and Lenz}}]{Niedermayer_2008}
\bibinfo{author}{\bibfnamefont{T.}~\bibnamefont{Niedermayer}},
  \bibinfo{author}{\bibfnamefont{B.}~\bibnamefont{Eckhardt}}, \bibnamefont{and}
  \bibinfo{author}{\bibfnamefont{P.}~\bibnamefont{Lenz}},
  \bibinfo{journal}{Chaos: An Interdisciplinary Journal of Nonlinear Science}
  \textbf{\bibinfo{volume}{18}}, \bibinfo{pages}{037128}
  (\bibinfo{year}{2008}), \urlprefix\url{https://doi.org/10.1063%2F1.2956984}.

\bibitem[{\citenamefont{Solovev and
  Friedrich}(2022{\natexlab{a}})}]{Solovev_2022}
\bibinfo{author}{\bibfnamefont{A.}~\bibnamefont{Solovev}} \bibnamefont{and}
  \bibinfo{author}{\bibfnamefont{B.~M.} \bibnamefont{Friedrich}},
  \bibinfo{journal}{Chaos} \textbf{\bibinfo{volume}{32}},
  \bibinfo{pages}{013124} (\bibinfo{year}{2022}{\natexlab{a}}).

\bibitem[{\citenamefont{Brumley et~al.}(2012)\citenamefont{Brumley, Polin,
  Pedley, and Goldstein}}]{Brumley_2012}
\bibinfo{author}{\bibfnamefont{D.~R.} \bibnamefont{Brumley}},
  \bibinfo{author}{\bibfnamefont{M.}~\bibnamefont{Polin}},
  \bibinfo{author}{\bibfnamefont{T.~J.} \bibnamefont{Pedley}},
  \bibnamefont{and} \bibinfo{author}{\bibfnamefont{R.~E.}
  \bibnamefont{Goldstein}}, \bibinfo{journal}{Physics Review Letters}
  \textbf{\bibinfo{volume}{109}}, \bibinfo{pages}{268102}
  (\bibinfo{year}{2012}).

\bibitem[{\citenamefont{Ma et~al.}(2014)\citenamefont{Ma, Klindt, Riedel-Kruse,
  J\"ulicher, and Friedrich}}]{Ma_2014}
\bibinfo{author}{\bibfnamefont{R.}~\bibnamefont{Ma}},
  \bibinfo{author}{\bibfnamefont{G.}~\bibnamefont{Klindt}},
  \bibinfo{author}{\bibfnamefont{I.}~\bibnamefont{Riedel-Kruse}},
  \bibinfo{author}{\bibfnamefont{F.}~\bibnamefont{J\"ulicher}},
  \bibnamefont{and}
  \bibinfo{author}{\bibfnamefont{B.}~\bibnamefont{Friedrich}},
  \bibinfo{journal}{Phys Rev Lett.} \textbf{\bibinfo{volume}{113}},
  \bibinfo{pages}{048101} (\bibinfo{year}{2014}).

\bibitem[{\citenamefont{Acebron et~al.}(2005)\citenamefont{Acebron, Bonilla,
  Vicente, Ritort, and Spigler}}]{Acebron_2005}
\bibinfo{author}{\bibfnamefont{J.~A.} \bibnamefont{Acebron}},
  \bibinfo{author}{\bibfnamefont{L.~L.} \bibnamefont{Bonilla}},
  \bibinfo{author}{\bibfnamefont{C.~J.~P.} \bibnamefont{Vicente}},
  \bibinfo{author}{\bibfnamefont{F.}~\bibnamefont{Ritort}}, \bibnamefont{and}
  \bibinfo{author}{\bibfnamefont{R.}~\bibnamefont{Spigler}},
  \bibinfo{journal}{Reviews of Modern Physics} \textbf{\bibinfo{volume}{77}},
  \bibinfo{pages}{137} (\bibinfo{year}{2005}).

\bibitem[{\citenamefont{Ermentrout and Kopell}(1986)}]{Ermentrout_1986}
\bibinfo{author}{\bibfnamefont{G.}~\bibnamefont{Ermentrout}} \bibnamefont{and}
  \bibinfo{author}{\bibfnamefont{N.}~\bibnamefont{Kopell}},
  \bibinfo{journal}{Comm. Pure Appl. Math.} \textbf{\bibinfo{volume}{49}},
  \bibinfo{pages}{623} (\bibinfo{year}{1986}).

\bibitem[{\citenamefont{Ermentrout and Kopell}(1990)}]{Ermentrout_1990}
\bibinfo{author}{\bibfnamefont{G.}~\bibnamefont{Ermentrout}} \bibnamefont{and}
  \bibinfo{author}{\bibfnamefont{N.}~\bibnamefont{Kopell}},
  \bibinfo{journal}{SIAM J. Appl. Math.} \textbf{\bibinfo{volume}{50}},
  \bibinfo{pages}{1014} (\bibinfo{year}{1990}).

\bibitem[{\citenamefont{Ren and Ermentrout}(2000)}]{Ren_2000}
\bibinfo{author}{\bibfnamefont{L.}~\bibnamefont{Ren}} \bibnamefont{and}
  \bibinfo{author}{\bibfnamefont{B.}~\bibnamefont{Ermentrout}},
  \bibinfo{journal}{Physica D: Nonlinear Phenomena}
  \textbf{\bibinfo{volume}{143}}, \bibinfo{pages}{56} (\bibinfo{year}{2000}),
  \urlprefix\url{https://doi.org/10.1016%2Fs0167-2789%2800%2900096-8}.

\bibitem[{\citenamefont{Elgeti and Gompper}(2013)}]{Elgeti_2013}
\bibinfo{author}{\bibfnamefont{J.}~\bibnamefont{Elgeti}} \bibnamefont{and}
  \bibinfo{author}{\bibfnamefont{G.}~\bibnamefont{Gompper}},
  \bibinfo{journal}{PNAS; Proceedings of the National Academy of Sciences}
  \textbf{\bibinfo{volume}{110}}, \bibinfo{pages}{4470} (\bibinfo{year}{2013}).

\bibitem[{\citenamefont{Muruganandam et~al.}(2008)\citenamefont{Muruganandam,
  Ferreira, El-Nashar, and Cerdeira}}]{Muruganandam_2008}
\bibinfo{author}{\bibfnamefont{P.}~\bibnamefont{Muruganandam}},
  \bibinfo{author}{\bibfnamefont{F.~F.} \bibnamefont{Ferreira}},
  \bibinfo{author}{\bibfnamefont{H.~F.} \bibnamefont{El-Nashar}},
  \bibnamefont{and} \bibinfo{author}{\bibfnamefont{H.~A.}
  \bibnamefont{Cerdeira}}, \bibinfo{journal}{Pramana}
  \textbf{\bibinfo{volume}{70}}, \bibinfo{pages}{1143} (\bibinfo{year}{2008}),
  \urlprefix\url{https://doi.org/10.1007%2Fs12043-008-0119-8}.

\bibitem[{\citenamefont{Aeyels and Rogge}(2004)}]{Aeyels_2004}
\bibinfo{author}{\bibfnamefont{D.}~\bibnamefont{Aeyels}} \bibnamefont{and}
  \bibinfo{author}{\bibfnamefont{J.~A.} \bibnamefont{Rogge}},
  \bibinfo{journal}{Progress of Theoretical Physics}
  \textbf{\bibinfo{volume}{112}}, \bibinfo{pages}{921} (\bibinfo{year}{2004}).

\bibitem[{\citenamefont{Zheng et~al.}(1998)\citenamefont{Zheng, Hu, and
  Hu}}]{Zheng_1998}
\bibinfo{author}{\bibfnamefont{Z.}~\bibnamefont{Zheng}},
  \bibinfo{author}{\bibfnamefont{G.}~\bibnamefont{Hu}}, \bibnamefont{and}
  \bibinfo{author}{\bibfnamefont{B.}~\bibnamefont{Hu}}, \bibinfo{journal}{Phys.
  Rev. Lett.} \textbf{\bibinfo{volume}{81}}, \bibinfo{pages}{5318}
  (\bibinfo{year}{1998}).

\bibitem[{\citenamefont{Ottino-L{\"o}ffler and
  Strogatz}(2016)}]{Ottino_Loffler_2016}
\bibinfo{author}{\bibfnamefont{B.}~\bibnamefont{Ottino-L{\"o}ffler}}
  \bibnamefont{and} \bibinfo{author}{\bibfnamefont{S.~H.}
  \bibnamefont{Strogatz}}, \bibinfo{journal}{Physical Review E}
  \textbf{\bibinfo{volume}{94}} (\bibinfo{year}{2016}),
  \urlprefix\url{https://doi.org/10.1103%2Fphysreve.94.062203}.

\bibitem[{\citenamefont{Cross and Greenside}(2009)}]{Cross_2009}
\bibinfo{author}{\bibfnamefont{M.}~\bibnamefont{Cross}} \bibnamefont{and}
  \bibinfo{author}{\bibfnamefont{H.}~\bibnamefont{Greenside}},
  \emph{\bibinfo{title}{Pattern Formation and Dynamics in Nonequilibrium
  Systems}} (\bibinfo{publisher}{Cambridge Univ. Press, Cambridge, U. K.},
  \bibinfo{year}{2009}).

\bibitem[{\citenamefont{Barab\'asi and Stanley}(1995)}]{Barabasi_1995}
\bibinfo{author}{\bibfnamefont{A.-L.} \bibnamefont{Barab\'asi}}
  \bibnamefont{and} \bibinfo{author}{\bibfnamefont{H.~E.}
  \bibnamefont{Stanley}}, \emph{\bibinfo{title}{Fractal Concepts in Surface
  Growth}} (\bibinfo{publisher}{Cambridge University Press, Cambridge},
  \bibinfo{year}{1995}).

\bibitem[{\citenamefont{Solovev and
  Friedrich}(2022{\natexlab{b}})}]{Solovev_2022a}
\bibinfo{author}{\bibfnamefont{A.}~\bibnamefont{Solovev}} \bibnamefont{and}
  \bibinfo{author}{\bibfnamefont{B.~M.} \bibnamefont{Friedrich}},
  \bibinfo{journal}{New Journal of Physics} \textbf{\bibinfo{volume}{24}},
  \bibinfo{pages}{013015} (\bibinfo{year}{2022}{\natexlab{b}}).

\bibitem[{\citenamefont{Chat\'e et~al.}(1999)\citenamefont{Chat\'e, Pikovsky,
  and Rudzick}}]{Chate_1999}
\bibinfo{author}{\bibfnamefont{H.}~\bibnamefont{Chat\'e}},
  \bibinfo{author}{\bibfnamefont{A.}~\bibnamefont{Pikovsky}}, \bibnamefont{and}
  \bibinfo{author}{\bibfnamefont{O.}~\bibnamefont{Rudzick}},
  \bibinfo{journal}{Physica D} \textbf{\bibinfo{volume}{131}},
  \bibinfo{pages}{17} (\bibinfo{year}{1999}).

\bibitem[{\citenamefont{Narematsu et~al.}(2015)\citenamefont{Narematsu, Quek,
  Chiam, and Iwadate}}]{Narematsu_2015}
\bibinfo{author}{\bibfnamefont{N.}~\bibnamefont{Narematsu}},
  \bibinfo{author}{\bibfnamefont{R.}~\bibnamefont{Quek}},
  \bibinfo{author}{\bibfnamefont{K.-H.} \bibnamefont{Chiam}}, \bibnamefont{and}
  \bibinfo{author}{\bibfnamefont{Y.}~\bibnamefont{Iwadate}},
  \bibinfo{journal}{Cytoskeleton} \textbf{\bibinfo{volume}{72}},
  \bibinfo{pages}{633} (\bibinfo{year}{2015}).

\end{thebibliography}

\end{document}